\documentclass[12pt]{amsart}

\usepackage{subfiles}
\usepackage{fullpage} 
\usepackage{graphicx} 
\usepackage{prefix}
\usepackage{amsmath}
\usepackage{amsfonts}
\usepackage{tikz-cd}
\usepackage{braket} 
\usepackage{quiver}

\title{Transformations of Hilbert Module Reference Frames}
\author{Lucus Brady}

\address{Department of Mathematical Sciences\\Montana State University\\Bozeman, MT 59717}
\email{lucusbrady@montana.edu}
\date{\today}

\begin{document}

\maketitle

\begin{abstract}

In this paper, we provide a functorial characterization of Hilbert module classical and quantum reference frames. Using the functorial characterization, we construct a category for equivariant couplings and Hilbert module reference frames. We prove the construction of the reference frame system from the equivariant coupling data yields a functor. We also deduce a ``unitary" groupoid as frame transformations for  Hilbert module reference frames of a fixed Hilbert module dynamical system. We provide some examples of these frame transformations in the case of the indistinguishable classical two state switch and in the case of Hilbert spaces.

\end{abstract}

\tableofcontents

\section{Introduction}

In this paper, we continue the development of a Hilbert module framework for reference frames as described in \cite{brady2026hilbertmoduleapproachclassical}. In the prior work, we motivated and developed a framework for reference frames utilizing Hilbert modules and correspondences of $C^*$-dynamical systems. Conceptually, the framework treats a reference frame as a physical system which acts as a background for another physical system. Due to the asymmetric relationship between the two systems, our model is designed to remember the  reference system's role so that observables in the joint system are relative to the background system. Motivated in the classical setting, the relationship between the two systems is naturally codified using a non-degenerate positive equivariant correspondence and Hilbert module dynamical system. We refer to this data as an equivariant coupling. Applying an equivariant version of both the KSGNS construction and interior tensor product to the data yields the relation joint system which we refer to as the reference frame system. In addition to the equivariant coupling, a choice of relativization map is required to describe how observables in the main system are conditioned/relativized to invariant relational observables in the reference frame system. We refer to the data of an equivariant coupling and relativization map as a Hilbert module reference frame.

For a fixed equivariant coupling, multiple relativization maps can exist (see the examples in Section 3.5 of \cite{brady2026hilbertmoduleapproachclassical}). Furthermore, for any two fixed $C^*$-algebras, multiple equivariant couplings can exist between the two systems (see the examples in Section 3.5 of \cite{brady2026hilbertmoduleapproachclassical}). Due to the degeneracy of the data, we want to understand how these different choices are related, and we want to understand in what sense these different choices describe the same reference frame system. To answer both of these structural questions, we develop frame transformations for Hilbert module reference frames.

In our framework, frame transformations are defined at the level of the equivariant couplings and relativization map. As a result, a frame transformation codifies the ways of changing the reference system and equivariant coupling while preserving the relational observables. Due to the data involved for defining a Hilbert module reference frame, there is no obvious choice for defining frame transformations or even morphisms between two reference frames.

To resolve the ambiguity of morphisms, we use a functorial characterization of an equivariant coupling (Corollary \ref{Cor: functorial characterization of equivariant couplings}) to identify equivariant couplings as objects in a functor category, denoted as $\textnormal{EqCpl}$. While these morphisms describe relationships between equivariant couplings, we still need to include the relativization map data. To describe how the morphisms of $\textnormal{EqCpl}$ interact with the relativization maps, we prove the construction of a reference frame system from the data of an equivariant coupling yields a functor on the subcategory $\textnormal{EqCpl}_{*\textnormal{-iso}}$ of $\textnormal{EqCpl}$ to $\textnormal{C}^*_{\textnormal{CP}}$ (see Theorem \ref{Theorem: functoriality of the RF construction}), the category of $C^*$-algebras with morphisms given by strict completely positive maps. Note that the restriction to the subcategory is a result of the functoriality of the KSGNS construction. Using the functor from the reference frame construction, we bootstrap our way to define a category, denoted by $\textnormal{HMRF}$, whose objects are Hilbert module reference frames.

Fixing a Hilbert module dynamical system, we pass to a subcategory of  $\textnormal{HMRF}$ whose objects represent Hilbert module reference frames for the fixed dynamical system. As a first pass, the groupoid/core of this subcategory is taken to define frame transformations. By requiring the induced morphisms between reference frame systems from a frame transformation to induce a map at the level of states, we are required to restrict to a sub-groupoid whose morphisms induce unitary maps (Theorem \ref{Theorem: unitary result 1} and Theorem \ref{Theorem: unitary result 2}). Thus, we led to a ``unitary" groupoid for frame transformations. These unitary frame transformations induce $*$-algebra isomorphisms at the level of the reference frame systems (Corollary \ref{Cor: properties of the induced maps from FT}).

While the abstract view point of equivariant couplings and Hilbert module reference frames is  introduced to define morphisms and frame transformations, there are concrete ways of obtaining both morphisms of equivariant couplings as well as unitary frame transformations. In fact, we prove a general template for constructing both morphisms (Corollary \ref{Cor: building morphism in EqCpl} and Corollary \ref{Cor: construction unitary frame transformations}). Using these templates, we describe two classes of examples of unitary frame transformation. The first class of examples come from when both the main system and reference system are given by a classical indistinguishable two state system. The second class of examples come from when both the modules for the main system and correspondence are Hilbert spaces. In both settings, we provide a characterization of unitary frame transformations. From the characterization in the two state switch setting, we see how the choice of relativization map can allow for additional symmetry at the level of unitary frame transformations. This presents the question how relativization maps relate to the ideas of spontaneous symmetry breaking that we hope to answer in a future paper.

We now outline the layout of this paper. In Section 2, we build the category $\textnormal{HMod}$, and we prove Hilbert dynamical systems are in one-to-one correspondence with unitary valued functors $BG\to \textnormal{HMod}$. In Section 3, we utilize the functorial characterization of equivariant correspondences and Hilbert module dynamical systems to develop the category $\textnormal{EqCpl}$ which contains all equivariant couplings as objects. In Section 4, we prove the reference frame construction from an equivariant coupling provides a functor from the subcategory $\textnormal{EqCpl}_{*\textnormal{-iso}}$ to $\textnormal{C}^*_{\textnormal{CP}}$. Finally, in Section 5, we use the functoriality of the reference frame construction to develop the category $\textnormal{HMRF}$ which contains Hilbert module reference frames as objects. Using this category and properties of the induced morphisms from the reference frame construction, we deduce unitary frame transformation as an appropriate notion of transformations for Hilbert module reference frames.

We conclude the introduction with a comment on how this paper fits into a research program for building a $C^*$-algebra model of quantum reference frames for algebraic quantum field theories. First, in the paper \cite{brady2026functorialityksgnsconstructionintertwiners}, we built a category to view strict positive $C^*$-correspondences between $C^*$-algebras as objects with morphisms given by intertwiners. Using this category, we proved the KSGNS construction defines an idempotent endofunctor. As an application, we proved an equivariant KSGNS construction for strict positive equivariant $C^*$-correspondences. In the paper \cite{brady2026hilbertmoduleapproachclassical}, we built a model of classical and quantum reference frames using Hilbert modules and equivariant correspondences. Using this framework and the categorical formulation in \cite{brady2026functorialityksgnsconstructionintertwiners}, we provide in this paper a functorial/categorical perspective for Hilbert module reference frames leading to a notion frame transformation. In a future paper, we will utilize the categorical construction from this paper to build examples of Hilbert module reference frames for algebraic quantum field theories.

\subsection*{Acknowledgments} I would like to thank my advisor, Ryan Grady, for his support and encouragement as I have worked on this project.

\subsection*{Funding} Part of this project was completed while I was sponsored in part by the Air Force Research Laboratory under Agreement Number FA8750-24-1-1019. The U.S.Government is authorized to reproduce and distribute reprints for government purposes not withstanding any copyright notation thereon. Any opinions, findings, and conclusions or recommendations expressed in this material are those of the author and do not necessarily reflect the view of the funder.

\subsection*{Background Assumptions} Throughout this paper, we assume the reader is familiar with completely positive maps and Hilbert modules. For information about completely positive maps, see \cite{Paulsen1982} or \cite{blackadar2006operator}. For information about Hilbert modules, see \cite{lance1995hilbert}, \cite{raeburn1998morita}, or \cite{blackadar2006operator}. We also assume the reader is familiar with correspondences of $C^*$-algebras. For a review of these objects, see Section 2 of \cite{brady2026hilbertmoduleapproachclassical}. Additionally, we assume the reader is familiar with the Hilbert module description of classical and quantum reference frames as described in \cite{brady2026hilbertmoduleapproachclassical}. Lastly, we assume the reader is familiar with the categorical development of $\textnormal{PosCor}$ and $\textnormal{PosCor}_{*\textnormal{-alg}}$ from \cite{brady2026functorialityksgnsconstructionintertwiners}, as well as the categorical formulation of equivariant correspondences from \cite{brady2026functorialityksgnsconstructionintertwiners}.

\subsection*{Notation and Conventions}

\begin{itemize}
    \item  We take a $C^*$-dynamical system as the data $(A, G, \alpha)$ where $A$ is a $C^*$-algebra, $G$ is a group, and $\alpha:G\to \textnormal{Aut}_{*\textnormal{-alg}}(A)$ is a group homomorphism. When $G$ is a topological group and $\alpha$ is continuous with respect to the point-norm topology, we refer to the triple as a continuous $C^*$-dynamical system.
    \item Given a Hilbert $A$-module $E$, a Hilbert $B$-module $F$,  and a $*$-algebra homomorphism $\rho:A\to \mathcal{L}(F)$, we denote the interior tensor product of $E$ and $F$ with respect to $\rho$ as $E\otimes_\rho F$. Given $x\in E$ and $y\in F$, we denote the equivalence class of $x\otimes y$ in $E\otimes_\rho F$ as $x\dot{\otimes}y$.
    \item Given a $C^*$-algebra $A$, a Hilbert $B$-module $F$,  and a strict completely positive map $\phi:A\to \mathcal{L}(F)$, we denote the triple from the KSGNS construction as $(F_\phi, \pi_\phi, V_\phi)$ where $F_\phi=A\otimes_\phi F$. Given $a\in A$ and $x\in F$, we denote the equivalence class of $a\otimes x$ in $F_\phi$ as $a\dot{\otimes}x$. 
    \item We denote $\textnormal{NDPosCor}_{*\textnormal{-alg}}$ as the full subcategory of $\textnormal{PosCor}_{*\textnormal{-alg}}$ from \cite{brady2026functorialityksgnsconstructionintertwiners} where objects are triples $(A, B, (E_B, \phi))$ where $(E_B, \phi)$ is a non-degenerate positive $C^*$-correspondence from $A$ to $B$.
    \item Given a $C^*$-algebra $A$, we denote the category of Hilbert $A$-modules as $\textnormal{HMod}_A$; that is, the category whose objects are given by Hilbert $A$-modules, and whose morphisms are adjointable $A$-linear maps.
    \item  We denote $C^*\textnormal{-Alg}$ as the category whose objects are $C^*$-algebras, and where a morphism $A\to B$ is given by a non-degenerate $*$-algebra homomorphism $\rho:A\to M(B)$ with $\rho(A)\subset B$.
    \item We denote $\textnormal{C}^*_{\textnormal{CP}}$ as the category whose objects are $C^*$-algebras, and whose morphisms are strict completely positive maps.
    \item Given a group $G$, we let $BG$ denote the category given by a single object (denoted by $\textnormal{pt}$), morphisms given by $G$, and composition given by the group law of $G$. If $G$ is a topological group, we enrich $BG$ over topological spaces so that the topology on $\textnormal{hom}(\textnormal{pt}, \textnormal{pt})=G$ agrees with the topology on $G$. 
    \item Given a group homomorphism $f:G\to H$, we denote the induced functor $BG\to BH$ also as $f$.
\end{itemize}

\section{The Category $\textnormal{HMod}$}

To develop a category encoding equivariant couplings for Hilbert modules, we first build the category $\textnormal{HMod}$ whose objects are Hilbert modules over any $C^*$-algebra. To construct this category, we utilize the Grothendieck construction on the pseudo-functor which associates a $C^*$-algebra with its corresponding category of Hilbert modules. Using the strong operator topology and point-norm topology, we enrich $\textnormal{HMod}$ over topological spaces and provide sufficient conditions for composition to be continuous. Finally, leveraging this enrichment as well as the categorical structure, we conclude with a functorial characterization of Hilbert module dynamical systems.

\subsection{Building the category $\textnormal{HMod}$}

At the level of objects, we have an assignment from
$C^*\textnormal{-Alg}$ to $\textnormal{Cat}$ which associates each $C^*$-algebra to its category of Hilbert modules. Using the interior tensor product, we obtain an assignment of morphisms. Leveraging properties of non-degenerate $*$-algebra maps, the assignment on objects and morphisms yields a pseudo-functor $\textnormal{HMod}_{(-)}:\textnormal{C}^*\textnormal{-Alg}\to \textnormal{Cat}$. Applying the Grothendieck construction to this pseudo-functor yields our desired category $\textnormal{HMod}$.

\begin{proposition}\label{Prop: induced map from interior}\

\noindent Let $A$ and $B$ be $C^*$-algebras, and let $(F, \pi)$ be a $C^*$-correspondence from $A$ to $B$. Let $E_1$ and $E_2$ be Hilbert $A$-modules. For each $\eta\in \mathcal{L}(E_1, E_2)$, there exists a unique adjointable map $\eta\otimes_\pi I\in \mathcal{L}(E_1\otimes_\pi F, E_2\otimes_\pi F)$ such that
    $$(\eta\otimes_\pi I)(x\dot{\otimes} f)=\eta(x)\dot{\otimes}f\quad\quad (\eta\otimes_\pi I)^*(y\dot{\otimes}f)=\eta^*(y)\dot{\otimes} f$$
    for all $x\dot{\otimes}f\in E_1\otimes_\pi F$ and $y\dot{\otimes}f\in E_2\otimes_\pi F$.
		
\end{proposition}

\begin{proof}\

\noindent The construction of $\eta\otimes_\pi I$ follows from the construction $\hat{\eta}$ in Proposition 3.3 in \cite{brady2026functorialityksgnsconstructionintertwiners}.

\end{proof}

\begin{proposition}\

\noindent Let $A$ and $B$ be $C^*$-algebras. If $\rho:A\to M(B)$ is a non-degenerate $*$-algebra homomorphism, then the assignment 
\begin{align*}
    (-)\otimes_\rho B:\textnormal{HMod}_A&\to \textnormal{HMod}_B\\
    E&\mapsto E\otimes_\rho B\\
    (\eta:E\to F)&\mapsto (\eta\otimes_\rho I:E\otimes_\rho B\to F\otimes_\rho B)
\end{align*}
defines a covariant functor. 

\end{proposition}

\begin{proof}\

\noindent The interior tensor product yields a well-defined assignment on the objects. From Proposition \ref{Prop: induced map from interior}, the assignment on morphisms is well-defined. From the construction of $\eta\otimes_\rho I$ in Proposition \ref{Prop: induced map from interior}, it is clear the assignment forms a covariant functor.

\end{proof}

\begin{proposition}\label{Prop: induced maps for HMod}\

\begin{enumerate}
    \item Let $A$ be a $C^*$-algebra, and let $\textnormal{inc}:A\to M(A)$ be the inclusion map. Then $(-)\otimes_{\textnormal{inc}}A$ is naturally isomorphic to $1_{\textnormal{HMod}_A}$; in particular, for each object $E\in \textnormal{HMod}_A$ the map
$$\iota:E\otimes_\textnormal{inc}A\to E\quad\quad \iota(x\dot{\otimes}a)=xa$$
defines a unitary isomorphism.
    \item Let $A, B,$ and $C$ be $C^*$-algebras, and let $\rho:A\to M(B)$ and $\chi:B\to M(C)$ be non-degenerate $*$-algebra homomorphisms. Denote the unique extension of $\chi$ to $M(B)$ by $\tilde{\chi}$. Then $((-)\otimes_{\rho}B)\otimes_\chi C$ is naturally isomorphic to $(-)\otimes_{\tilde{\chi}\circ \rho} C$ ; in particular, for each object $E\in \textnormal{HMod}_A$ the map
$$U:(E\otimes_\rho B)\otimes_\chi C\to E\otimes_{\tilde{\chi}\circ\rho}C\quad\quad U((x\dot{\otimes}b)\dot{\otimes}c)=x\dot{\otimes}\chi(b)c$$
defines a unitary isomorphism.
\end{enumerate}

\end{proposition}

\begin{proof}\

\noindent The construction of $\iota$ and naturality of the isomorphisms follows the same argument in Proposition 3.5 of \cite{brady2026functorialityksgnsconstructionintertwiners}. Similarly, the construction of $U$ and naturality of the isomorphisms follows the same argument in Proposition 3.6 of \cite{brady2026functorialityksgnsconstructionintertwiners}

\end{proof}

\begin{remark}\

\noindent The adjoint of $\iota$ is given by
$$\iota^*(x)=\lim\limits_{\lambda\to\infty}x\dot{\otimes} a_\lambda$$
where $(a_\lambda)_{\lambda\in\Lambda}$ is any approximate unit for $A$. Similarly, the adjoint of $U$ on simple tensors is given by
$$U^*(x\dot{\otimes}c)=\lim\limits_{\lambda\to\infty} (x\dot{\otimes}b_\lambda)\dot{\otimes}c$$
where $(b_\lambda)_{\lambda\in\Lambda}$ is any approximate unit for $B$.

\end{remark}

\begin{corollary}\

\noindent The assignment
\begin{align*}
    \textnormal{HMod}_{(-)}:C^*\textnormal{-Alg}&\to \textnormal{Cat}\\
    A&\mapsto \textnormal{HMod}_A\\
    (\rho:A\to B)&\mapsto \left((-)\otimes_\rho B:\textnormal{HMod}_A\to \textnormal{HMod}_B\right) 
\end{align*}
defines a covariant pseudo-functor.

\end{corollary}

\begin{definition}\

\noindent Define $\textnormal{HMod}=\int \textnormal{HMod}_{(-)}$; that is, $\textnormal{HMod}$ is the category where
\begin{itemize}
    \item objects are given by pairs $(A, E_A)$ where $A$ is a $C^*$-algebra and $E_A$ is a Hilbert $A$-module.
    \item a morphism from $(A, E_A)\to (B, E_B)$ is given by the data $(\rho, \eta)$ where $\rho:A\to M(B)$ is a non-degenerate $*$-algebra homomorphism with $\rho(A)\subset B$ and $\eta:E_A\otimes_\rho B\to E_B$ is an adjointable $B$-linear map.
    \item  given morphisms $(\rho, \eta):(A, E_A)\to (B, E_B)$ and $(\chi, \xi):(B, E_B)\to (C, E_C)$
$$(\chi, \xi)\circ (\rho, \eta)=(\chi\circ \rho, \xi\circ (\eta\otimes_\chi I)\circ U^{-1})$$ where
$$U:(E_A\otimes_\rho B)\otimes_\chi C\to E_A\otimes_{\chi\circ\rho} C\quad\quad U((x\dot{\otimes}b)\dot{\otimes c})=x\dot{\otimes}\chi(b)c$$ is the unitary from Proposition \ref{Prop: induced maps for HMod}.
\end{itemize}

\end{definition}

\subsection{Enrichment of $\textnormal{HMod}$ over $\textnormal{Top}$}

The morphisms in the categories $\textnormal{C}^*\textnormal{-Alg}$ and $\textnormal{HMod}_A$ carry topologies that we want to extend to $\textnormal{HMod}$; in particular, we want to topologize the hom-sets of $\textnormal{HMod}$ to encode convergence in both the point-norm topology and strong operator topology. While we can easily take care of the point-norm topology for the $*$-algebra map data, we run into issues for making sense of strong convergence on the adjointable map data. To see this issue explicitly, let us fix objects $(A, E_A), (B, E_B)\in\textnormal{HMod}$. Given morphisms $(\rho, \eta),(\chi, \xi)\in\textnormal{hom}((A, E_A), (B, E_B))$, we may not be able to compare $\eta$ and $\xi$ in the strong operator topology for the domains of the functions are $E_A\otimes_\rho B$ and $E_A\otimes_\chi B$, respectively. However, as $\rho(A), \chi(A)\subset B$, we can pullback the domains of $\eta$ and $\xi$ to $E_A$ enabling us to compare the maps pointwise.

\begin{lemma}[Lemma 5.2 in \cite{brady2026functorialityksgnsconstructionintertwiners}]\label{Lemma:C linear maps}\

\noindent Let $A$ and $B$ be $C^*$-algebras, and let $E$ be a Hilbert $A$-module. Let $\rho:A\to M(B)$ be a non-degenerate $*$-algebra homomorphism with $\rho(A)\subset B$. Let $(b_\lambda)_{\lambda\in\Lambda}$ be an approximate unit in $B$. For each $\lambda\in\Lambda$, the function
		$$V_\lambda:E\to E\otimes_\rho B\quad\quad V_\lambda(x)=x\dot{\otimes} b_\lambda $$
is a $\mathbb{C}$-linear contraction. Furthermore, the net  $(V_\lambda)_{\lambda\in\Lambda}$ converges in the strong operator topology on $\mathcal{B}(E, E\otimes_\rho B)$ to a $\mathbb{C}$-linear contraction $V_\rho$ which is independent of $(b_\lambda)_{\lambda\in\Lambda}$.
		
	\end{lemma}
	
	\begin{definition}\

		\noindent Let $(A, E_A), (B, E_B)\in \textnormal{HMod}$. For each $a\in A$ and $x\in E_A$ define the pseudo-metrics
		\begin{align*}
			d_{x, a}:\textnormal{hom}((A, E_A), (B, E_B))&\times\textnormal{hom}((A, E_A), (B, E_B))\to [0, \infty)\\
			d_{x, a}((\rho, \eta), (\chi, \xi))=&||\rho(a)-\chi(a)||_{B}+||(\eta\circ V_\rho)(x)-(\xi\circ V_{\chi})(x)||_{E_B}
		\end{align*}
		Define the topology on $\textnormal{hom}((A, E_A), (B, E_B))$ as the weakest topology such that each $d_{x, a}$ is continuous.
		
	\end{definition}

Let us make a few remarks. First, using properties of norms, it immediately follows each $d_{x, a}$ is a pseudo-metric. Second, the topology on the hom-sets is generated by a subbasis of balls from the pseudo-metrics. Third, as the topology is generated by pseudo-metrics, we can utilize the pseudo-metrics to characterize convergence of nets. In particular, from how the pseudo-metrics are defined, a net of morphisms $((\rho_\lambda, \eta_\lambda))_{\lambda\in\Lambda}$ converges to $(\rho, \eta)$ if and only if $\rho_\lambda\to \rho$ in the point norm topology  and $\eta_\lambda\circ V_{\rho_\lambda}\to \eta\circ V_\rho$ in the strong operator topology. Using this convergence property as well as a similar argument in Proposition 5.7 of \cite{brady2026functorialityksgnsconstructionintertwiners}, we obtain the following sufficient condition for continuity of composition.

\begin{proposition}\

		\noindent For $i=1, 2, 3$, let $O_i\in\textnormal{HMod}$. Then the function
		\begin{align*}
			\textnormal{Maps}(O_1, O_2)\times \textnormal{Maps}(O_2, O_3)&\to \textnormal{Maps}(O_1, O_3)\\
			((\rho, \eta), (\chi, \xi))&\mapsto (\chi\circ\rho, \xi\circ(\eta\otimes_\chi I)\circ U^{-1})
		\end{align*}
is jointly continuous when the first factor is restricted to a subspace for which the adjointable maps are uniformly norm bounded.
		
\end{proposition}

\subsection{Functorial characterization of Hilbert module dynamical systems }

We now prove a functorial characterization of Hilbert module dynamical systems. We begin with a reminder of the terminology needed for making sense of a Hilbert module dynamical systems. For motivation of the following definitions, see either \cite{brady2026functorialityksgnsconstructionintertwiners} or \cite{brady2026hilbertmoduleapproachclassical}.

\begin{definition}[Definition 6.3 in \cite{brady2026functorialityksgnsconstructionintertwiners}]\

		\noindent Let $E_1$ and $E_2$ be Hilbert $A$-modules, and let $\alpha\in\textnormal{Aut}_{*\textnormal{-alg}}(A)$. 
		\begin{itemize}
			\item A (bounded) $\alpha$-linear map from $E_1$ to $E_2$ is a (bounded) $\mathbb{C}$-linear map $T:E_1\to E_2$ such that for all $x\in E_1$ and $a\in A$, $T(xa)=T(x)\alpha(a)$.
			\item A function $T:E_1\to E_2$  is $\alpha$-adjointable if there exists a map $T^*:E_2\to E_1$ such that for all $x\in E_1$ and $y\in E_2$, $\langle T(x), y\rangle_{E_2}=\alpha(\langle x, T^*(y)\rangle_{E_1})$. Denote the set of $\alpha$-adjointable maps from $E_1$ to $E_2$ as $\mathcal{L}^\alpha(E_1, E_2)$.
			\item An $\alpha$-adjointable map $T:E_1\to E_2$ is unitary if $TT^*=1_{E_2}$ and $T^*T=1_{E_1}$. Denote the set of $\alpha$-adjointable unitaries from $E_1$ to $E_2$ as $\mathcal{U}^\alpha(E_1, E_2)$.
			
		\end{itemize}
		
\end{definition}

\begin{remark}\

\noindent While not assumed, it follows $\alpha$-adjointable maps are $\alpha$-linear and have unique adjoints. Using the Uniform Boundedness Principle, it follows $\alpha$-adjointable maps are bounded. 

\end{remark}

\begin{lemma}[Lemma 6.1 in \cite{brady2026functorialityksgnsconstructionintertwiners}]\label{Lemma: beta unitary}\

\noindent Let $E_A$ be a Hilbert $A$-module, and let $\alpha$ be a $*$-automorphism of $A$. Then the map
$$U_\alpha:E_A\otimes_{\textnormal{inc}\circ\alpha} A\to E_A\quad\quad U_\alpha(x\dot{\otimes}a)=x\alpha^{-1}(a)$$
defines an $\alpha^{-1}$-adjointable unitary.
		
\end{lemma}

\begin{remark}\

\noindent We observe the map $V_{\textnormal{inc}\circ\alpha}$ given in Lemma \ref{Lemma:C linear maps} is the adjoint of $U_\alpha$.

\end{remark}
    
\begin{proposition}[Proposition 6.4 in \cite{brady2026functorialityksgnsconstructionintertwiners}]\label{Prop: rel between B-linear and beta-linear}\

\noindent Let $E_1$ and $E_2$ be Hilbert $A$-modules, and let $\alpha\in\textnormal{Aut}_{*\textnormal{-alg}}(A)$. Let $U_\alpha$ be the $\alpha^{-1}$-adjointable unitary given by
$$U_\alpha:E_1\otimes_{\textnormal{inc}\circ\alpha} A\to E_1\quad\quad U_\alpha(x\dot{\otimes} a)=x\alpha^{-1}(a)$$
Then the map
$$\mathcal{L}^\alpha(E_1, E_2)\to \mathcal{L}(E_1\otimes_{\textnormal{inc}\circ\alpha} A, E_2)\quad\quad T\mapsto T\circ U_\alpha$$
defines a linear bijection. Furthermore, the map restricts to a bijection $$\mathcal{U}^\alpha(E_1, E_2)\to \mathcal{U}(E_1\otimes_{\textnormal{inc}\circ\alpha} A, E_2).$$
		
\end{proposition}

\begin{definition}\

\noindent A Hilbert module dynamical system is given by the data $((A, G, \alpha), E_A, U)$ where
\begin{itemize}
    \item $(A, G, \alpha)$ is a $C^*$-dynamical system,
    \item $E_A$ is a Hilbert $A$-module, 
    \item and $U:G\to \mathcal{B}(E_A)$ is a group homomorphism,
\end{itemize}
and the data satisfies the following condition: for all $g\in G$, $U_g$ is an $\alpha_g$-adjointable unitary. When $G$ is a topological group, the Hilbert module dynamical system is continuous if $(A, G, \alpha)$ is a continuous $C^*$-dynamical system and $U$ is continuous with respect to the strong operator topology on $\mathcal{B}(E_A)$.
    
\end{definition}

\begin{definition}\

\noindent Let $G$ be a group. A functor $\mathcal{F}:BG\to \textnormal{HMod}$ is unitary valued if for all $g\in G$, the morphism $\mathcal{F}(g)=(\alpha_g, \eta_g)$ satisfies the condition that $\eta_g$ is unitary.
    
\end{definition}

\begin{theorem}\label{Theorem: functorial characterization of hilbert module dynamical systems}\

\noindent A Hilbert module dynamical system $((A, G, \alpha), E, W)$ determines and is determined by a unitary valued functor $\mathcal{F}:BG\to \textnormal{HMod}$ where
\begin{enumerate}
    \item $\mathcal{F}(\textnormal{pt})=(A, E)$,
    \item and $\mathcal{F}(g)=(\textnormal{inc}\circ\alpha_g, W_g\circ U_{\alpha_g})$ for all $g\in G$.
\end{enumerate}
Furthermore, when $G$ is a topological group, the Hilbert module dynamical system is continuous if and only if $\mathcal{F}$ is topological.

\end{theorem}

\begin{proof}\

\noindent This follows a similar argument as that for Theorem 6.7 in  \cite{brady2026functorialityksgnsconstructionintertwiners}.

\end{proof}

\section{The Category of Equivariant Couplings}

We now combine together the categories $\textnormal{HMod}$ and $\textnormal{NDPosCor}_{*\textnormal{-alg}}$ to build the category $\textnormal{Cpl}$ codifying the algebraic data of an equivariant coupling. Using this category as well as the functorial characterizations of a Hilbert module dynamical system and a non-degenerate positive equivariant $C^*$-correspondence, we prove a functorial characterization of $G$-equivariant couplings as unitary valued functors $BG\to \textnormal{Cpl}$. From this characterization, we organize the data of $G$-equivariant couplings as a functor category. Using this perspective as well as the Grothendeik construction, we build a category, called $\textnormal{EqCpl}$, organizing all equivariant couplings as objects. 

\subsection{The category of couplings: $\textnormal{Cpl}$}

\begin{definition}\

\noindent Define the category $\textnormal{Cpl}$ as the subcategory of $\textnormal{NDPosCor}_{*\textnormal{-alg}}\times \textnormal{HMod}$ where
\begin{itemize}
    \item objects are pairs $((A, B, (E_B, \phi)), (C, E_C))$ satisfying the condition of $A=C$.
    \item morphisms
$$((\rho, \chi, \eta), (\sigma, \xi)):((A, B, (E_B, \phi)), (A, E_A))\to ((C, D, (E_E, \psi)), (C, E_C))$$
satisfy the condition of $\rho=\sigma$.
\end{itemize}
    
\end{definition}

\begin{remark}\

\noindent As both $\textnormal{NDPosCor}_{*\textnormal{-alg}}$ and $\textnormal{HMod}$ are locally small categories, we can form the cartesian product category $\textnormal{NDPosCor}_{*\textnormal{-alg}}\times\textnormal{HMod}$. Therefore, the category $\textnormal{Cpl}$ is well-defined.

\end{remark}

Using the enrichment of $\textnormal{NDPosCor}_{*\textnormal{-alg}}$ and $\textnormal{HMod}$, we enrich the hom-sets of the category $\textnormal{NDPosCor}_{*\textnormal{-alg}}\times \textnormal{HMod}$ with the product topology, and we enrich the hom-sets of $\textnormal{Cpl}$ with the subspace topology. Therefore, a net of morphisms $(((\rho_\lambda, \chi_\lambda, \eta_\lambda), (\rho_\lambda, \xi_\lambda)))_{\lambda\in\Lambda}$ between two objects in $\textnormal{Cpl}$ converges to $((\rho, \chi, \eta), (\rho , \xi))$ if and only if
$$(\rho_\lambda, \chi_\lambda, \eta_\lambda)\to (\rho, \chi, \eta)\quad\quad (\rho_\lambda, \xi_\lambda)\to (\rho, \chi)$$
Furthermore, using the sufficient conditions for composition being continuous in $\textnormal{NDPosCor}_{*\textnormal{-alg}}$  (see Proposition 5.9 in \cite{brady2026functorialityksgnsconstructionintertwiners}) and $\textnormal{HMod}$, we immediately obtain the following sufficient condition for composition to be continuous in $\textnormal{Cpl}$.

\begin{proposition}\

\noindent For $i=1, 2, 3$, let $O_i\in\textnormal{Cpl}$. Then the function
\begin{align*}
			\textnormal{Maps}(O_1, O_2)\times \textnormal{Maps}(O_2, O_3)&\to \textnormal{Maps}(O_1, O_3)\\
			((\rho, \chi, \eta), (\rho, \xi)), (\mu, \sigma, \gamma), (\mu, \lambda))&\mapsto ((\mu\circ\rho, \sigma\circ\chi, \gamma\circ\hat{\eta}\circ U^{-1}_1), (\mu\circ\rho, \lambda\circ (\xi\otimes_{\mu} I)\circ  U_2^{-1})
		\end{align*}
		is jointly continuous when the first factor is restricted to a subspace for which the adjointable maps are uniformly norm bounded.

\end{proposition}

While the construction of $\textnormal{Cpl}$ seems arbitrary, the category naturally arises as a pullback in $\textnormal{Cat}$. Define the assignments
\begin{align*}
    \mathcal{P}_1:\textnormal{NDPosCor}_{*\textnormal{-alg}}&\to \textnormal{C}^*\textnormal{-Alg} & \mathcal{P}_2:\textnormal{HMod}&\to \textnormal{C}^*\textnormal{-Alg} \\
    (A, B, (E_B, \phi))&\mapsto A & (A, E_A)&\mapsto A\\
    (\rho, \chi, \eta)&\mapsto \rho & (\sigma, \xi)&\mapsto \sigma
\end{align*}
which form covariant functors. Note that $\mathcal{P}_2$ is the projection from the Grothendieck construction.

\begin{proposition}\

\noindent The category $\textnormal{Cpl}$ together with the projection functors
$$\mathcal{Q}_1:\textnormal{Cpl}\to\textnormal{NDPosCor}_{*\textnormal{-alg}}\quad\quad\mathcal{Q}_2:\textnormal{Cpl}\to \textnormal{HMod}$$
from the cartesian product $\textnormal{NDPosCor}_{*\textnormal{-alg}}\times\textnormal{HMod}$ satisfy the universal property of the pullback in $\textnormal{Cat}$ of the following diagram:
\begin{center}
    \begin{tikzcd}
	&& {\textnormal{HMod}} \\
	\\
	{\textnormal{NDPosCor}_{*\textnormal{-alg}}} && {\textnormal{C}^*\textnormal{-Alg}}
	\arrow["{\mathcal{P}_2}"{description}, from=1-3, to=3-3]
	\arrow["{\mathcal{P}_1}"{description}, from=3-1, to=3-3]
\end{tikzcd}
\end{center}
    
\end{proposition}

\begin{proof}\

\noindent It is clear that the category $\textnormal{Cpl}$ together with the projection functors make the following diagram commute:
\begin{center}
    \begin{tikzcd}
	\textnormal{Cpl} && {\textnormal{HMod}} \\
	\\
	{\textnormal{NDPosCor}_{*\textnormal{-alg}}} && {\textnormal{C}^*\textnormal{-Alg}}
	\arrow["{\mathcal{Q}_2}"{description}, from=1-1, to=1-3]
	\arrow["{\mathcal{Q}_1}"{description}, from=1-1, to=3-1]
	\arrow["{\mathcal{P}_2}"{description}, from=1-3, to=3-3]
	\arrow["{\mathcal{P}_1}"{description}, from=3-1, to=3-3]
\end{tikzcd}
\end{center}
Therefore, we only verify the data satisfies the universal property of the pullback. Let $\mathcal{C}$ be a category, and let
$$\mathcal{F}_1:\mathcal{C}\to\textnormal{NDPosCor}_{*\textnormal{-alg}}\quad\quad \mathcal{F}_2:\mathcal{C}\to \textnormal{HMod}$$
be functors making the following diagram commute:
\begin{center}
    \begin{tikzcd}
	\mathcal{C} && {\textnormal{HMod}} \\
	\\
	{\textnormal{NDPosCor}_{*\textnormal{-alg}}} && {\textnormal{C}^*\textnormal{-Alg}}
	\arrow["{\mathcal{F}_2}"{description}, from=1-1, to=1-3]
	\arrow["{\mathcal{F}_1}"{description}, from=1-1, to=3-1]
	\arrow["{\mathcal{P}_2}"{description}, from=1-3, to=3-3]
	\arrow["{\mathcal{P}_1}"{description}, from=3-1, to=3-3]
\end{tikzcd}
\end{center}
Define the functor
$$\mathcal{F}:\mathcal{C}\to \textnormal{NDPosCor}_{*\textnormal{-alg}}\times \textnormal{HMod}\quad\quad \mathcal{F}=(\mathcal{F}_1,\mathcal{F}_2) $$

We claim $\mathcal{F}$ takes value in $\textnormal{Cpl}$. Given an object $O\in\mathcal{C}$, we have
$$\mathcal{F}(O)=(\mathcal{F}_1(O), \mathcal{F}_2(O))=((A, B, (E_B, \phi)), (C, E_C))$$
As $\mathcal{P}_1\circ\mathcal{F}_1=\mathcal{P}_2\circ\mathcal{F}_2$, we have $A=C$. Thus $\mathcal{F}(O)$ is an object in $\textnormal{Cpl}$. Given a morphism $m:O_1\to O_2$ in $\mathcal{C}$, we have 
$$\mathcal{F}(m)=(\mathcal{F}_1(m), \mathcal{F}_2(m))=((\rho, \chi, \eta), (\sigma, \xi))$$
As $\mathcal{P}_1\circ\mathcal{F}_1=\mathcal{P}_2\circ\mathcal{F}_2$, we have $\rho=\sigma$. Thus $\mathcal{F}(m)$ is a morphism in the subcategory $\textnormal{Cpl}$. Hence the functor $\mathcal{F}$ does take value in the subcategory $\textnormal{Cpl}$.  By construction of $\mathcal{F}$, it is clear we have $\mathcal{Q}_i\circ\mathcal{F}=\mathcal{F}_i$. 

We now claim $\mathcal{F}$ is the unique functor into $\textnormal{Cpl}$ such that $\mathcal{Q}_i\circ\mathcal{F}=\mathcal{F}_i$. If $\mathcal{F}'$ is another functor satisfying this property, then, by viewing $\mathcal{F}'$ as a functor into the cartesian product $\textnormal{NDPosCor}_{*\textnormal{-alg}}\times\textnormal{HMod}$, we can apply the  universal property of the cartesian product to write $\mathcal{F}'=(\mathcal{F}'_1, \mathcal{F}'_2)$ where $\mathcal{F}'_i=\pi_i\circ\mathcal{F}'$. As $\mathcal{F}'$ takes value in $\textnormal{Cpl}$ and $\mathcal{Q}_i$ is the restriction of $\pi_i$ to the subcategory $\textnormal{Cpl}$,  it follows 
$$\mathcal{F}_i'=\pi_i\circ\mathcal{F}'=\mathcal{Q}_i\circ\mathcal{F}'=\mathcal{F}_i.$$
Thus $\mathcal{F}'=\mathcal{F}$ showing $\mathcal{F}$ is unique. 
    
\end{proof}

\begin{definition}\

\noindent Let $G$ be a group. A functor $\mathcal{F}:BG\to \textnormal{Cpl}$ is unitary valued if for all $g\in G$, the morphism $\mathcal{F}(g)=((\rho_g, \chi_g, \eta_g), (\rho_g, \xi_g))$ satisfies the two following conditions:
\begin{enumerate}
    \item $\eta_g$ is unitary.
    \item $\xi_g$ is unitary.
\end{enumerate}
    
\end{definition}

\begin{corollary}\label{Cor: unitary value functors into Cpl}\

\noindent A unitary valued functor $\mathcal{F}:BG\to\textnormal{Cpl}$ determines and is determined by unitary valued functors $\mathcal{F}_1:BG\to \textnormal{HMod}$ and $\mathcal{F}_2:BG\to \textnormal{NDPosCor}_{*\textnormal{-alg}}$
which make the following diagram commute:
\begin{center}
    \begin{tikzcd}
	BG && {\textnormal{HMod}} \\
	\\
	{\textnormal{NDPosCor}_{*\textnormal{-alg}}} && {\textnormal{C}^*\textnormal{-Alg}}
	\arrow["{\mathcal{F}_1}"{description}, from=1-1, to=1-3]
	\arrow["{\mathcal{F}_2}"{description}, from=1-1, to=3-1]
	\arrow["{\mathcal{P}_1}"{description}, from=1-3, to=3-3]
	\arrow["{\mathcal{P}_2}"{description}, from=3-1, to=3-3]
\end{tikzcd}
\end{center}
Furthermore, $\mathcal{F}$ is topological if and only if $\mathcal{F}_1$ and $\mathcal{F}_2$ are topological.

\end{corollary}

\begin{proof}\

\noindent Using the universal property of the pullback, a functor $\mathcal{F}:BG\to \textnormal{Cpl}$ determines and is determined by functors $\mathcal{F}_1:BG\to \textnormal{HMod}$ and $\mathcal{F}_2:BG\to \textnormal{NDPosCor}_{*\textnormal{-alg}}$
which makes the following diagram commute:
\begin{center}
    \begin{tikzcd}
	BG && {\textnormal{HMod}} \\
	\\
	{\textnormal{NDPosCor}_{*\textnormal{-alg}}} && {\textnormal{C}^*\textnormal{-Alg}}
	\arrow["{\mathcal{F}_1}"{description}, from=1-1, to=1-3]
	\arrow["{\mathcal{F}_2}"{description}, from=1-1, to=3-1]
	\arrow["{\mathcal{P}_1}"{description}, from=1-3, to=3-3]
	\arrow["{\mathcal{P}_2}"{description}, from=3-1, to=3-3]
\end{tikzcd}
\end{center}
Writing
$$\mathcal{F}(g)=((\rho_g, \chi_g, \eta_g), (\rho_g, \xi_g))$$
we have
$$\mathcal{F}_1(g)=(\rho_g, \chi_g, \eta_g)\quad\quad \mathcal{F}_2(g)=(\rho_g, \xi_g)$$
Thus, we see $\mathcal{F}$ is unitary valued if and only if both $\mathcal{F}_1$ and $\mathcal{F}_2$ are unitary valued. The last claim follows from the convergence property of nets in the hom-sets of $\textnormal{Cpl}$.
    
\end{proof}

\begin{corollary}\label{Cor: functorial characterization of equivariant couplings}\

\noindent Let $E_A$ be a Hilbert $A$-module. A $G$-equivariant coupling
$$(((A, G, \alpha), (B, G, \beta), ((E_B, \phi), W)), ((A, G, \alpha), E_A, U))$$
for $E_A$ determines and is determined by a unitary valued functor $\mathcal{F}:BG\to\textnormal{Cpl}$ such that
\begin{enumerate}
    \item $\mathcal{F}(\textnormal{pt})=((A, B, (E_B, \phi)), (A, E_A))$
    \item For each $g\in G$, $\mathcal{F}(g)=((\alpha_g, \textnormal{inc}\circ\beta_g, W_g\circ U_{\beta_g}), (\alpha_g, U_g\circ U_{\alpha_g} ))$.
\end{enumerate}

\end{corollary}

\begin{proof}\

\noindent This immediately follows from Corollary \ref{Cor: unitary value functors into Cpl}, Theorem \ref{Theorem: functorial characterization of hilbert module dynamical systems},
and Corollary 6.8 in \cite{brady2026functorialityksgnsconstructionintertwiners}

\end{proof}

\subsection{The category of equivariant couplings: $\textnormal{EqCpl}$}

Using the functorial characterization of $G$-equivariant couplings, we have been led to the category $\textnormal{Fun}(BG, \textnormal{Cpl})$. Recall, this category has objects as functors $BG\to \textnormal{Cpl}$, and  morphisms are given by natural transformations. As $BG$ has one object, a natural transformation $\tau:\mathcal{F}_1\to \mathcal{F}_2$ is equivalent to the data of a morphism $((\rho, \chi, \eta), (\rho, \xi)):\mathcal{F}_1(\textnormal{pt})\to \mathcal{F}_2(\textnormal{pt})$ in $\textnormal{Cpl}$ which makes the following diagram commute in $\textnormal{Cpl}$ for all $g\in G$:
\begin{center}
    \begin{tikzcd}
	{\mathcal{F}_1(\textnormal{pt})} && {\mathcal{F}_1(\textnormal{pt})} \\
	\\
	{\mathcal{F}_2(\textnormal{pt})} && {\mathcal{F}_2(\textnormal{pt})}
	\arrow["{\mathcal{F}_1(g)}"{description}, from=1-1, to=1-3]
	\arrow["{((\rho, \chi, \eta), (\rho, \xi))}"{description}, from=1-1, to=3-1]
	\arrow["{((\rho, \chi, \eta), (\rho, \xi))}"{description}, from=1-3, to=3-3]
	\arrow["{\mathcal{F}_2(g)}"{description}, from=3-1, to=3-3]
\end{tikzcd}
\end{center}
Restricting our attention to the functors which are unitary valued per Corollary \ref{Cor: functorial characterization of equivariant couplings}, we make the following definition.

\begin{definition}\

\noindent Let $G$ be a group. Define the category $G\textnormal{-EqCpl}$ as the full subcategory of $\textnormal{Fun}(BG, \textnormal{Cpl})$ where objects are given by unitary valued functors.

\end{definition}

Given a group homomorphism $f:G\to H$, we obtain a functor $f:BG\to BH$. Using this functor, we can pullback functors defined on $BH$ to functors defined on $BG$. This leads to the assignment
$$ f^*:H\textnormal{-EqCpl}\to G\textnormal{-EqCpl} \quad\quad \mathcal{F}\mapsto f^*\mathcal{F}$$

In addition to pulling back functors, we can also pullback natural transformations. Suppose $\tau:\mathcal{F}_1\to\mathcal{F}_2$ is a natural transformation for functors $\mathcal{F}_1, \mathcal{F}_2:BH\to \mathcal{C}$ for some category $\mathcal{C}$. Again, the data of a natural transformation $\tau$ is equivalent to the choice of morphism $\mathcal{F}_1(\text{pt})\to \mathcal{F}_2(\text{pt})$ such that the following diagram commutes in $\mathcal{C}$ for all $h\in H$:
\begin{center}
    \begin{tikzcd}
	{\mathcal{F}_1(\textnormal{pt})} && {\mathcal{F}_1(\textnormal{pt})} \\
	\\
	{\mathcal{F}_2(\textnormal{pt})} && {\mathcal{F}_2(\textnormal{pt})}
	\arrow["{\mathcal{F}_1(h)}"{description}, from=1-1, to=1-3]
	\arrow[{description}, from=1-1, to=3-1]
	\arrow[{description}, from=1-3, to=3-3]
	\arrow["{\mathcal{F}_2(h)}"{description}, from=3-1, to=3-3]
\end{tikzcd}
\end{center}
As $(f^*\mathcal{F}_i)(g)=\mathcal{F}_i(f(g))$, it follows the following diagram commutes in $\mathcal{C}$ for all $g\in G$:
\begin{center}
    \begin{tikzcd}
	{\mathcal{F}_1(\textnormal{pt})} && {\mathcal{F}_1(\textnormal{pt})} \\
	\\
	{\mathcal{F}_2(\textnormal{pt})} && {\mathcal{F}_2(\textnormal{pt})}
	\arrow["{\mathcal{F}_1(f(g))}"{description}, from=1-1, to=1-3]
	\arrow[{description}, from=1-1, to=3-1]
	\arrow[{description}, from=1-3, to=3-3]
	\arrow["{\mathcal{F}_2(f(g))}"{description}, from=3-1, to=3-3]
\end{tikzcd}
\end{center}
Thus we obtain a natural transformation $f^*\tau:f^*\mathcal{F}_1\to f^*\mathcal{F}_2$. 

From the construction of $f^*\tau$, it immediately follows that $f^*(\tau\circ\mu)=f^*\tau\circ f^*\mu$ and $f^*(1_{\mathcal{F}})=1_{f^*\mathcal{F}}$. Therefore the assignment
\begin{align*}
    f^*:H\textnormal{-EqCpl}&\to G\textnormal{-EqCpl}\\
    \mathcal{F}&\mapsto f^*\mathcal{F}\\
    \left(\tau:\mathcal{F}_1\to \mathcal{F}_2\right)&\mapsto\left(f^*\tau:f^*\mathcal{F}_1\to f^*\mathcal{F}_2\right)
\end{align*}
defines a covariant functor. Furthermore, from how $f^*$ is constructed, it immediately follows $1_G^*=1_{G\textnormal{-EqCpl}}$ and $(g\circ f)^*=f^*\circ g^*$.  We record these properties as well as the assignment in the follow proposition.

\begin{proposition}\

\noindent Let $G$ and $H$ be groups. For each group homomorphism $f:G\to H$, the assignment
\begin{align*}
    f^*:H\textnormal{-EqCpl}&\to G\textnormal{-EqCpl}\\
    \mathcal{F}&\mapsto f^*\mathcal{F}\\
    \left(\tau:\mathcal{F}_1\to \mathcal{F}_2\right)&\mapsto\left(f^*\tau:f^*\mathcal{F}_1\to f^*\mathcal{F}_2\right)
\end{align*}
defines a covariant functor. Furthermore
\begin{enumerate}
    \item If $1_G:G\to G$ is the identity map, then $1_G^*=1_{G\textnormal{-EqCpl}}$.
    \item If $f:G\to H$ and $g:H\to K$ are group homomorphisms, then $(f\circ g)^*=g^*\circ f^*$.
\end{enumerate}

\end{proposition}

\begin{corollary}\

\noindent The assignment
\begin{align*}
    (-)\textnormal{-EqCpl}:\textnormal{Grp}&\to \textnormal{Cat}\\
    G&\mapsto G\textnormal{-EqCpl}\\
    (f:G\to H)&\mapsto (f^*:H\textnormal{-EqCpl}\to G\textnormal{-EqCpl})
\end{align*}
defines a contravariant functor.

\end{corollary}

\begin{definition}\

\noindent Define the category $\textnormal{EqCpl}=\int (-)\textnormal{-EqCpl}$; that is, $\textnormal{EqCpl}$ is the category where
\begin{itemize}
    \item  objects are pairs $(G, \mathcal{F})$ where $G$ is a group and $\mathcal{F}:BG\to \textnormal{Cpl}$ is a unitary valued functor.
    \item a morphism $(G_1, \mathcal{F}_1)\to (G_2, \mathcal{F}_2)$ is given by the data $(f, \tau)$ where $f:G_1\to G_2$ is a group homomorphism and $\tau:\mathcal{F}_1\to f^*\mathcal{F}_2$ is a natural transformation.
    \item given morphisms
$$(f, \tau):(G_1, \mathcal{F}_1)\to (G_2, \mathcal{F}_2)\quad\quad (g, \mu):(G_2, \mathcal{F}_2)\to (G_3, \mathcal{F}_3)$$ the composition is given by the morphism
$$(g\circ f, f^*\mu\circ \tau):(G_1, \mathcal{F}_1)\to (G_3, \mathcal{F}_3)$$

\end{itemize}
Define the subcategory $\textnormal{EqCpl}_{*\textnormal{-iso}}$ of $\textnormal{EqCpl}$ where morphisms $(f, \tau):(G_1, \mathcal{F}_1)\to (G_2, \mathcal{F}_2)$ satisfy the property that if $((\rho, \chi, \eta), (\rho, \xi))$ is the data associated to $\tau$, then $\rho$ is a $*$-algebra isomorphism.

\end{definition}

Thinking of objects in $\text{EqCpl}$ as unitary representations of a group, the natural transformations should behave as intertwiners of the representations. Due to the composition rule in both $\textnormal{NDPosCor}_{*\textnormal{-alg}}$ and  $\textnormal{HMod}$, it is not apparent that commutativity of the diagram 
\begin{center}
    \begin{tikzcd}
	{\mathcal{F}_1(\textnormal{pt})} && {\mathcal{F}_1(\textnormal{pt})} \\
	\\
	{\mathcal{F}_2(\textnormal{pt})} && {\mathcal{F}_2(\textnormal{pt})}
	\arrow["{\mathcal{F}_1(g)}"{description}, from=1-1, to=1-3]
	\arrow[{description}, from=1-1, to=3-1]
	\arrow[{description}, from=1-3, to=3-3]
	\arrow["{\mathcal{F}_2(f(g))}"{description}, from=3-1, to=3-3]
\end{tikzcd}
\end{center}
is intertwining the morphism data of $\mathcal{F}_1(g)$ and $\mathcal{F}_2(f(g))$. However, this is indeed the case, and we make this explicit in the next Theorem. 

\begin{lemma}\

\noindent Let $((B_1, G_1, \beta_1), E_{B_1}, W)$ be a Hilbert module dynamical system, and let $(B_2, G_2, \beta_2)$ be a $C^*$-dynamical system. If $f:G_1\to G_2$ is a group homomorphism and $\chi:B_1\to M(B_2)$ is a non-degenerate $*$-algebra homomorphism such that $\chi(B_1)\subset B_2$ and for all $g\in G_1$, $\chi\circ \beta_{1, g}=\text{inc}\circ\beta_{2, f(g)}\circ \chi$, then
\begin{enumerate}
    \item for each $g\in G_1$, there exists a unique $\beta_{2, f(g)}$-adjointable unitary 
    $$W_g\otimes_\chi \beta_{2, f(g)}:E_{B_1}\otimes_\chi B_2\to E_{B_1}\otimes_\chi B_2\quad\quad(W_g\otimes_\chi \beta_{2, f(g)})(x\dot{\otimes}b_2)=W_{1, g}(x)\dot{\otimes}\beta_{2, f(g)}(b_2). $$
\item  the map
$$W\otimes_\chi f^*\beta_{2}:G\to \mathcal{B}(E_{B_1}\otimes_\chi B_2)\quad\quad g\mapsto W_g\otimes_\chi \beta_{2, f(g)}$$ 
defines a group homomorphism.
\end{enumerate}
Furthermore, if $(A, G_1, \alpha)$ is another $C^*$-dynamical system and $((E_{B_1}, \phi), W)$ defines a non-degenerate positive equivariant $C^*$-correspondence from $(A, G_1, \alpha)$ to $(B_1, G_1, \beta_1)$, then the data $((E_{B_1}\otimes_\chi B_2, \phi_1\otimes_\chi I), W\otimes_\chi f^*\beta_2)$ forms a non-degenerate positive equivariant $C^*$-correspondence from $(A, G_1, \alpha)$ to $(B_2, G_1, f^*\beta_{2})$.

\end{lemma}

\begin{proof}\

\noindent We have a Hilbert module dynamical system $((B_1, G_1, \beta), E_{B_1}, W)$ as well as an equivariant $C^*$-correspondence $((B_2, \chi), f^*\beta_{2})$ from $(B_1, G_1, \beta_1)$ to $(B_2, G_1, f^*\beta_{2})$. Therefore, using the equivariant interior tensor product (Proposition 2.7 in \cite{brady2026hilbertmoduleapproachclassical}), we obtain, for each $g\in G_1$, a unique $\beta_{2, f(g)}$-adjointable unitary
$$W_g\otimes_\chi \beta_{2, f(g)}:E_{B_1}\otimes_\chi B_2\to E_{B_1}\otimes_\chi B_2\quad\quad (W_g\otimes_\chi \beta_{2, f(g)})(x\dot{\otimes}b)=W_g(x)\dot{\otimes}\beta_{2, f(g)}(b).$$
Furthermore, from the equivariant interior tensor product, the assignment
$$W\otimes_\chi f^*\beta_{2}:G\to \mathcal{B}(E_{B_1}\otimes_\chi B_2)\quad\quad g\mapsto W_g\otimes_\chi \beta_{2, f(g)}$$
defines a group homomorphism. 

To prove the last claim, we note that as $(E_{B_1}\otimes_\chi B_2, \phi_1\otimes_\chi I)$ is a non-degenerate positive $C^*$-correspondence from $A$ to $B_2$, we only need to verify that for all $a\in A$ and $g\in G_1$
$$(W_{1, g}\otimes_\chi \beta_{2, f(g)})\circ (\phi_1(a)\otimes_\chi I)=(\phi_1(\alpha_{1, g}(a))\otimes_\chi I)\circ (W_{1, g}\otimes_\chi \beta_{2, f(g)})$$
Fix $a\in A$ and $g\in G_1$. As $(W_{1, g}\otimes_\chi \beta_{2, f(g)})\circ (\phi_1(a)\otimes_\chi I)$ and $(\phi_1(\alpha_{1, g}(a))\otimes_\chi I)\circ (W_{1, g}\otimes_\chi \beta_{2, f(g)})$ are continuous and $\beta_{2, f(g)}$-linear, it suffices to prove the maps agree on elements of the form $x\dot{\otimes}b_2\in E_{B_1}\otimes_{B_1}B_2$. Given such an element, we have
\begin{align*}
   ( (W_{1, g}\otimes_\chi \beta_{2, f(g)})\circ (\phi_1(a)\otimes_\chi I))(x\dot{\otimes}b_2)&=(W_{1, g}\otimes_\chi \beta_{2, f(g)})(\phi_1(a)x\dot{\otimes}b_2)\\
&=W_{1, g}(\phi_1(a)x)\dot{\otimes}\beta_{2, f(g)}(b_2)\\
   &=\phi_1(\alpha_{1, g}(a))W_{1, g}(x)\dot{\otimes}\beta_{2, f(g)}(b_2)\\
   &=(\phi_1(\alpha_{1, g}(a))\otimes_\chi I)(W_{1, g}(x)\dot{\otimes}\beta_{2, f(g)}(b_2))\\
   &=((\phi_1(\alpha_{1, g}(a))\otimes_\chi I)\circ (W_{1, g}\otimes_\chi \beta_{2, f(g)}))(x\dot{\otimes}b_2)
\end{align*}

\end{proof}

\begin{theorem}\label{Theorem: characterization of morphisms}\

\noindent For $i=1, 2$, let $(G_i, \mathcal{F}_i)\in\textnormal{EqCpl}$ with corresponding $G_i$-equivariant couplings given by
$$(((A_i, G_i, \alpha_i), (B_i, G_i, \beta_i), ((E_{B_i}, \phi_i), W_i)), ((A_i, G_i, \alpha_i), E_{A_i}, U_i)).$$
Let $f:G_1\to G_2$ be a group homomorphism, and let $((\rho, \chi, \eta), (\rho, \xi)):\mathcal{F}_1(\textnormal{pt})\to \mathcal{F}_2(\textnormal{pt})$ be a morphism in $\textnormal{Cpl}$.
The diagram 
\begin{center}
\begin{tikzcd}[every label/.append
style={font=\fontsize{9}{11}}]
	{\mathcal{F}_1(\textnormal{pt})} &&& {\mathcal{F}_1(\textnormal{pt})} \\
	\\
	\\
	{\mathcal{F}_2(\textnormal{pt})} &&& {\mathcal{F}_2(\textnormal{pt})}
	\arrow["{\mathcal{F}_1(g)}"{description}, from=1-1, to=1-4]
	\arrow["{((\rho, \chi, \eta), (\rho, \xi))}"{description}, from=1-1, to=4-1]
	\arrow["{((\rho, \chi, \eta), (\rho, \xi))}"{description}, from=1-4, to=4-4]
	\arrow["{\mathcal{F}_2(f(g))}"{description}, from=4-1, to=4-4]
\end{tikzcd}
\end{center}
commutes in $\textnormal{Cpl}$ for all $g\in G_1$ if and only if for all $g\in G_1$
\begin{enumerate}
    \item $\rho\circ \alpha_{1, g}=\alpha_{2, f(g)}\circ \rho$
    \item $\chi\circ \beta_{1, g}=\textnormal{inc}\circ\beta_{2, f(g)}\circ \chi$
    \item $\xi\circ (U_{1, g}\otimes_\rho \alpha_{2, f(g)})=U_{2, f(g)}\circ \xi$.
    \item $\eta\circ (W_{1, g}\otimes_\chi \beta_{2, f(g)})=W_{2, f(g)}\circ \eta$
\end{enumerate}

\end{theorem}

\begin{proof}\

\noindent Assume the diagram commutes for all $g\in G_1$. The assignment $\tau:\mathcal{F}_1\to f^*\mathcal{F}_2$ determined by the data $((\rho, \chi, \eta), (\rho, \xi))$ defines a natural transformation. Thus $(f, \tau)$ forms a morphism from $(G_1, \mathcal{F}_1)$ to $(G_2, \mathcal{F}_2)$.  As $(f, \tau)$ is a morphism, we have for all $g\in G_1$
$$\rho\circ \alpha_{1, g}=\alpha_{2, f(g)}\circ\rho\quad\quad \chi\circ\beta_{1, g}=\text{inc}\circ \beta_{2, f(g)}\circ\chi$$
Thus $(1)$ and $(2)$ hold. Note, as $(1)$ and $(2)$ hold, the prior lemma implies $U_{1, g}\otimes_\rho \alpha_{2, f(g)}$ and $W_{1, g}\otimes_\chi \beta_{2, f(g)}$ are well defined.

We now prove $(3)$. We observe $U_{1, g}\otimes_\rho \alpha_{2, f(g)}$ makes the following diagram commute:
\begin{center}
    \begin{tikzcd}
	{(E_{A_1}\otimes_\rho A_2)\otimes_{\text{inc}\circ\alpha_{2, f(g)}}A_2} && {E_{A_1}\otimes_{\text{inc}\circ\alpha_{2, f(g)}\circ\rho}A_2=E_{A_1}\otimes_{\rho\circ\alpha_{1, g}}A_2} \\
	\\
	&& {(E_{A_1}\otimes_{\text{inc}\circ\alpha_{1,g}} A_1)\otimes_{\rho}A_2} \\
	\\
	{E_{A_1}\otimes_\rho A_2} && {E_{A_1}\otimes_\rho A_2}
	\arrow["{U_1}"{description}, from=1-1, to=1-3]
	\arrow["{U_2^{-1}}"{description}, from=1-3, to=3-3]
	\arrow["{(U_{1, g}\circ U_{\alpha_{1, g}})\otimes_\rho I}"{description}, from=3-3, to=5-3]
	\arrow["{V_{\text{inc}\circ\alpha_{2, f(g)}}}"{description}, from=5-1, to=1-1]
	\arrow["{U_{1 ,g}\otimes_\rho \alpha_{2, f(g)}}"{description}, from=5-1, to=5-3]
\end{tikzcd}
\end{center}
Note, $U_1$ and $U_2$ are the unitaries from Proposition 3.6 in \cite{brady2026functorialityksgnsconstructionintertwiners}. Furthermore, $(f, \tau)$ being a morphism implies
$$\xi\circ ((U_{1, g}\circ U_{\alpha_{1, g}})\otimes_\rho I)\circ U_2^{-1}=(U_{2, f(g)}\circ U_{\alpha_{2, f(g)}})\circ (\xi\otimes_\rho I)\circ U_1^{-1}$$
Therefore
$$\xi\circ(U_{1, g}\otimes_\rho \alpha_{2, f(g)})=U_{2, f(g)}\circ U_{\alpha_{2, f(g)}}\circ(\xi\otimes_\rho I)\circ V_{\text{inc}\circ\alpha_{2, f(g)}}.$$

We claim $\xi=U_{\alpha_{2, f(g)}}\circ(\xi\otimes_\rho I)\circ V_{\text{inc}\circ\alpha_{2, f(g)}}$. Using continuity and linearity of the maps as well as density of $E_{A_1}\otimes_{A_1}A_2$ in $E_{A_1}\otimes_\rho A_2$, it suffices to show the maps agree on elements of the form $x\dot{\otimes}a_2\in E_{A_1}\otimes_{A_1}A_2$. Given such an element and an approximate unit $(a_\lambda)_{\lambda\in\Lambda}$ for $A_2$, we have
\begin{align*}
    (U_{\alpha_{2, f(g)}}\circ(\xi\otimes_\rho I)\circ V_{\text{inc}\circ\alpha_{2, f(g)}})(x\dot{\otimes}a_2)&=\lim\limits_{\lambda\to\infty}(U_{\alpha_{2, f(g)}}\circ (\eta\otimes_\rho I))((x\dot{\otimes} a_2)\dot{\otimes}a_\lambda)\\
    &=\lim\limits_{\lambda\to\infty}U_{\alpha_{2, f(g)}}(\xi(x\dot{\otimes}a_2)\dot{\otimes}a_\lambda)\\
    &=\lim\limits_{\lambda\to\infty}\xi(x\dot{\otimes} a_2)\alpha_{2, f(g)}^{-1}(a_\lambda)=\xi(x\dot{\otimes}b_2)
\end{align*}
Hence $\xi=U_{\alpha_{2, f(g)}}\circ(\xi\otimes_\rho I)\circ V_{\text{inc}\circ\alpha_{2, f(g)}}$ showing $\xi\circ (U_{1, g}\otimes_{\rho}\alpha_{2, f(g)})=U_{2, f(g)}\circ \xi$.

Now we prove $(4)$. We observe $W_{1, g}\otimes_\chi \beta_{2, f(g)}$ makes the following diagram commute:
\begin{center}
    \begin{tikzcd}
	{(E_{B_1}\otimes_\chi B_2)\otimes_{\text{inc}\circ\beta_{2, f(g)}}B_2} && {E_{B_1}\otimes_{\text{inc}\circ\beta_{2, f(g)}\circ\chi}B_2=E_{B_1}\otimes_{\chi\circ\beta_{1, g}}B_2} \\
	\\
	&& {(E_{B_1}\otimes_{\text{inc}\circ\beta_{1,g}} B_1)\otimes_{\chi}B_2} \\
	\\
	{E_{B_1}\otimes_\chi B_2} && {E_{B_1}\otimes_\chi B_2}
	\arrow["{U_3}"{description}, from=1-1, to=1-3]
	\arrow["{U_4^{-1}}"{description}, from=1-3, to=3-3]
	\arrow["{\widehat{W_{1, g}\circ U_{\beta_{1, g}}}}"{description}, from=3-3, to=5-3]
	\arrow["{V_{\text{inc}\circ\beta_{2, f(g)}}}"{description}, from=5-1, to=1-1]
	\arrow["{W_{1 ,g}\otimes_\chi \beta_{2, f(g)}}"{description}, from=5-1, to=5-3]
\end{tikzcd}
\end{center}
Note, $U_3$ and $U_4$ are the unitaries from Proposition \ref{Prop: induced map from interior}. Furthermore, $(f, \tau)$ being a morphism implies
$$\eta\circ \widehat{(W_{1, g}\circ U_{\beta_{1, g}})}\circ U_4^{-1}=W_{2, f(g)}\circ U_{\beta_{2, f(g)}}\circ \hat{\eta}\circ U_3^{-1}.$$
Therefore
$$\eta\circ(W_{1, g}\otimes_\chi \beta_{2, f(g)})=W_{2, f(g)}\circ U_{\beta_{2, f(g)}}\circ\hat{\eta}\circ V_{\text{inc}\circ\beta_{2, f(g)}}.$$

We claim $\eta=U_{\beta_{2, f(g)}}\circ\hat{\eta}\circ V_{\text{inc}\circ\beta_{2, f(g)}}$. Using continuity and linearity of the maps as well as density of $E_{B_1}\otimes_{B_1}B_2$ in $E_{B_1}\otimes_\chi B_2$, it suffices to show the maps agree on elements of the form $x\dot{\otimes}b_2\in E_{B_1}\otimes_{B_1}B_2$. Given such an element and an approximate unit $(b_\lambda)_{\lambda\in\Lambda}$ for $B_2$, we have
\begin{align*}
    (U_{\beta_{2, f(g)}}\circ\hat{\eta}\circ V_{\text{inc}\circ\beta_{2, f(g)}})(x\dot{\otimes}b_2)&=\lim\limits_{\lambda\to\infty}(U_{\beta_{2, f(g)}}\circ \hat{\eta})((x\dot{\otimes} b_2)\dot{\otimes}b_\lambda)\\
    &=\lim\limits_{\lambda\to\infty}U_{\beta_{2, f(g)}}(\eta(x\dot{\otimes}b_2)\dot{\otimes}b_\lambda)\\
    &=\lim\limits_{\lambda\to\infty}\eta(x\dot{\otimes} b_2)\beta_{2, f(g)}^{-1}(b_\lambda)=\eta(x\dot{\otimes}b_2)
\end{align*}
Hence $\eta=U_{\beta_{2, f(g)}}\circ\hat{\eta}\circ V_{\text{inc}\circ\beta_{2, f(g)}}$ showing $\eta\circ (W_{1, g}\otimes_\chi \beta_{2, f(g)})=W_{2, f(g)}\circ \eta$. 

Now assume $(1)-(4)$ holds. We observe commutativity of the diagram is equivalent to saying the morphism
$$((\rho\circ\alpha_{1, g}, \chi\circ\beta_{1, g}, \eta\circ \widehat{(W_{1, g}\circ U_{\beta_{1, g}})}\circ U_2^{-1}), (\rho\circ\alpha_{1, g}, \xi\circ ((U_{1, g}\circ U_{\alpha_{1, g}})\otimes_\rho I)\circ U_4^{-1}))$$
is the same as the morphism
$$((\alpha_{2, f(g)}\circ\rho, \textnormal{inc}\circ\beta_{2, f(g)}\circ\chi,  (W_{2, f(g)}\circ U_{\beta_{2, f(g)}})\circ\hat{\eta} \circ U_1^{-1}), (\alpha_{2, f(g)}\circ\rho, (U_{2, f(g)}\circ U_{\alpha_{2, f(g)}})\circ (\xi\otimes_\rho I)\circ U_3^{-1}))$$
for all $g\in G_1$. As we have assumed 
$$\rho\circ \alpha_{1, g}=\alpha_{2, f(g)}\circ \rho\quad\quad \chi\circ \beta_{1, g}=\textnormal{inc}\circ\beta_{2, f(g)}\circ \chi,$$
we just need to show
$$\eta\circ \widehat{(W_{1, g}\circ U_{\beta_{1, g}})}\circ U_1^{-1}=(W_{2, f(g)}\circ U_{\beta_{2, f(g)}})\circ\hat{\eta} \circ U_3^{-1}$$
and
$$\xi\circ ((U_{1, g}\circ U_{\alpha_{1, g}})\otimes_\rho I)\circ U_2^{-1}=(U_{2, f(g)}\circ U_{\alpha_{2, f(g)}})\circ (\xi\otimes_\rho I)\circ U_4^{-1}.$$

We start by verifying 
$$\eta\circ \widehat{(W_{1, g}\circ U_{\beta_{1, g}})}\circ U_1^{-1}=(W_{2, f(g)}\circ U_{\beta_{2, f(g)}})\circ\hat{\eta} \circ U_3^{-1}.$$
We observe the following diagram commutes for each $g\in G_1$:
\begin{center}
    \begin{tikzcd}
	{(E_{B_1}\otimes_\chi B_2)\otimes_{\text{inc}\circ\beta_{2, f(g)}}B_2} && {E_{B_1}\otimes_{\text{inc}\circ\beta_{2, f(g)}\circ\chi}B_2=E_{B_1}\otimes_{\chi\circ\beta_{1, g}}B_2} \\
	\\
	&& {(E_{B_1}\otimes_{\text{inc}\circ\beta_{1,g}} B_1)\otimes_{\chi}B_2} \\
	\\
	{E_{B_1}\otimes_\chi B_2} && {E_{B_1}\otimes_\chi B_2}
	\arrow["{U_1}"{description}, from=1-1, to=1-3]
	\arrow["{U_2^{-1}}"{description}, from=1-3, to=3-3]
	\arrow["{\widehat{W_{1, g}\circ U_{\beta_{1, g}}}}"{description}, from=3-3, to=5-3]
	\arrow["{V_{\text{inc}\circ\beta_{2, f(g)}}}"{description}, from=5-1, to=1-1]
	\arrow["{W_{1 ,g}\otimes_\chi \beta_{2, f(g)}}"{description}, from=5-1, to=5-3]
\end{tikzcd}
\end{center}
As we have
$$\eta=U_{\beta_{2, f(g)}}\circ\hat{\eta}\circ V_{\text{inc}\circ\beta_{2, f(g)}}\quad\quad \eta\circ (W_{1, g}\otimes_\chi \beta_{2, f(g)})=W_{2, f(g)}\circ \eta,$$commutativity of the diagram implies
$$\eta\circ \widehat{(W_{1, g}\circ U_{\beta_{1, g}})}\circ U_2^{-1}=(W_{2, f(g)}\circ U_{\beta_{2, f(g)}})\circ\hat{\eta} \circ U_1^{-1}.$$

We now verify
$$\xi\circ ((U_{1, g}\circ U_{\alpha_{1, g}})\otimes_\rho I)\circ U_4^{-1}=(U_{2, f(g)}\circ U_{\alpha_{2, f(g)}})\circ (\xi\otimes_\rho I)\circ U_3^{-1}.$$
We observe the following diagram commtues for each $g\in G_1$:
\begin{center}
    \begin{tikzcd}
	{(E_{A_1}\otimes_\rho A_2)\otimes_{\text{inc}\circ\alpha_{2, f(g)}}A_2} && {E_{A_1}\otimes_{\text{inc}\circ\alpha_{2, f(g)}\circ\rho}A_2=E_{A_1}\otimes_{\rho\circ\alpha_{1, g}}A_2} \\
	\\
	&& {(E_{A_1}\otimes_{\text{inc}\circ\alpha_{1,g}} A_1)\otimes_{\rho}A_2} \\
	\\
	{E_{A_1}\otimes_\rho A_2} && {E_{A_1}\otimes_\rho A_2}
	\arrow["{U_3}"{description}, from=1-1, to=1-3]
	\arrow["{U_4^{-1}}"{description}, from=1-3, to=3-3]
	\arrow["{(U_{1, g}\circ U_{\alpha_{1, g}})\otimes_\rho I}"{description}, from=3-3, to=5-3]
	\arrow["{V_{\text{inc}\circ\alpha_{2, f(g)}}}"{description}, from=5-1, to=1-1]
	\arrow["{U_{1 ,g}\otimes_\rho \alpha_{2, f(g)}}"{description}, from=5-1, to=5-3]
\end{tikzcd}
\end{center}
As we have
$$\xi=U_{\alpha_{2, f(g)}}\circ(\xi\otimes_\rho I)\circ V_{\text{inc}\circ\beta_{2, f(g)}}\quad\quad \xi\circ (U_{1, g}\otimes_\rho \alpha_{2, f(g)})=U_{2, f(g)}\circ \xi,$$
commutativity of the diagram implies
$$\xi\circ ((U_{1, g}\circ U_{\alpha_{1, g}})\otimes_\rho I)\circ U_4^{-1}=(U_{2, f(g)}\circ U_{\alpha_{2, f(g)}})\circ (\xi\otimes_\rho I)\circ U_3^{-1}.$$

\end{proof}

\begin{corollary}\label{Cor: building morphism in EqCpl}\

\noindent For $i=1, 2$, let $(G_i, \mathcal{F}_i)\in\textnormal{EqCpl}$ with corresponding $G_i$-equivariant couplings given by
$$(((A_i, G_i, \alpha_i), (B_i, G_i, \beta_i), ((E_{B_i}, \phi_i), W_i)), ((A_i, G_i, \alpha_i), E_{A_i}, U_i)).$$ 
Assume we have the following data:
\begin{enumerate}
    \item a group homomorphism $f:G_1\to G_2$ 
    \item a non-degenerate $*$-algebra homomorphism $\rho:A_1\to A_2$.
    \item a non-degenerate $*$-algebra homomorphism $\chi:B_1\to M(B_2)$ with $\chi(B_1)\subset B_2$.
    \item an adjointable map $\xi:E_{A_1}\otimes_\rho A_2\to E_{A_2}$.
    \item an adjointable map $\eta:E_{B_1}\otimes_{\chi}B_2\to E_{B_2}$.
\end{enumerate}
Then the data $(f, ((\rho, \chi, \eta), (\rho, \xi)))$ determines a morphism $(f, \tau):(G_1, \mathcal{F}_1)\to (G_2, \mathcal{F}_2)$ for which the associated data to $\tau$ is $((\rho, \chi, \eta), (\rho, \xi))$ if and only if
\begin{enumerate}
    \item for all $a\in A_1$, $\phi_{2}(\rho(a))\circ \eta=\eta\circ (\phi_1(a)\otimes_\chi I)$.
    \item for all $g\in G_1$, $\rho\circ \alpha_{1, g}=\alpha_{2, f(g)}\circ \rho$
    \item for all $g\in G_1$, $\chi\circ \beta_{1, g}=\textnormal{inc}\circ\beta_{2, f(g)}\circ \chi$
    \item for all $g\in G_1$, $\xi\circ (U_{1, g}\otimes_\rho \alpha_{2, f(g)})=U_{2, f(g)}\circ \xi$.
    \item for all $g\in G_1$, $\eta\circ (W_{1, g}\otimes_\chi \beta_{2, f(g)})=W_{2, f(g)}\circ \eta$
\end{enumerate}

\end{corollary}

\begin{proof}\

\noindent If the data determines a morphism $(f, \tau):(G_1, \mathcal{F}_1)\to (G_2, \mathcal{F}_2)$ for which the associated data to $\tau$ is $((\rho, \chi, \eta), (\rho, \xi))$, then the prior Theorem implies $(2)-(5)$ holds. Furthermore, as $(\rho, \chi, \eta)$ is a morphism from $(A_1, B_1, (E_{B_1}, \phi_1))$ to $(A_2, B_2, (E_{B_2}, \phi_2))$ in $\textnormal{NDPosCor}_{*\textnormal{-alg}}$, then $(1)$ holds. Now suppose $(1)-(5)$ holds. Condition $(1)$ implies 
$$(\rho, \chi, \eta):(A_1, B_1, (E_{B_1}, \phi_1))\to (A_2, B_2, (E_{B_2}, \phi_2))$$
is a morphism in $\textnormal{NDPosCor}_{*\textnormal{-alg}}$. Therefore  we obtain a morphism
$$((\rho, \chi, \eta), (\rho, \xi)):((A_1, B_1, (E_{B_1}, \phi_1)), (A_1, E_{A_1}))\to ((A_2, B_2, (E_{B_2}, \phi_2)), (A_2, E_{A_2}))$$
in $\textnormal{Cpl}$. Using the prior Theorem, conditions $(2)-(5)$ implies this morphism defines a natural transformation $\tau:\mathcal{F}_1\to f^*\mathcal{F}_2$. Hence we obtain a morphism $(f, \tau):(G_1, \mathcal{F}_1)\to (G_2, \mathcal{F}_2)$ for which the associated data to $\tau$ is $((\rho, \chi, \eta), (\rho, \xi))$

\end{proof}

\section{Functoriality of Constructing The Reference Frame System}

Our goal in this section is to prove the construction of the reference frame system from the data of a $G$-equivariant coupling yields a functor $\textnormal{EqCpl}_{*\textnormal{-iso}}\to \textnormal{C}^*_{\textnormal{CP}}$. To get started, we recall the reference frame system for a $G$-equivariant coupling is constructed in three steps:
\begin{enumerate}
    \item we apply the equivariant KSGNS construction to dilate the non-degenerate positive equivariant $C^*$-correspondence to an equivariant $C^*$-correspondence.
    \item we utilize the equivariant interior tensor product with the dilated correspondence to construct a Hilbert module dynamical system.
    \item we assemble the reference frame system using the data of the constructed Hilbert module dynamical system.
\end{enumerate}
Our goal for proving functoriality of the entire construction is to show that each of these three steps is functorial. Once this is established, we can compose the three functors to conclude the entire construction is functorial. 

\subsection{Functoriality of Assembling the Reference Frame System}

Working backwards, we start by showing the third step is functorial. Let us first describe the argument for proving this. 

The goal is convert the data of a Hilbert module dynamical system to the data of a $C^*$-algebra, and to convert the data of an adjointable map to the data of a strict completely positive map. Using Theorem \ref{Theorem: functorial characterization of hilbert module dynamical systems}, this assignment should yield a functor
$$\int_{G\in\textnormal{Grp}}\mathcal{U}\textnormal{Fun}(BG, \textnormal{HMod})\to \textnormal{C}^*_{\textnormal{CP}}$$
where $\mathcal{U}\textnormal{Fun}(BG, \textnormal{HMod})$ is the full subcategory of $\textnormal{Fun}(BG, \textnormal{HMod})$ for which the functors are unitary valued. 

Given objects $(G_1, \mathcal{F}_1), (G_2, \mathcal{F}_2)\in \int_{G\in\textnormal{Grp}}\mathcal{U}\textnormal{Fun}(BG, \textnormal{HMod})$ with associated Hilbert module dynamical systems $((A_i, G_i, \alpha_i), E_{A_i}, U_i)$, we obtain an assignment on objects given by $(G_i, \mathcal{F}_i)\mapsto \mathcal{L}(E_{A_i})^{G_i}$. Given a morphism $(f, \tau):(G_1, \mathcal{F}_1)\to (G_2, \mathcal{F}_2)$ with associated data $(\rho, \eta)$ to $\tau$, we utilize the interior tensor product and conjugation by $\eta$ to obtain the strict completely positive map
$$\Phi_{(f, \tau)}:\mathcal{L}(E_{A_1})\to \mathcal{L}(E_{A_2})\quad\quad \Phi_{(f, \tau)}(T)=\eta (T\otimes_\rho I)\eta^*.$$
By taking the assignment
\begin{align*}
\int_{G\in\textnormal{Grp}}\mathcal{U}\textnormal{Fun}(BG, \textnormal{HMod})&\to \textnormal{C}^*_{\textnormal{CP}}\\
    (G, \mathcal{F})&\mapsto \mathcal{L}(E)\\
    ((f, \tau):(G_1, \mathcal{F}_1)\to (G_2, \mathcal{F}_2))&\mapsto (\Phi_{(f, \tau)}:\mathcal{L}(E_1)\to \mathcal{L}(E_2))
\end{align*}
we obtain a covariant functor (see Theorem \ref{Theorem: functor for adjointable operators} below).

To get a functor whose assignment on objects is the $C^*$-algebra $\mathcal{L}(E_{A})^{G}$ rather than $\mathcal{L}(E_{A})$, we verify the map $\Phi_{(f, \tau)}$ is equivariant (see Proposition \ref{Prop: properties of the induce CP map} below); that is,
$$\textnormal{Ad}(U_{2, f(g)})(\Phi_{(f, \tau)}(T))=\Phi_{(f, \tau)}(\textnormal{Ad}(U_{1, g})(T))$$
for all $T\in \mathcal{L}(E_{A_1})$ and $g\in G_1$. Once we have established the equivariance, we can utilize functoriality of the assignment given in Theorem \ref{Theorem: functor for adjointable operators} to know the assignment 
\begin{align*}
    \int_{G\in\textnormal{Grp}}\mathcal{U}\textnormal{Fun}(BG, \textnormal{HMod})&\to \textnormal{C}^*_{\textnormal{CP}}\\
    (G, \mathcal{F})&\mapsto \mathcal{L}(E)^G\\
    ((f, \tau):(G_1, \mathcal{F}_1)\to (G_2, \mathcal{F}_2))&\mapsto (\Phi_{(f, \tau)}:\mathcal{L}(E_1)^G\to \mathcal{L}(E_2)^G)
\end{align*}
defines a covariant functor.

\begin{theorem}\label{Theorem: functor for adjointable operators}\

\noindent  The assignment
\begin{align*}
    \int_{G\in\textnormal{Grp}}\mathcal{U}\textnormal{Fun}(BG, \textnormal{HMod})&\to \textnormal{C}^*_{\textnormal{CP}}\\
    (G, \mathcal{F})&\mapsto \mathcal{L}(E)\\
    ((f, \tau):(G_1, \mathcal{F}_1)\to (G_2, \mathcal{F}_2))&\mapsto (\Phi_{(f, \tau)}:\mathcal{L}(E_1)\to \mathcal{L}(E_2))
\end{align*}
defines a covariant functor.
    
\end{theorem}

\begin{proof}\

\noindent We only verify the assignment is functorial. We begin by checking the assignment on morphisms preserves the identity morphism. Let $(G, \mathcal{F})\in\int_{G\in\textnormal{Grp}}\mathcal{U}\textnormal{Fun}(BG, \textnormal{HMod})$ with associated Hilbert module dynamical system $((A, G, \alpha), E, U)$, and let $(1_G, 1_{\mathcal{F}})$ be the identity morphism for $(G, \mathcal{F})$. Note, $(1_A, \iota)$ is the data associated to $1_\mathcal{F}$ where
$$\iota:E\otimes_{1_A}A\to E\quad\quad x\dot{\otimes}a\mapsto xa$$
is the unitary isomorphism from Proposition \ref{Prop: induced maps for HMod}. Let $x\in E$, and let $(a_\lambda)_{\lambda\in\Lambda}$ be an approximate unit for $A$. For all $T\in \mathcal{L}(E)$,
\begin{align*}
\iota(T\otimes_{1_A}I)\iota^*x=&\lim\limits_{\lambda\to\infty}\iota(T\otimes_{1_A}I)(x\dot{\otimes}a_\lambda)=\lim\limits_{\lambda\to\infty}\iota(Tx\dot{\otimes}a_\lambda)\\
&=\lim\limits_{\lambda\to\infty}(Tx)a_\lambda=Tx
\end{align*}
Hence $\Phi_{(1_G, 1_{\mathcal{F}})}(T)=T$ proving $\Phi_{(1_G, 1_\mathcal{F})}=1_{\mathcal{L}(E)}$.

Now we check the assignment on morphisms preserves composition. For $i=1, 2, 3$, let $(G_i, \mathcal{F}_i)\in \int_{G\in\textnormal{Grp}}\mathcal{U}\textnormal{Fun}(BG, \textnormal{HMod})$ with associated Hilbert module dynamical system $((A_i, G_i, \alpha_i), E_{A_i}, U_i)$, and for $i=1, 2$, let $(f_i, \tau_i):(G_i, \mathcal{F}_i)\to (G_{i+1}, \mathcal{F}_{i+1})$ be morphisms with data $(\rho_i, \eta_i)$ associated to $\tau_i$. Note the data associated to $f_1^*\tau_2\circ\tau_1$ is given by
$$(\rho_2\circ\rho_1, \eta_2\circ (\eta_1\otimes_{\rho_2}I)\circ U^{-1})$$
where
$$U:(E_{A_1}\otimes_{\rho_1}A_2)\otimes_{\rho_2}A_3\to E_{A_1}\otimes_{\rho_2\circ\rho_1}A_3\quad\quad U((x\dot{\otimes}a)\dot{\otimes}a')=x\dot{\otimes}\rho_2(a)a'$$ is the unitary from Proposition \ref{Prop: induced maps for HMod}. Given $T\in \mathcal{L}(E_1)$,
$$U^{-1}(T\otimes_{\rho_2\circ\rho_1}I)U=(T\otimes_{\rho_1}I)\otimes_{\rho_2}I.$$ 
Therefore
\begin{align*}
    \Phi_{(f_2\circ f_1, f_1^*\tau_2\circ\tau_1)}(T)&=\eta_2(\eta_1\otimes_{\rho_2}I)U^{-1}(T\otimes_{\rho_2\circ\rho_1}I)U(\eta_1^*\otimes_{\rho_2}I)\eta_2\\
    &=\eta_2(\eta_1\otimes_{\rho_2}I)((T\otimes_{\rho_1}I)\otimes_{\rho_2}I)(\eta_1^*\otimes_{\rho_2}I)\eta_1^*\\
    &=\eta_2(\eta(T\otimes_{\rho_1}I)\eta^*\otimes_{\rho_2}I)\eta_2^*=(\Phi_{(f_2, \tau_2)}\circ\Phi_{(f_1, \tau_1)})(T)
\end{align*}
Hence the assignment preserves composition.

\end{proof}

\begin{proposition}\

\noindent For $i=1, 2$, let $(G_i, \mathcal{F}_i)\in\int_{G\in\textnormal{Grp}}\mathcal{U}\textnormal{Fun}(BG, \textnormal{HMod})$ with associated Hilbert module dynamical system $((A_i, G_i, \alpha_i), E_i, U_i)$. Let $f:G_1\to G_2$ be a group homomorphism, and let\\ $(\rho, \eta):\mathcal{F}_1(\textnormal{pt})\to \mathcal{F}_2(\textnormal{pt})$ be a morphism in $\textnormal{HMod}$. Then the diagram
\begin{center}
    \begin{tikzcd}
	{\mathcal{F}_1(\textnormal{pt})} &&& {\mathcal{F}_1(\textnormal{pt})} \\
	\\
	{\mathcal{F}_2(\textnormal{pt})} &&& {\mathcal{F}_2(\textnormal{pt})}
	\arrow["{\mathcal{F}_1(g)}"{description}, from=1-1, to=1-4]
	\arrow["{(\rho, \eta)}"{description}, from=1-1, to=3-1]
	\arrow["{(\rho, \eta)}"{description}, from=1-4, to=3-4]
	\arrow["{\mathcal{F}_2(f(g))}"{description}, from=3-1, to=3-4]
\end{tikzcd}
\end{center}
commutes in $\textnormal{HMod}$ for all $g\in G_1$ if and only if for all $g\in G_1$
\begin{enumerate}
    \item $\rho\circ\alpha_{1, g}=\alpha_{2, f(g)}\circ\rho$
    \item $\eta\circ (U_{1, g}\otimes_\rho \alpha_{2, f(g)})=U_{2, f(g)}\circ\eta$
\end{enumerate}

\end{proposition}

\begin{proof}\

\noindent The proof follows the same argument that was presented in Theorem \ref{Theorem: characterization of morphisms}.

\end{proof}

\begin{proposition}\label{Prop: properties of the induce CP map}\

\noindent For $i=1, 2$, let $(G_i, \mathcal{F}_i)\in\int_{G\in\textnormal{Grp}}\mathcal{U}\textnormal{Fun}(BG, \textnormal{HMod})$ with associated Hilbert module dynamical system $((A_i, G_i, \alpha_i), E_i, U_i)$, and let $(f, \tau):(G_1, \mathcal{F}_1)\to (G_2, \mathcal{F}_2)$ be a morphism with associated data $(\rho, \eta)$ to $\tau$.
\begin{enumerate}
    \item $||\Phi_{(f, \tau)}||=||\eta||^2$.
    \item $\Phi_{(f, \tau)}$ is unital if and only if $\eta$ is a coisometry.
    \item If $\eta$ is an isometry, then $\Phi_{(f, \tau)}$ is a $*$-algebra homomorphism.
    \item For all $g\in G_1$, $\textnormal{Ad}(U_{2, f(g)})\circ \Phi_{(f, \tau)}=\Phi_{(f, \tau)}\circ\textnormal{Ad}(U_{1, g})$.
\end{enumerate}

\end{proposition}

\begin{proof}\

\begin{enumerate}
    \item As $\Phi_{(f, \tau)}$ is a completely positive map between unital $C^*$-algebras, we have
$$||\Phi_{(f, \tau)}||=||\Phi_{(f, \tau)}(I)||=||\eta\eta^*||=||\eta||^2.$$
    \item As $\Phi_{(f, \tau)}(I)=\eta\eta^*$, it follows $\Phi_{(f, \tau)}$ is unital if and only if $\eta$ is a coisometry.
    \item Assume $\eta$ is an isometry. As $\Phi_{(f, \tau)}$ is a completely positive map, we just need to verify $\Phi_{(f, \tau)}$ is multiplicative. Given $S, T\in\mathcal{L}(E_1)$, we have
\begin{align*}
    \Phi_{(f, \tau)}(ST)&=\eta(ST\otimes_\rho I)\eta^*=\eta(S\otimes_\rho I)(T\otimes_\rho I)\eta^*\\
    &=\eta(S\otimes_\rho I)\eta^*\eta(T\otimes_\rho I)\eta^*=\Phi_{(f, \tau)}(S)\Phi_{(f, \tau)}(T)
\end{align*}
\item Let $g\in G_1$ and $T\in \mathcal{L}(E_1)$, then, using the prior proposition, we have
\begin{align*}
    \Phi_{(f, \tau)}(&\textnormal{Ad}(U_{1, g})(T))=\eta((U_{1, g}TU_{1, g^{-1}})\otimes_\chi I)\eta^*\\
    &=\eta(U_{1, g}\otimes_{\rho}\alpha_{2, f(g)})(T\otimes_\rho I)(U_{1, g^{-1}}\otimes_\rho \alpha_{2, f(g^{-1})})\eta^*\\
    &=U_{2, f(g)}\eta(T\otimes_\rho I)\eta^*U_{2, f(g^{-1})}=\textnormal{Ad}(U_{2, f(g)})(\Phi_{(f, \tau)}(T))
\end{align*}
\end{enumerate}

\end{proof}

\begin{corollary}\label{Cor: functor for step 3}\

\noindent  The assignment
\begin{align*}
    \int_{G\in\textnormal{Grp}}\mathcal{U}\textnormal{Fun}(BG, \textnormal{HMod})&\to \textnormal{C}^*_{\textnormal{CP}}\\
    (G, \mathcal{F})&\mapsto \mathcal{L}(E)^G\\
    ((f, \tau):(G_1, \mathcal{F}_1)\to (G_2, \mathcal{F}_2))&\mapsto (\Phi_{(f, \tau)}:\mathcal{L}(E_1)^G\to \mathcal{L}(E_2)^G)
\end{align*}
defines a covariant functor.

\end{corollary}

\subsection{Functoriality of the Equivariant Interior Tensor Product}

We now prove the second step defines a functor. First, let us describe the argument for proving this. 

Our goal is to convert the data of a $G$-equivariant coupling via the equivariant interior tensor product to the data of a Hilbert module dynamical system. In order to apply the equivariant interior tensor product, the non-degenerate positive equivariant $C^*$-correspondences for the $G$-equivariant couplings must be an equivariant $C^*$-correspondences. Letting $\textnormal{EqCpl}_{*\textnormal{-alg}}$ denote the full subcategory of $\textnormal{EqCpl}$ for which the objects have this given property, we look to build a functor
$$\textnormal{EqCpl}_{*\textnormal{-alg}}\to \int_{G\in\textnormal{Grp}}\mathcal{U}\textnormal{Fun}(BG, \textnormal{HMod})$$

On objects, the assignment is given by the equivariant interior tensor product. Given an object $(G, \mathcal{F})\in \textnormal{EqCpl}_{*\textnormal{-alg}}$ with associated $G$-equivariant coupling
$$(((A, G, \alpha), (B, G, \beta), ((E_B, \phi), W)), ((A, G, \alpha), E_A, U)),$$
we obtain the assignment $(G, \mathcal{F})\mapsto (G, \mathcal{F}_\otimes)$ where $\mathcal{F}_\otimes:BG\to \textnormal{HMod}$ is the unitary valued functor corresponding to the Hilbert module dynamical system
$$((B, G, \beta), E_{A}\otimes_\phi E_B, U\otimes_\phi W).$$

The assignment on morphisms is a bit more complicated. Let us fix objects $(G_1, \mathcal{F}_1), (G_2, \mathcal{F}_2)$ in  $\textnormal{EqCpl}_{*\textnormal{-alg}}.$ with corresponding $G_i$-equivariant couplings
$$(((A_i, G_i, \alpha_i), (B_i, G_i, \beta_i), ((E_{B_i}, \phi_i), W_i)), ((A_i, G_i, \alpha_i), E_{A_i}, U_i)),$$
Given a morphism $(f, \tau):(G_1, \mathcal{F}_1)\to (G_2, \mathcal{F}_2)$ with associated data $((\rho, \chi, \eta), (\rho, \xi))$ to $\tau$, we look to build a morphism
$$(f, \tau_{\otimes}):(G_1, \mathcal{F}_{1, \otimes})\to (G_2, \mathcal{F}_{2, \otimes}).$$
To specify the natural transformation $\tau_{\otimes}$, it suffices to specify a morphism
$$(\gamma, \zeta):\mathcal{F}_{1, \otimes}(\textnormal{pt})\to \mathcal{F}_{2, \otimes}(\textnormal{pt})$$
which makes the following diagram commute in $\textnormal{HMod}$ for all $g\in G_1$:
\begin{center}
\begin{tikzcd}[every label/.append
style={font=\fontsize{9}{11}}]
	{\mathcal{F}_{1, \otimes}(\textnormal{pt})} &&& {\mathcal{F}_{1, \otimes}(\textnormal{pt})} \\
	\\
	\\
	{\mathcal{F}_{2, \otimes}(\textnormal{pt})} &&& {\mathcal{F}_{2, \otimes}(\textnormal{pt})}
	\arrow["{\mathcal{F}_{1, \otimes}(g)}"{description}, from=1-1, to=1-4]
	\arrow["{(\gamma, \zeta)}"{description}, from=1-1, to=4-1]
	\arrow["{(\gamma, \zeta)}"{description}, from=1-4, to=4-4]
	\arrow["{\mathcal{F}_{2, \otimes}(f(g))}"{description}, from=4-1, to=4-4]
\end{tikzcd}
\end{center}
By taking $\gamma=\chi$, we only need to construct the map $\zeta$. 

The idea for $\zeta$ is to take $\zeta=\xi\otimes \eta$; however, due to domain issues, we need to modify $\xi\otimes \eta$. In particular, we take $\zeta:=\Lambda_{(f, \tau)}$ as the adjointable map making the following diagram commute: 
\begin{center}
    \begin{tikzcd}
	{(E_{A_1}\otimes_{\phi_1}E_{B_1})\otimes_\chi B_2} &&& {E_{A_1}\otimes_{\phi_1\otimes_\chi I}(E_{B_1}\otimes_\chi B_2)} \\
	\\
	&&& {E_{A_1}\otimes_{\phi_2\circ\rho}E_{B_2}} \\
	\\
	{E_{A_2}\otimes_{\phi_2}E_{B_2}} &&& {(E_{A_1}\otimes_\rho A_2)\otimes_{\phi_2}E_{B_2}}
	\arrow["\mu"{description}, from=1-1, to=1-4]
	\arrow["{\Lambda_{(f, \tau)}}"{description}, dashed, from=1-1, to=5-1]
	\arrow["{I\otimes\eta}"{description}, from=1-4, to=3-4]
	\arrow["\sigma^*"{description}, from=3-4, to=5-4]
	\arrow["{\xi\otimes_{\phi_2}I}", from=5-4, to=5-1]
\end{tikzcd}
\end{center}
Note, in the diagram, we have
$$\mu:(E_{A_1}\otimes_{\phi_1} E_{B_1})\otimes_\chi B_2\to E_{A_1}\otimes_{\phi_1\otimes_\chi I}(E_{B_1}\otimes_\chi B_2)\quad\quad \mu((x\dot{\otimes}y)\dot{\otimes}b)=x\dot{\otimes}(y\dot{\otimes}b),$$
$$I\otimes\eta:E_{A_1}\otimes_{\phi_1\otimes_\chi I}(E_{B_1}\otimes_\chi B_2)\to E_{A_1}\otimes_{\phi_2\circ\rho} E_{B_2} \quad\quad (I\otimes\eta)(x\dot{\otimes}(y\dot{\otimes}b))=x\dot{\otimes}\eta(y\dot{\otimes}b),$$
$$\sigma:(E_{A_1}\otimes_{\phi_1} E_{B_1})\otimes_\chi B_2\to E_{A_1}\otimes_{\phi_1\otimes_\chi I}(E_{B_1}\otimes_\chi B_2)\quad\quad \sigma((x\dot{\otimes}a)\dot{\otimes}y)=x\dot{\otimes}\phi_2(a)y,$$
and
$$\xi\otimes_{\phi_2}I:(E_{A_1}\otimes_{\rho}A_2)\otimes_{\phi_2} E_{B_2}\to E_{A_2}\otimes_{\phi_2}E_{B_2}\quad\quad (\xi\otimes_{\phi_2}I)((x\dot{\otimes}a)\dot{\otimes}y)=\xi(x\dot{\otimes}a)\dot{\otimes}y.$$
As verified in Lemma \ref{Lemma: Check of Map 1} , Lemma \ref{Lemma: Check of Map 2}, Lemma \ref{Lemma: Check of Map 3}
, and Proposition \ref{Prop: induced map from interior}, respectively, all four maps are well-defined, and the maps $\mu$ and $\sigma$ are unitary isomorphisms. 

Using the data $(\sigma, \Lambda_{(f, \tau)})$, we verify in Proposition \ref{Prop: Lambda is a morphism} that we obtain a morphism
$$(f, \tau_\otimes):(G_1, \mathcal{F}_{1, \otimes})\to (G_2, \mathcal{F}_{2, \otimes}).$$
As verified in Theorem \ref{Theorem: functor for step 2}, the assignments on objects and morphisms yield a covariant functor
$$\textnormal{EqCpl}_{*\textnormal{-alg}}\to\int_{G\in\textnormal{Grp}}\mathcal{U}\textnormal{Fun}(BG, \textnormal{HMod}).$$

We now fill in the details for showing step two is functorial. We begin by verifying the maps needed to define $\Lambda_{(f, \tau)}$ are well-defined adjointable maps as well as $\mu$ and $\sigma$ are unitary isomorphisms.

\begin{lemma}\label{Lemma: Check of Map 1}\

\noindent For $i=1, 2$, let $(G_i, \mathcal{F}_i)\in\textnormal{EqCpl}_{*\textnormal{-alg}}$ with associated $G_i$-equivariant coupling given by
$$(((A_i, G_i, \alpha_i), (B_i, G_i, \beta_i), ((E_{B_i}, \phi_i), W_i), ((A_i, G_i, \alpha_i), E_{A_i}, U_i)).$$
If $(f, \tau):(G_1, \mathcal{F}_1)\to (G_2, \mathcal{F}_2)$ is a morphism with associated data $((\rho, \chi, \eta), (\rho, \xi))$ to $\tau$,  then
\begin{align*}
  \mu:(E_{A_1}\otimes_{\phi_1} E_{B_1})\otimes_{\chi} B_2\to E_{A_1}\otimes_{\phi_1\otimes_{\chi}I}(E_{B_1} \otimes_{\chi}B_2)\quad\quad \mu((x\dot{\otimes}y)\dot{\otimes}b)=x\dot{\otimes}(y\dot{\otimes}b)
\end{align*}
is a well-defined unitary. 

\end{lemma}

\begin{proof}\

\noindent First, the map
$$(E_{A_1}\otimes_{A_1}E_{B_1})\otimes_{B_1}B_2\to E_{A_1}\otimes_{\phi_1\otimes_{\chi}I}(E_{B_1} \otimes_{\chi}B_2)\quad\quad (x\dot{\otimes}y)\dot{\otimes}b\mapsto x\dot{\otimes}(y\dot{\otimes}b)$$
is a well-defined $B_2$-linear map. Furthermore, since
\begin{align*}
    \left\|\sum\limits_{k=1}^n\sum\limits_{\ell=1}^{m_k} x_{k, \ell}\dot{\otimes} (y_{k, \ell}\dot{\otimes}b_k)\right\|^2
    &=\left\|\sum\limits_{i, j=1}^n\sum\limits_{k=1}^{m_i}\sum\limits_{ \ell=1}^{m_j} \langle y_{i, k}\dot{\otimes}b_k, (\phi_1\otimes_\chi I)(\langle x_{i, k}, x_{j, \ell}\rangle)(y_{j, \ell}\dot{\otimes}b_j)\rangle\right\|_{B_2}\\
    &=\left\|\sum\limits_{i, j=1}^n\sum\limits_{k=1}^{m_i}\sum\limits_{ \ell=1}^{m_j} b_i^*\chi\left(\langle y_{i, k}, \phi_1(\langle x_{i, k}, x_{j, \ell}\rangle)y_{j, \ell}\right)b_j\right\|_{B_2}\\
    &=\left\|\sum\limits_{k=1}^n\left(\sum\limits_{\ell=1}^{m_k} x_{k, \ell}\dot{\otimes} y_{k, \ell}\right)\dot{\otimes} b_{k}\right\|^2
\end{align*}
the map defines a $B_2$-linear isometry. As $(E_{A_1}\otimes_{A_1}E_{B_1})\otimes_{\chi}B_2$ is dense in $(E_{A_1}\otimes_{\phi_1} E_{B_1})\otimes_\chi B_2$, we can extend the map to a $B_2$-linear isometry 
$$\mu:(E_{A_1}\otimes_{\phi_1}E_{B_1})\otimes_{\chi}B_2\to E_{A_1}\otimes_{\phi_1\otimes_{\chi}I}(E_{B_1} \otimes_{\chi}B_2)\quad\quad (x\dot{\otimes}y)\dot{\otimes}b\mapsto x\dot{\otimes}(y\dot{\otimes}b)$$
Since the image of $\mu$ contains $E_{A_1}\otimes_{A_1}(E_{B_1}\otimes_{B_1}B_2)$ which is dense in $E_{A_1}\otimes_{\phi_1\otimes_{\chi}I}(E_{B_1}\otimes_\chi B_2)$ and the image of $\mu$ is closed (isometry from a complete metric space), it follows $\mu$ is surjective. As $\mu$ is a bijective $B_2$-linear isometry, it follows $\mu$ is an adjointable unitary.

\end{proof}

\begin{lemma}\label{Lemma: Check of Map 3}\

\noindent For $i=1, 2$, let $(G_i, \mathcal{F}_i)\in \textnormal{EqCpl}_{*\textnormal{-alg}}$ with associated $G_i$-equivariant coupling given by
$$(((A_i, G_i, \alpha_i), (B_i, G_i, \beta_i), ((E_{B_i}, \phi_i), W_i), ((A_i, G_i, \alpha_i), E_{A_i}, U_i)).$$
If $(f, \tau):(G_1, \mathcal{F}_1)\to (G_2, \mathcal{F}_2)$ is a morphism with associated data $((\rho, \chi, \eta), (\rho, \xi))$ to $\tau$, then
$$\sigma:(E_{A_1}\otimes_{\rho}A_2)\otimes_{\phi_2} E_{B_2}\to E_{A_1}\otimes_{\phi_2\circ\rho} E_{B_2}\quad\quad \sigma((x\dot{\otimes}a)\dot{\otimes}y)=x\dot{\otimes}\phi_2(a)y$$
is a well-defined unitary. 

\end{lemma}

\begin{proof}\

\noindent First, the map
$$(E_{A_1}\otimes_{A_1}A_2)\otimes_{A_2}E_{B_2}\to E_{A_1}\otimes_{\phi_2\circ\rho}E_{B_2}\quad\quad (x\dot{\otimes}a)\dot{\otimes}y\mapsto x\dot{\otimes}\phi_2(a)y$$
is a well-defined $B_2$-linear map. Furthermore, since
\begin{align*}
    \left\|\sum\limits_{k=1}^n\sum\limits_{\ell=1}^{m_k} x_{k, \ell}\dot{\otimes} \phi_{2}(a_{k, \ell})y_k\right\|^2&=\left\|\sum\limits_{i, j=1}^n\sum\limits_{k=1}^{m_i}\sum\limits_{ \ell=1}^{m_j}\langle \phi_2(a_{i, k})y_k, \phi_2(\rho(\langle x_{i, k}, x_{j, \ell}\rangle))\phi_2(a_{j, \ell})y_j\rangle\right\|_{B_2}\\
    &=\left\|\sum\limits_{i, j=1}^n\sum\limits_{k=1}^{m_i}\sum\limits_{ \ell=1}^{m_j}\langle y_k, \phi_2(a^*_{i, k}\rho(\langle x_{i, k}, x_{j, \ell}\rangle)a_{j, \ell})y_j\rangle\right\|_{B_2}\\
    &=\left\|\sum\limits_{k=1}^n\left(\sum\limits_{\ell=1}^{m_k} x_{k, \ell}\dot{\otimes} a_{k, \ell}\right)\dot{\otimes} y_{k}\right\|^2
\end{align*}
the map defines a $B_2$-linear isometry. Therefore we can extend the map to a $B_2$-linear isometry
$$\sigma:(E_{A_1}\otimes_{\rho}A_2)\otimes_{\phi_2} E_{B_2}\to E_{A_1}\otimes_{\phi_2\circ\rho} E_{B_2}\quad\quad \sigma((x\dot{\otimes}a)\dot{\otimes}y)=x\dot{\otimes}\phi_2(a)y$$

We claim $\sigma$ is surjective. To prove this, we show $E_{A_1}\otimes_{A_1}E_{B_2}$ is contained in the image of $\sigma$. As $E_{A_1}\otimes_{A_1}E_{B_2}$ is dense in $E_{A_1}\otimes_{\phi_2\circ\rho}E_{B_2}$ and the image of $\sigma$ is closed, it will follow that $\sigma$ is surjective. As $\sigma$ is linear, it suffices to show the every simple tensor $x\dot{\otimes}y\in E_{A_1}\otimes_{A_1}E_{B_2}$ is contained in the image of $\sigma$. 

Fix $x\dot{\otimes}y\in E_{A_1}\otimes_{A_1}E_{B_2}$ and let $(a_\lambda)_{\lambda\in\Lambda}$ be an approximate unit in $A_2$. As $\phi_2$ is a non-degenerate $*$-algebra homomorphism, we know $(\phi(a_\lambda))_{\lambda\in\Lambda}$ converges to the identity operator in the strict topology; i.e, $\phi_2(a_\lambda)y\to y$ with respect to the norm on $E_{B_2}$. Using this as well as
$$||x\dot{\otimes}(\phi_2(a_\lambda)y-y)||\leq \left\|\phi_2(\rho(\langle x, x\rangle))\right\|^{\frac{1}{2}}\left\|\phi(a_\lambda)y-y\right\|,$$
it follows $x\dot{\otimes}\phi_2(a_\lambda)y\to x\dot{\otimes}y$. Using that $\sigma$ is an isometry, it follows $((x\dot{\otimes}a_\lambda)\dot{\otimes}y)_{\lambda\in\Lambda}$ is a Cauchy net in $(E_{A_1}\otimes_\rho A_2)\otimes_{\phi_2}E_{B_2}$ and therefore converges to a point $z$. Using continuity of $\sigma$, it follows $\sigma(z)=x\dot{\otimes}y$. Hence $\sigma$ is surjective.

As $\sigma$ is a bijective $B_2$-linear isometry, it follows $\sigma$ is an adjointable unitary map proving the claim.
    
\end{proof}

\begin{remark}\

\noindent We observe the adjoint of $\sigma$ on simple tensors is given by
$$ \sigma^*(x\dot{\otimes}y)=\lim\limits_{\lambda\to\infty}(x\dot{\otimes}a_\lambda)\dot{\otimes} y$$
where $(a_\lambda)_{\lambda\in\Lambda}$ is any approximate unit for $A_2$.

\end{remark}

\begin{lemma}\label{Lemma: Check of Map 2}\

\noindent For $i=1, 2$, let $(G_i, \mathcal{F}_i)\in\textnormal{EqCpl}_{*\textnormal{-alg}}$ with associated $G_i$-equivariant coupling given by
$$(((A_i, G_i, \alpha_i), (B_i, G_i, \beta_i), ((E_{B_i}, \phi_i), W_i), ((A_i, G_i, \alpha_i), E_{A_i}, U_i)).$$
If $(f, \tau):(G_1, \mathcal{F}_1)\to (G_2, \mathcal{F}_2)$ is a morphism with associated data $((\rho, \chi, \eta), (\rho, \xi))$ to $\tau$,  then
$$I\otimes\eta:E_{A_1}\otimes_{\phi_1\otimes_\chi I}(E_{B_1}\otimes_\chi B_2)\to E_{A_1}\otimes_{\phi_2\circ\rho} E_{B_2}\quad\quad (I\otimes \eta)(x\dot{\otimes} z)=x\dot{\otimes}\eta(z)$$
is a well-defined adjointable $B_2$-linear map.

\end{lemma}

\begin{proof}\

\noindent First, as $\eta\circ (\phi_1(a)\otimes_\chi I)=(\phi_2\circ \rho)(a)\circ \eta$ for all $a\in A_1$, it follows
$$E_{A_1}\otimes_{A_1}(E_{B_1}\otimes_{\chi}B_2)\to E_{A_1}\otimes_{\phi_2\circ\rho} E_{B_2}\quad\quad x\dot{\otimes} z\mapsto x\dot{\otimes}\eta(z)$$
is a well-defined $B_2$-linear map. Furthermore, as
\begin{align*}
    \left\|\sum\limits_{k=1}^nx_{k}\dot{\otimes} \eta(z_k)\right\|^2&=\left\|\sum\limits_{i, j=1}^n\langle \eta(z_i), \phi_2(\rho(\langle x_{i}, x_{ j}\rangle))\eta(z_j)\rangle\right\|_{B_2}\\
    &=\left\|\sum\limits_{i, j=1}^n\langle z_{i}, \eta^*\phi_2(\rho(\langle x_{i}, x_{ j}\rangle))\eta(z_j)\rangle\right\|_{B_2}\\
    &=\left\|\sum\limits_{i, j=1}^n\langle z_i, (\phi_1\otimes_\chi I)(\langle x_{i}, x_{ j}\rangle)\eta^*\eta(z_j)\rangle\right\|_{B_2}\\
    &\leq ||\eta||^2*\left\|\sum\limits_{i, j=1}^n\langle z_i, (\phi_1\otimes_\chi I)(\langle x_{i}, x_{j}\rangle)z_j\rangle\right\|_{B_2}\\
    &=||\eta||^2*\left\|\sum\limits_{k=1}^nx_k\dot{\otimes}z_k\right\|^2
\end{align*}
where the inequality follows from an analogous argument for the proof of $(1)$ in Lemma 4.2 of \cite{brady2026functorialityksgnsconstructionintertwiners}. Therefore the map extends to a bounded $B_2$-linear map 
$$I\otimes \eta: E_{A_1}\otimes_{\phi_1\otimes_\chi I}(E_{B_1}\otimes_{\chi}B_2)\to E_{A_1}\otimes_{\phi_2\circ\rho} E_{B_2}\quad\quad x\dot{\otimes}(y\dot{\otimes} b)\mapsto x\dot{\otimes}\eta(y\dot{\otimes}b)$$
By a similar argument in the construction of $I\otimes\eta$ and using an analogous argument in the proof of $(2)$ in Lemma $4.2$ of \cite{brady2026functorialityksgnsconstructionintertwiners}, it follows
$$I\otimes\eta^*:E_{A_1}\otimes_{\phi_2\circ\rho}E_{B_2}\to E_{A_1}\otimes_{\phi_1\otimes_\chi}I(E_{B_1}\otimes_\chi B_2)\quad\quad x\dot{\otimes}z\mapsto x\dot{\otimes}\eta^*(z)$$ is a well defined bounded $B_2$-linear with $||I\otimes\eta^*||\leq ||\eta^*||$.

We claim $I\otimes \eta^*$ is the adjoint of $I\otimes\eta$. As both maps are linear and continuous, it suffices to prove the claim on elements $x_1\dot{\otimes}z_1\in E_{A_1}\otimes_{\phi_1\otimes_\chi I}(E_{B_1}\otimes_{\chi} B_2)$ and $x_2\dot{\otimes}z_2\in E_{A_1}\otimes_{\phi_2\circ\rho} E_{B_2}$. Given such elements, we have
\begin{align*}
    \langle (I\otimes \eta^*)(x_2\dot{\otimes}z_2), x_1\dot{\otimes} z_1\rangle&=\langle x_2\dot{\otimes}\eta^*(z_2), x_1\dot{\otimes}z_1\rangle=\langle \eta^*(z_2), (\phi_1\otimes_\chi I)(\langle x_2, x_1\rangle)z_1\rangle\\
    &=\langle z_2, \eta(\phi_1\otimes_\chi I)(\langle x_2, x_1\rangle)z_1\rangle=\langle z_2, \phi_2(\rho(\langle x_1, x_2\rangle))\eta(z_2)\rangle\\
    &=\langle x_2\dot{\otimes}z_2, x_1\dot{\otimes}\eta(z_1)\rangle=\langle x_2\dot{\otimes}z_2, (I\otimes \eta)(x_1\dot{\otimes}z_1)\rangle
\end{align*}
Thus $I\otimes\eta$ is adjointable with adjoint $I\otimes\eta^*$.

\end{proof}

\begin{proposition}\label{Prop: Induced adjointable map Lambda}\

\noindent For $i=1, 2$, let $(G_i, \mathcal{F}_i)\in\textnormal{EqCpl}_{ *\textnormal{-alg}}$ with associated $G_i$-equivariant coupling given by
$$(((A_i, G_i, \alpha_i), (B_i, G_i, \beta_i), ((E_{B_i}, \phi_i), W_i), ((A_i, G_i, \alpha_i), E_{A_i}, U_i)).$$
If $(f, \tau):(G_1, \mathcal{F}_1)\to (G_2, \mathcal{F}_2)$ is a morphism with associated data $((\rho, \chi, \eta), (\rho, \xi))$ to $\tau$, then there exists an adjointable map
$$\Lambda_{(f, \tau)}:(E_{A_1}\otimes_{\phi_1} E_{B_1})\otimes_{\rho}B_2\to E_{A_2}\otimes_{\phi_2}E_{B_2}$$
which makes the following diagram commute:
\begin{center}
    \begin{tikzcd}
	{(E_{A_1}\otimes_{\phi_1}E_{B_1})\otimes_\chi B_2} &&& {E_{A_1}\otimes_{\phi_1\otimes_\chi I}(E_{B_1}\otimes_\chi B_2)} \\
	\\
	&&& {E_{A_1}\otimes_{\phi_2\circ\rho}E_{B_2}} \\
	\\
	{E_{A_2}\otimes_{\phi_2}E_{B_2}} &&& {(E_{A_1}\otimes_\rho A_2)\otimes_{\phi_2}E_{B_2}}
	\arrow["\mu"{description}, from=1-1, to=1-4]
	\arrow["{\Lambda_{(f, \tau)}}"{description}, dashed, from=1-1, to=5-1]
	\arrow["{I\otimes\eta}"{description}, from=1-4, to=3-4]
	\arrow["\sigma^*"{description}, from=3-4, to=5-4]
	\arrow["{\xi\otimes_{\phi_2}I}", from=5-4, to=5-1]
\end{tikzcd}
\end{center}

\end{proposition}

\begin{remark}\

\noindent Using the limit characterization of $\sigma^*$, we have on simple tensors
$$\Lambda_{(f, \tau)}((x\dot{\otimes}y)\dot{\otimes}b))=\lim\limits_{\lambda\to\infty}\xi(x\dot{\otimes}a_\lambda)\dot{\otimes}\eta(y\dot{\otimes}b)$$
where $(a_\lambda)_{\lambda\in\Lambda}$ is any approximate unit for $A_2$.

\end{remark}

We now verify $(f, (\chi, \Lambda_{(f, \tau)}))$ defines a morphism $(G_1, \mathcal{F}_{1, \otimes})\to (G_2, \mathcal{F}_{2, \otimes})$. 

\begin{lemma}\

\noindent For $i=1, 2$, let $(G_i, \mathcal{F}_i)\in\textnormal{EqCpl}_{*\textnormal{-alg}}$ with associated $G_i$-equivariant coupling given by
$$(((A_i, G_i, \alpha_i), (B_i, G_i, \beta_i), ((E_{B_i}, \phi_i), W_i), ((A_i, G_i, \alpha_i), E_{A_i}, U_i)).$$
If $(f, \tau):(G_1, \mathcal{F}_1)\to (G_2, \mathcal{F}_2)$ is a morphism with associated data $((\rho, \chi, \eta), (\rho, \xi))$ to $\tau$, then the adjointable map $\Lambda_{(f, \tau)}$ satisfies the following intertwining property: for all $g\in G_1$, $$\Lambda_{(f, \tau)}\circ (U_{1, g}\otimes_{\phi_1} W_{1, g})\otimes_\chi \beta_{2, f(g)}=(U_{2, f(g)}\otimes_{\phi_2} W_{2, f(g)})\circ \Lambda_{(f, \tau)}.$$
    
\end{lemma}

\begin{proof}\

\noindent It suffices to show the two maps agree on the simple tensors in the dense submodule $(E_{A_1}\otimes_{A_1} E_{B_1})\otimes_{B_1} B_2$. Let $(a_\lambda)_{\lambda\in\Lambda}$ be an approximate unit in $A_2$, then $(\alpha_{2, f(g)}(a_\lambda))_{\lambda\in\Lambda}$ is an approximate unit in $A_2$. Given $(x\dot{\otimes}y)\dot{\otimes}b\in (E_{A_1}\otimes_{A_1} E_{B_1})\otimes_{B_1}B_2$, we have
\begin{align*}
    (\Lambda_{(f, \tau)}\circ (U_{1, g}\otimes_{\phi_1} W_{1, g})&\otimes_\chi \beta_{2, f(g)})((x\dot{\otimes}y)\dot{\otimes}b)\\
    &=\Lambda_{(f, \tau)}((U_{1, g}x\dot{\otimes}W_{1, f}y)\dot{\otimes}\beta_{2, f(g)}(b))\\
    &=(\xi\otimes_{\phi_2}I\circ\sigma^*)(U_{1, g}x\dot{\otimes}\eta(W_{1, g}y\dot{\otimes} \beta_{2, f(g)}(b)))\\
    &=(\xi\otimes_{\phi_2}I\circ\sigma^*)(U_{1, g}x\dot{\otimes}W_{2, f(g)}\eta(y\dot{\otimes}b))\\
    &=\lim\limits_{\lambda\to\infty}(\xi\otimes_{\phi_2}I)((U_{1, g}x\dot{\otimes}\alpha_{2, f(g)}(a_\lambda))\dot{\otimes}W_{2, f(g)}\eta(y\dot{\otimes}b))\\
    &=\lim\limits_{\lambda\to\infty}\xi(U_{1, g}x\dot{\otimes}\alpha_{2, f(g)}(a_\lambda))\dot{\otimes}W_{2, f(g)}\eta(y\dot{\otimes}b))\\
    &=\lim\limits_{\lambda\to \infty}U_{2, f(g)}\xi(x\dot{\otimes}a_\lambda)\dot{\otimes}W_{2, f(g)}\eta(y\dot{\otimes}b)\\
    &=\lim\limits_{\lambda\to \infty}(U_{2, f(g)}\otimes_{\phi_2}W_{2, f(g)})(\xi(x\dot{\otimes}a_\lambda)\dot{\otimes}\eta(y\dot{\otimes}b))\\
    &=((U_{2, f(g)}\otimes_{\phi_2}W_{2, f(g)})\circ \Lambda_{(f, \tau)})((x\dot{\otimes}y)\dot{\otimes}b)
\end{align*}
Hence, for all $g\in G_1$,
$$\Lambda_{(f, \tau)}\circ (U_{1, g}\otimes_{\phi_1} W_{1, g})\otimes_\chi \beta_{2, f(g)}=(U_{2, f(g)}\otimes_{\phi_2} W_{2, f(g)})\circ \Lambda_{(f, \tau)}.$$

\end{proof}

\begin{proposition}\label{Prop: Lambda is a morphism}\

\noindent For $i=1, 2$, let $(G_i, \mathcal{F}_i)\in\textnormal{EqCpl}_{ *\textnormal{-alg}}$. If $(f, \tau):(G_1, \mathcal{F}_1)\to (G_2, \mathcal{F}_2)$ is a morphism with associated data $((\rho, \chi, \eta), (\rho, \xi))$ to $\tau$, then the following diagram commutes in $\textnormal{HMod}$ for all $g\in G_1$:
\begin{center}
    \begin{tikzcd}
	{E_{A_1}\otimes_{\phi_1}E_{B_1}} &&& {E_{A_1}\otimes_{\phi_1}E_{B_1}} \\
	\\
	\\
	{E_{A_2}\otimes_{\phi_2}E_{B_2}} &&& {E_{A_2}\otimes_{\phi_2}E_{B_2}}
	\arrow["{\mathcal{F}_{1, \otimes}(g)}"{description}, from=1-1, to=1-4]
	\arrow["{(\chi, \Lambda_{(f, \tau)})}"{description}, from=1-1, to=4-1]
	\arrow["{(\chi, \Lambda_{(f, \tau)})}"{description}, from=1-4, to=4-4]
	\arrow["{\mathcal{F}_{2, \otimes}(f(g))}"{description}, from=4-1, to=4-4]
\end{tikzcd}
\end{center}

\end{proposition}

\begin{proof}\

\noindent Fix $g\in G_1$, and write
$$\mathcal{F}_{1, \otimes}(g)=(\beta_{1, g}, (U_{1, g}\otimes_{\phi_1}W_{1, g})\circ U_{\beta_{1, g}})\quad\quad \mathcal{F}_{2, \otimes}((f(g))=(\beta_{2, f(g)}, (U_{2, f(g)}\otimes_{\phi_2}W_{2, f(g)})\circ U_{\beta_{2, f(g)}})$$
As $\chi\circ \beta_{1, g}=\beta_{2, f(g)}\circ\chi$, we just need to show
$$(U_{2, f(g)}\otimes_{\phi_2}W_{2, f(g)})\circ U_{\beta_{2, f(g)}}\circ (\Lambda_{(f, \tau)}\otimes_{\beta_{2, f(g)}}I)\circ U_1^{-1}=\Lambda_{(f, \tau)}\circ ((U_{1, g}\otimes_{\phi_1}W_{1, g})\circ U_{\beta_{1, g}})\otimes_\chi I\circ U_2^{-1}$$
To prove this, we first observe
\begin{align*}
    \Lambda_{(f, \tau)}\circ ((U_{1, g}&\otimes_{\phi_1}W_{1, g})\circ U_{\beta_{1, g}})\otimes_\chi I\circ U_2^{-1}\\
    &=\Lambda_{(f, \tau)}\circ ((U_{1, g}\otimes_{\phi_1}W_{1, g})\otimes_\chi \beta_{2, f(g)})\circ U_{\beta_{1, f(g)}}\otimes_{\chi}\beta_{2, f(g^{-1})}\circ U_2^{-1}\\
    &=(U_{2, f(g)}\otimes_{\phi_2} W_{2, f(g)})\circ \Lambda_{(f, \tau)}\circ U_{\beta_{1, f(g)}}\otimes_{\chi}\beta_{2, f(g^{-1})}\circ U_2^{-1}
\end{align*}
Now, we claim 
$$\Lambda_{(f, \tau)}\circ U_{\beta_{1, f(g)}}\otimes_{\chi}\beta_{2, f(g^{-1})}\circ U_2^{-1}=U_{\beta_{2, f(g)}}\circ (\Lambda_{(f, \tau)}\otimes_{\beta_{2, f(g)}}I)\circ U_1^{-1}.$$
To prove this, it suffices to show the maps agree on simple tensors 
$$(x\dot{\otimes}y)\dot{\otimes}b\in (E_{A_1}\otimes_{\phi_1}E_{B_1})\otimes_{\beta_{2, f(g)}\circ\chi}B_2=(E_{A_1}\otimes_{\phi_1}E_{B_1})\otimes_{\chi\circ\beta_{1, g}}B_2.$$
Given such an element and an approximate unit $(b_\lambda)_{\lambda\in\Lambda}$ for $B_1$, we have
\begin{align*}
    (\Lambda_{(f, \tau)}\circ U_{\beta_{1, f(g)}}\otimes_{\chi}\beta_{2, f(g^{-1})}\circ U_2^{-1})((x\dot{\otimes}y)\dot{\otimes}b)&=\lim\limits_{\lambda\to\infty}(\Lambda_{(f, \tau)}\circ U_{\beta_{1, f(g)}}\otimes_{\chi}\beta_{2, f(g^{-1})})(((x\dot{\otimes} y)\dot{\otimes}b_\lambda)\dot{\otimes}b)\\
    &=\lim\limits_{\lambda\to\infty}\Lambda_{(f, \tau)}((x\dot{\otimes}y)\beta_{1, f(g^{-1})}(b_\lambda)\dot{\otimes}b)\\
    &=\Lambda_{(f, \tau)}((x\dot{\otimes}y)\dot{\otimes}b)
\end{align*}
while
\begin{align*}
    (U_{\beta_{2, f(g)}}\circ (\Lambda_{(f, \tau)}\otimes_{\beta_{2, f(g)}}I)\circ U_1^{-1})((x\dot{\otimes}y)\dot{\otimes}b)&=\lim\limits_{\lambda\to\infty}(U_{\beta_{2, f(g)}}\circ (\Lambda_{(f, \tau)}\otimes_{\beta_{2, f(g)}}I))(((x\dot{\otimes}y)\dot{\otimes}b_\lambda)\dot{\otimes}b)\\
    &=\lim\limits_{\lambda\to\infty}U_{\beta_{2, f(g)}}(\Lambda((x\dot{\otimes}y)\dot{\otimes}b_\lambda)\dot{\otimes}b)\\
    &=\lim\limits_{\lambda\to\infty}\Lambda((x\dot{\otimes}y)\dot{\otimes}b_\lambda)\beta_{2, f(g^{-1})}(b)\\
    &=\lim\limits_{\lambda\to\infty}\Lambda((x\dot{\otimes}y)\dot{\otimes}b_\lambda\beta_{2, f(g^{-1})}(b))=\Lambda((x\dot{\otimes}y)\dot{\otimes}b)
\end{align*}
Hence the two maps agree. Therefore
$$(U_{2, f(g)}\otimes_{\phi_2} W_{2, f(g)})\circ \Lambda_{(f, \tau)}\circ U_{\beta_{1, f(g)}}\otimes_{\chi}\beta_{2, f(g^{-1})}\circ U_2^{-1}=(U_{2, f(g)}\otimes_{\phi_2}W_{2, f(g)})\circ U_{\beta_{2, f(g)}}\circ (\Lambda_{(f, \tau)}\otimes_{\beta_{2, f(g)}}I)\circ U_1^{-1}$$ proving the main claim.
    
\end{proof}

\begin{theorem}\label{Theorem: functor for step 2}\

\noindent The assignment
\begin{align*}
    \textnormal{EqCpl}_{*\textnormal{-alg}}&\to \int_{\textnormal{Grp}}\mathcal{U}\textnormal{Fun}(B-, \textnormal{HMod})\\
    (G, \mathcal{F})&\mapsto (G, \mathcal{F}_\otimes)\\
    ((f, \tau):(G_1, \mathcal{F}_1)\to (G_2, \mathcal{F}_2))&\mapsto ((f, \tau_\otimes): (G_1, \mathcal{F}_{1, \otimes})\to (G_2, \mathcal{F}_{2, \otimes}))
\end{align*}
defines a covariant functor.

\end{theorem}

\begin{proof}\

\noindent The assignment being well-defined on objects follows from the interior tensor product and Theorem \ref{Theorem: functorial characterization of hilbert module dynamical systems}. The assignment being well-defined on morphisms follows from Proposition \ref{Prop: Induced adjointable map Lambda} and Proposition \ref{Prop: Lambda is a morphism}. Therefore, we just need to verify the assignment is functorial.

We start by proving the assignment preserves the identity morphism. Given an object $(G_1, \mathcal{F}_1)\in \textnormal{EqCpl}_{*\textnormal{-alg}}$ with associated $G_1$-equivariant coupling given by
$$(((A_i, G_i, \alpha_i), (B_i, G_i, \beta_i), ((E_{B_i}, \phi_i), W_i), ((A_i, G_i, \alpha_i), E_{A_i}, U_i)).$$
First, the data associated to $1_{\mathcal{F}_1}$ is given by $((1_{A_1}, \text{inc}, \iota), (1_{A_1}, \iota'))$ where $\iota$ and $\iota'$ are the unitaries
$$\iota:E_{B_1}\otimes_{\text{inc}}B_1\xrightarrow{x\dot{\otimes}b\mapsto xb} E_{B_1}\quad\quad \iota':E_{A_1}\otimes_{\text{inc}}A_1\xrightarrow{x\dot{\otimes}a\mapsto xa} E_{A_1}$$
Second, the data associated to $1_{\mathcal{F}_{1, \otimes}}$ is given by $(\text{inc}, \iota'')$ where $\iota''$ is the unitary
$$\iota'':(E_{A_1}\otimes_{\phi_1}E_{B_1})\otimes_{\text{inc}}B_1\to E_{A_1}\otimes_{\phi_1}E_{B_1}\quad\quad \iota''((x\dot{\otimes}y)\dot{\otimes}b)= (x\dot{\otimes}y)b$$
Thus, we just need to show $\Lambda_{(1_{G_1}, 1_{\mathcal{F}_1})}=\iota''$. Given $(x\dot{\otimes} y)\dot{\otimes}b\in (E_{A_1}\otimes_{\phi_1}E_{B_1})\otimes_{\text{inc}}B_1$ and an approximate unit $(a_\lambda)_{\lambda\in\Lambda}$ for $A_1$, we have
\begin{align*}
    \Lambda_{(1_{G_1}, 1_{\mathcal{F}_1})}((x\dot{\otimes}y)\dot{\otimes}b)&=\lim\limits_{\lambda\to \infty}\iota'(x\dot{\otimes}a_\lambda)\dot{\otimes}\iota(y\dot{\otimes}b)=\lim\limits_{\lambda\to\infty}xa_\lambda\dot{\otimes}yb=x\dot{\otimes}yb
\end{align*}
Therefore $\Lambda_{(1_{G_1}, 1_{\mathcal{F}_1})}=\iota''$ showing $(1_{G_1}, 1_{\mathcal{F}_{1}})\mapsto (1_{G_1}, 1_{\mathcal{F}_{1, \otimes}})$.

Now we prove the assignment preserves composition. For $i=1, 2,3 $, let $(G_i, \mathcal{F}_i)$ be an object in $\textnormal{EqCpl}_{*\textnormal{-iso}, *\textnormal{-alg}}$ with associated $G_i$-equivariant coupling given by
$$(((A_i, G_i, \alpha_i), (B_i, G_i, \beta_i), ((E_{B_i}, \phi_i), W_i), ((A_i, G_i, \alpha_i), E_{A_i}, U_i)).$$
For $i=1, 2$, let $(f_i, \tau_i):(G_i, \mathcal{F}_i)\to (G_{i+1}, \mathcal{F}_{i+1})$ be a morphism with associated data $((\rho_i, \chi_i, \eta_i), (\rho_i, \xi_i))$. Since
$$(f_2, \tau_2)\circ (f_{1}, \tau_1)\mapsto (f_2\circ f_1, (f_1^*\tau_2\circ \tau_1)_\otimes)$$
where $(f_1^*\tau_2\circ \tau_1)_\otimes$ has associated data $(\chi_2\circ\chi_1, \Lambda_{(f_2, \tau_2)\circ (f_1, \tau_2)})$ while
$$(f_2, \tau_{2, \otimes})\circ (f_{1}, \tau_{1, \otimes})=(f_2\circ f_1, f_1^*\tau_{2, \otimes}\circ \tau_{1, \otimes})$$
with associated data $(\chi_2\circ\chi_1, \Lambda_{(f_2, \tau_2)}\circ (\Lambda_{(f_1, \tau_1)}\otimes_{\chi_2} I)\circ U^{-1})$ where $U$ is the unitary
\begin{align*}
    U:((E_{A_1}\otimes_{\phi_1} E_{B_1})\otimes_{\chi_1} B_2)\otimes_{\chi_2}B_3&\to (E_{A_1}\otimes_{\phi_1} E_{B_2})\otimes_{\chi_2\circ\chi_1}B_3\\ U(((x\dot{\otimes}y)\dot{\otimes}b_1)\dot{\otimes}b_2)&=(x\dot{\otimes}y)\dot{\otimes}\chi_2(b_1)b_2
\end{align*}
then we just need to show
$$\Lambda_{(f_2, \tau_2)\circ (f_1, \circ\tau_1)}=\Lambda_{(f_2, \tau_2)}\circ (\Lambda_{(f_1, \tau_1)}\otimes_{\chi_2} I)\circ U^{-1}.$$
As $U$ is a unitary, it suffices to prove
$$\Lambda_{(f_2, \tau_2)\circ (f_1, \circ\tau_1)}\circ U=\Lambda_{(f_2, \tau_2)}\circ (\Lambda_{(f_1, \tau_1)}\otimes_{\chi_2} I)$$

As both map are continuous and $B_3$-linear, it suffices to show the maps agree on simple tensors $((x\dot{\otimes}y)\dot{\otimes}b_1)\dot{\otimes}b_2\in ((E_{A_1}\otimes_{\phi_1} E_{B_1})\otimes_{\chi_1} B_2)\otimes_{\chi_2}B_3$. Given such an element, an approximate unit $(a_\lambda)_{\lambda\in\Lambda}$ for $A_3$, and an approximate unit $(a_\mu)_{\mu\in M}$ for $A_2$, we have
\begin{align*}
    (\Lambda_{(f_2, \tau_2)\circ (f_1, \tau_1)}\circ U)(((x\dot{\otimes}y)\dot{\otimes}b_1)\dot{\otimes}b_2)&=\Lambda_{(f_2, \tau_2)\circ (f_1, \tau_1)}((x\dot{\otimes} y)\dot{\otimes}\chi_2(b_1)b_2)\\
    &=\lim\limits_{\lambda\to\infty}((\xi_2\circ(\xi_1\otimes_{\chi_2}I)\circ U_2^{-1})(x\dot{\otimes}a_\lambda))\dot{\otimes}\eta_2(\eta_1(y\dot{\otimes}b_1)\dot{\otimes}b_2)
\end{align*}
and
\begin{align*}
    (\Lambda_{(f_2, \tau_2)}\circ (\Lambda_{(f_1, \tau_1)}\otimes_{\chi_2} I))((&(x\dot{\otimes}y)\dot{\otimes}b_1)\dot{\otimes}b_2)\\
    &=\lim\limits_{\mu\to \infty}\lim\limits_{\lambda\to\infty}\xi_2(\xi_1(x\dot{\otimes}a_\mu)\dot{\otimes}a_\lambda)\dot{\otimes}\eta_2(\eta_{1}(y\dot{\otimes}b_1)\dot{\otimes}b_2)\\
    &=\lim\limits_{\mu\to \infty}\lim\limits_{\lambda\to\infty}(\xi_2\circ (\xi_1\otimes_{\rho_2}I))((x\dot{\otimes}a_\mu)\dot{\otimes}a_\lambda)\dot{\otimes}\eta_2(\eta_{1}(y\dot{\otimes}b_1)\dot{\otimes}b_2)
\end{align*}
We claim that we can swap the order of the limits. Note, if this is the case, then 
\begin{align*}
    \lim\limits_{\mu\to \infty}\lim\limits_{\lambda\to\infty}(\xi_2\circ (\xi_1\otimes_{\rho_2}I))&(x\dot{\otimes}a_\mu)\dot{\otimes}a_\lambda)\dot{\otimes}\eta_2(\eta_{1}(y\dot{\otimes}b_1)\dot{\otimes}b_2)\\
    &=\lim\limits_{\lambda\to \infty}\lim\limits_{\mu\to\infty}(\xi_2\circ (\xi_1\otimes_{\rho_2}I))((x\dot{\otimes}a_\mu)\dot{\otimes}a_\lambda)\dot{\otimes}\eta_2(\eta_{1}(y\dot{\otimes}b_1)\dot{\otimes}b_2)\\
    &=\lim\limits_{\lambda\to \infty}(\xi_2\circ (\xi_1\otimes_{\rho_2}I)\circ U_2^{-1})(x\dot{\otimes}a_\lambda)\dot{\otimes}\eta_2(\eta_{1}(y\dot{\otimes}b_1)\dot{\otimes}b_2)
\end{align*}
Therefore, it will follow that
$$\Lambda_{(f_2, \tau_2)\circ (f_1, \circ\tau_1)}\circ U=\Lambda_{(f_2, \tau_2)}\circ (\Lambda_{(f_1, \tau_1)}\otimes_{\chi_2} I).$$

To show that we can swap the order of the limits, we first verify pointwise convergence if we fix either $\mu$ or $\lambda$. As we are working in a Banach space, it suffices to show for a fixed $\mu$ or $\lambda$, the net is a Cauchy net. Writing
$$(\xi_2\circ (\xi_1\otimes_{\rho_2}I))((x\dot{\otimes}a_\mu)\dot{\otimes}a_\lambda)\dot{\otimes}\eta_2(\eta_{1}(y\dot{\otimes}b_1)\dot{\otimes}b_2)=x_{\mu, \lambda}\dot{\otimes}z$$
we have for all $\mu, \mu'\in M$ and $\lambda, \lambda'\in \Lambda$ that
$$||x_{\mu, \lambda}\dot{\otimes}z-x_{\mu', \lambda'}\dot{\otimes}z||\leq ||z||^2||x_{\mu, \lambda}-x_{\mu', \lambda'}||\leq C||z||^2||(x\dot{\otimes}a_\mu)\dot{\otimes}a_\lambda-(x\dot{\otimes}a_{\mu'})\dot{\otimes}a_{\lambda'}||$$
Now suppose we fix $\lambda\in \Lambda$, then for any $\mu, \mu'\in M$, we have
\begin{align*}
    ||(x\dot{\otimes}a_\mu)\dot{\otimes}a_\lambda-(x\dot{\otimes}a_{\mu'})\dot{\otimes}a_{\lambda'}||&\leq ||a_\lambda||^2||\sqrt{\rho_1(\langle x, x\rangle)}a_\mu-\sqrt{\rho_1(\langle x, x\rangle}a_{\mu'}||^{\frac{1}{2}}\\
    &\leq ||\sqrt{\rho_2(\langle x, x\rangle)}a_\mu-\sqrt{\rho_1(\langle x, x\rangle)}a_{\mu'}||^{\frac{1}{2}}
\end{align*}
As $(a_{\mu})_{\mu\in M}$ is an approximate unit, we have $\sqrt{\rho_1(\langle x, x\rangle)}a_\mu\to \sqrt{\rho_1(\langle x, x\rangle)}$. Thus the net $(\sqrt{\rho_1(\langle x, x\rangle}a_{\mu})_{\mu\in M}$ is a Cauchy net proving convergence in $\mu$ for a fixed $\lambda$. Observe the last inequality is independent $\lambda$; thus, we have convergence in $\mu$ uniformly over $\lambda$.  Now suppose we fix $\mu\in M$, then for any $\lambda, \lambda'\in \Lambda$, we have
\begin{align*}
     ||(x\dot{\otimes}a_\mu)\dot{\otimes}a_\lambda-(x\dot{\otimes}a_{\mu'})\dot{\otimes}a_{\lambda'}||&\leq ||\sqrt{\rho_2(\langle x\dot{\otimes}a_\mu, x\dot{\otimes}a_\mu) }a_\lambda-\sqrt{\rho_2(\langle x\dot{\otimes}a_\mu, x\dot{\otimes}a_\mu) }a_{\lambda'}||^{\frac{1}{2}}
\end{align*}
As $(a_{\lambda})_{\lambda\in \Lambda}$ is an approximate unit, we have $\sqrt{\rho_2(\langle x\dot{\otimes}a_\mu, x\dot{\otimes}a_\mu\rangle)}a_\lambda\to \sqrt{\rho_2(\langle x\dot{\otimes}a_\mu, x\dot{\otimes}a_\mu\rangle)}$. Thus the net $(\sqrt{\rho_2(\langle x\dot{\otimes}a_\mu, x\dot{\otimes}a_\mu\rangle)}a_\lambda)_{\lambda\in \Lambda}$ is a Cauchy net proving convergence in $\lambda$ for a fixed $\mu$. 

Now we show
$$\lim\limits_{\lambda\to\infty}\lim\limits_{\mu\to \infty}x_{\mu, \lambda}\dot{\otimes}z=\lim\limits_{\mu\to\infty}\lim\limits_{\lambda\to \infty}x_{\mu, \lambda}\dot{\otimes}z$$
First, write
$$\lim\limits_{\mu\to \infty}x_{\mu, \lambda}\dot{\otimes} z=y_{\lambda}\quad\quad \lim\limits_{\lambda\to \infty}x_{\mu, \lambda}\dot{\otimes} z=y_{\mu}.$$
Note, both limits exists due to the pointwise convergence in the prior paragraph. We claim $(y_\lambda)_{\lambda\in\Lambda}$ is a Cauchy net. Fix $\epsilon>0$. We have for all $\lambda, \lambda'\in \Lambda$ 
$$||y_\lambda-y_{\lambda'}||\leq ||y_\lambda-x_{\mu, \lambda}\dot{\otimes}z||+||x_{\mu, \lambda}\dot{\otimes}z-x_{\mu, \lambda'}\dot{\otimes}z||+||x_{\mu, \lambda'}\dot{\otimes}z-y_{\lambda'}||$$
As $x_{\mu, \lambda}\to y_{\lambda}$ uniformly for any $\lambda\in \Lambda$, then we find $\mu\in M$ such that 
$$||y_\lambda-x_{\mu, \lambda}\dot{\otimes}z||<\frac{\epsilon}{3}$$
for all $\lambda\in\Lambda$. Furthermore, as $(x_{\mu, \lambda}\dot{\otimes}z)_{\lambda\in\Lambda}$ is a Cauchy, there exists $\lambda_0\in \Lambda$ such that for all $\lambda, \lambda'\geq \lambda_0$, 
$$||x_{\mu, \lambda}\dot{\otimes}z-x_{\mu, \lambda'}\dot{\otimes}z||<\frac{\epsilon}{3}$$
Thus, for $(y_{\lambda})_{\lambda\in\Lambda}$ is indeed a Cauchy net. As $E_{A_2}\otimes_{\phi_2}E_{B_2}$ is a Banach space, the net converges to a point $y\in E_{A_2}\otimes_{\phi_2}E_{B_2}$. We claim
$$\lim\limits_{\mu\to \infty}y_\mu =y$$
Let $\epsilon>0$, then for all $\mu\in M$, we have
$$||y-y_\mu||\leq ||y-y_{\lambda}||+||y_\lambda-x_{\mu, \lambda}||+||x_{\mu, \lambda}-y_\mu||$$
As $y_\lambda\to y$ and $x_{\mu, \lambda}\to y_\mu$, we can find $\lambda_\mu\in \Lambda$ such that 
$$||y-y_{\lambda_\mu}||<\frac{\epsilon}{3}\quad\quad ||x_{\mu, \lambda_\mu}-y_\mu||<\frac{\epsilon}{3}$$
As $x_{\mu, \lambda}\to y_\lambda$, there exists $\mu_0\in M$ such that for all $\mu\geq \mu_0$ and all $\lambda\in\Lambda$ 
$$||y_\lambda-x_{\mu, \lambda}||<\frac{\epsilon}{3}$$
Hence $y_\mu\to y$ proving
$$\lim\limits_{\lambda\to\infty}\lim\limits_{\mu\to \infty}x_{\mu, \lambda}\dot{\otimes}z=\lim\limits_{\mu\to\infty}\lim\limits_{\lambda\to \infty}x_{\mu, \lambda}\dot{\otimes}z.$$

\end{proof}

\subsection{Functoriality of the KSGNS Construction}

Lastly, we show how the equivariant KSGNS construction yields an endofunctor on $\textnormal{EqCpl}_{*\textnormal{-iso}}$. Using Theorem $4.7$ in \cite{brady2026functorialityksgnsconstructionintertwiners}, the KSGNS construction defines an idempotent endofunctor on the category $\textnormal{NDPosCor}_{*\textnormal{-iso}, *\textnormal{-alg}}$. Note, $\textnormal{NDPosCor}_{*\textnormal{-iso}, *\textnormal{-alg}}$ is the subcategory of $\textnormal{NDPosCor}_{*\textnormal{-alg}}$ where morphisms
$$(\rho, \chi, \eta):(A, B, (E_B, \phi))\to (C, D, (E_D, \psi))$$
have the property that $\rho:A\to C$ is a $*$-algebra isomorphism. 

Given a unitary valued functor $\mathcal{F}=(\mathcal{F}_1, \mathcal{F}_2):BG\to \textnormal{Cpl}$, the unitary valued functor $\mathcal{F}_1:BG\to \textnormal{NDPosCor}_{*\textnormal{-alg}}$ takes value in $\textnormal{NDPosCor}_{*\textnormal{-iso}, *\textnormal{-alg}}$. We apply the KSGNS endofunctor to obtain another functor
$\textnormal{KSGNS}\circ \mathcal{F}_1:BG\to \textnormal{NDPosCor}_{*\textnormal{-alg}}$. From the construction of $\textnormal{KSGNS}$, the functor $\textnormal{KSGNS}\circ\mathcal{F}_1$ is unitary valued and the pair 
$$\textnormal{KSGNS}(\mathcal{F}):=(\textnormal{KSGNS}\circ \mathcal{F}_1, \mathcal{F}_2)$$
defines a unitary valued functor $BG\to \textnormal{Cpl}$. This leads to an assignment on objects given by
$$\textnormal{EqCpl}_{*\textnormal{-iso}}\to \textnormal{EqCpl}_{*\textnormal{-iso}}\quad\quad (G, \mathcal{F})\mapsto (G, \textnormal{KSGNS}(\mathcal{F}))$$

For the assignment on morphisms, fix a morphism $(f, \tau):(G_1, \mathcal{F}_1)\to (G_2, \mathcal{F}_2)$ in $\textnormal{EqCpl}_{*\textnormal{-iso}}$ with associated data $((\rho, \chi, \eta), (\rho, \xi))$ to $\tau$. As $\rho$ is a $*$-algebra isomorphism, we can utilize the endofunctor $\textnormal{KSGNS}$ on $\textnormal{NDPosCor}_{*\textnormal{-iso}, *\textnormal{-alg}}$ to obtain a morphism
$$\textnormal{KSGNS}(\rho, \chi, \eta):(\textnormal{KSGNS}\circ\mathcal{F}_1)(\textnormal{pt})\to (\textnormal{KSGNS}\circ\mathcal{F}_2)(\textnormal{pt})$$
As $\textnormal{Cpl}$ is a subcategory of the cartesian product, commutativity of the diagram 
 \begin{center}
    \begin{tikzcd}
	{\mathcal{F}_1(\textnormal{pt})} && {\mathcal{F}_1(\textnormal{pt})} \\
	\\
	{\mathcal{F}_2(\textnormal{pt})} && {\mathcal{F}_2(\textnormal{pt})}
	\arrow["{\mathcal{F}_1(g)}"{description}, from=1-1, to=1-3]
	\arrow["{((\rho, \chi, \eta), (\rho, \xi))}"{description}, from=1-1, to=3-1]
	\arrow["{((\rho, \chi, \eta), (\rho, \xi))}"{description}, from=1-3, to=3-3]
	\arrow["{\mathcal{F}_2(f(g))}"{description}, from=3-1, to=3-3]
\end{tikzcd}
\end{center}
 for all $g\in G_1$ as well as functoriality the endofunctor $\textnormal{KSGNS}$ imply the following diagram commutes for all $g\in G_1$:
 \begin{center}
    \begin{tikzcd}
	{\textnormal{KSGNS}(\mathcal{F}_1(\textnormal{pt}))} &&&& {\mathcal{F}_1(\textnormal{pt})} \\
	\\
	{\textnormal{KSGNS}(\mathcal{F}_2(\textnormal{pt}))} &&&& {\textnormal{KSGNS}(\mathcal{F}_2(\textnormal{pt}))}
	\arrow["{\textnormal{KSGNS}(\mathcal{F}_1)(g)}"{description}, from=1-1, to=1-5]
	\arrow["{(\textnormal{KSGNS}(\rho, \chi, \eta), (\rho, \xi))}"{description}, from=1-1, to=3-1]
	\arrow["{(\textnormal{KSGNS}(\rho, \chi, \eta), (\rho, \xi))}"{description}, from=1-5, to=3-5]
	\arrow["{\textnormal{KSGNS}(\mathcal{F}_2)(f(g))}"{description}, from=3-1, to=3-5]
\end{tikzcd}
\end{center}
 Letting $\textnormal{KSGNS}(\tau)$ denote the natural transformation to the pair $(\textnormal{KSGNS}(\rho, \chi, \eta), (\rho, \xi))$, we obtain a morphism
$$(f, \textnormal{KSGNS}(\tau)):(G_1, \textnormal{KSGNS}(\mathcal{F}_1))\to (G_2, \textnormal{KSNGS}(\mathcal{F}_2)).$$
This leads to the assignment on morphisms given by
$$\textnormal{EqCpl}_{*\textnormal{-iso}}\to \textnormal{EqCpl}_{*\textnormal{-iso}}\quad\quad (f, \tau)\mapsto (f, \textnormal{KSGNS}(\tau))$$
Using functoriality of $\textnormal{KSGNS}$, it immediately follows the assignment on morphisms preserves composition and the identity morphism. Thus we obtain covariant endofunctor on $\textnormal{EqCpl}_{*\textnormal{-iso}}$ which we record below.

\begin{theorem}\label{Theorem: functor for step 1}\

\noindent The assignment
\begin{align*}
    \textnormal{EqCpl}_{*\textnormal{-iso}}&\to \textnormal{EqCpl}_{*\textnormal{-iso}}\\
    (G, \mathcal{F})&\mapsto (G, \textnormal{KSGNS}(\mathcal{F}))\\
    ((f, \tau):(G_1, \mathcal{F}_1)\to (G_2, \mathcal{F}_2))&\mapsto ( (f, \textnormal{KSGNS}(\tau)):(G_1, \textnormal{KSGNS}(\mathcal{F}_1))\to (G_2, \textnormal{KSNGS}(\mathcal{F}_2)))
\end{align*}
defines a covariant functor.

\end{theorem}

\begin{remark}\

\noindent Using that $\textnormal{KSGNS}$ is an idempotent on $\textnormal{NDPosCor}_{*\textnormal{-alg}}$, it follows the functor in the theorem is idempotent. Furthermore, $\textnormal{KSGNS}$ takes value in $\textnormal{EqCpl}_{*\textnormal{-alg}}$.

\end{remark}

\subsection{Functoriality of the Reference Frame Construction}

By combining together Corollary \ref{Cor: functor for step 3}, Theorem  \ref{Theorem: functor for step 2}, and Theorem \ref{Theorem: functor for step 1}, we immediately obtain the following functorial statement proving the reference frame construction is functorial.

\begin{theorem}\label{Theorem: functoriality of the RF construction}\

\noindent The assignment
\begin{align*}
    \mathscr{RF}:\textnormal{EqCpl}_{*\textnormal{-iso}}&\to \textnormal{C}^*_{\textnormal{CP}}\\
    (G, \mathcal{F})&\mapsto \mathcal{L}(E_{A}\otimes_{\pi_\phi} F_\phi)^G\\
    ((f, \tau):(G_1, \mathcal{F}_1)\to (G_2, \mathcal{F}_2))&\mapsto (\Phi_{(f, \textnormal{KSGNS}(\tau)_\otimes)}:\mathcal{L}(E_{A_1}\otimes_{\pi_{\phi_1}} F_{\phi_1})^{G_1}\to \mathcal{L}(E_{A_2}\otimes_{\pi_{\phi_2}} F_{\phi_2})^{G_2})
\end{align*}
defines a covariant functor.

\end{theorem}

\section{The Category of Hilbert Module Reference Frames}

We conclude the paper by building the category of Hilbert module reference frames, denoted by $\textnormal{HMRF}$. Using the categorical structure of $\textnormal{HMRF}$, we deduce the transformation group of a Hilbert module reference frame as a ``unitary" groupoid. We then conclude the section with some examples of such transformations.

\subsection{The category $\textnormal{HMRF}$}

A Hilbert module reference frame is given by the data of a $G$-equivariant coupling together with a relativization map. As the relativization map is additional structure for the $G$-equivariant coupling, we look to utilize $\textnormal{EqCpl}_{*\textnormal{-iso}}$ as a foundation for building the category of Hilbert module reference frames. 

For $i=1, 2$, let $((G_i, \mathcal{F}_i), \pounds_i)$ be a pair where $(G_i, \mathcal{F}_i)\in \textnormal{EqCpl}_{*\textnormal{-iso}}$ and $\pounds_i$ is a relativization map for the $G_i$-equivariant coupling associated to $(G_i, \mathcal{F}_i)$. The data of a morphism $((G_1, \mathcal{F}_1), \pounds_1)\to ((G_2, \mathcal{F}_2), \pounds_2)$ should correspond to intertwining the maps
$$\mathcal{L}(E_{A_1})\xrightarrow{\pounds_1}\mathcal{L}(E_{A_1}\otimes_{\pi_{\phi_1}}F_{\phi_1})^{G_1}\quad\quad \mathcal{L}(E_{A_2})\xrightarrow{\pounds_2}\mathcal{L}(E_{A_2}\otimes_{\pi_{\phi_2}}F_{\phi_2})^{G_2}.$$
Given the data of a morphism $(f, \tau):(G_1, \mathcal{F}_1)\to (G_2, \mathcal{F}_2)$ in $\textnormal{EqCpl}_{*\textnormal{-iso}}$, we can utilize Theorem \ref{Theorem: functoriality of the RF construction} to obtain a strict completely positive map
$$\Phi_{(f, \textnormal{KSGNS}(\tau)_\otimes)}: \mathcal{L}(E_{A_1}\otimes_{\pi_1}F_{\phi_1})^{G_1}\to \mathcal{L}(E_{A_2}\otimes_{\pi_2}F_{\phi_2})^{G_2}.$$
Using the projection functor $\mathcal{Q}_2:\textnormal{Cpl}\to \textnormal{HMod}$ as well as the functor from Theorem \ref{Theorem: functor for adjointable operators}, we obtain a strict completely positive map
$$\Phi_{(f, \mathcal{Q}_2(\tau))}:\mathcal{L}(E_{A_1})\to \mathcal{L}(E_{A_2}).$$
Combining the two completely positive maps together, morphisms for Hilbert module reference frames are  taken as morphisms 
$$(f, \tau):(G_1, \mathcal{F}_1)\to (G_2, \mathcal{F}_2)$$
in $\textnormal{EqCpl}_{*\textnormal{-iso}}$ which make the following diagram commute:
\begin{center}
    \begin{tikzcd}
	{\mathcal{L}(E_{A_1}\otimes_{\pi_{\phi_1}}F_{\phi_1})^{G_1}} &&& {\mathcal{L}(E_{A_2}\otimes_{\pi_{\phi_2}}F_{\phi_2})^{G_2}} \\
	\\
	\\
	{\mathcal{L}(E_{A_1})} &&& {\mathcal{L}(E_{A_2})}
	\arrow["{\Phi_{(f, \textnormal{KSGNS}(\tau)_\otimes)}}"{description}, from=1-1, to=1-4]
	\arrow["{\pounds_1}"{description}, from=4-1, to=1-1]
	\arrow["{\Phi_{(f, \mathcal{Q}_2(\tau))}}"{description}, from=4-1, to=4-4]
	\arrow["{\pounds_2}"{description}, from=4-4, to=1-4]
\end{tikzcd}
\end{center}

Using the functor from Theorem \ref{Theorem: functor for adjointable operators}, we obtain functoriality of the completely positive map $\Phi_{(f, \mathcal{Q}_2(\tau))}$. Functoriality of the completely positive map $\Phi_{(f, \textnormal{KSGNS}(\tau)_\otimes)}$ follows from functoriality of the reference frame construction.

\begin{definition}\

\noindent Define the category of Hilbert module reference frames, denoted by $\textnormal{HMRF}$, as the category where
\begin{itemize}
    \item objects are given by the data $((G, \mathcal{F}), \pounds)$ where $(G, \mathcal{F})\in \textnormal{EqCpl}_{*\textnormal{-iso}}$ and $\pounds$ is a relativization for the $G$-equivariant coupling associated to $(G, \mathcal{F})$.
    \item  a morphism from $((G_1, \mathcal{F}_1), \pounds_1)\to ((G_2, \mathcal{F}_2), \pounds_2)$ is given by the data of a morphism $(f, \tau):(G_1, \mathcal{F}_1)\to (G_2, \mathcal{F}_2)$ in $\textnormal{EqCpl}_{*\textnormal{-iso}}$ which makes the following diagram commute:
\begin{center}
    \begin{tikzcd}
	{\mathcal{L}(E_{A_1}\otimes_{\pi_{\phi_1}}F_{\phi_1})^{G_1}} &&& {\mathcal{L}(E_{A_2}\otimes_{\pi_{\phi_2}}F_{\phi_2})^{G_2}} \\
	\\
	\\
	{\mathcal{L}(E_{A_1})} &&& {\mathcal{L}(E_{A_2})}
	\arrow["{\Phi_{(f, \textnormal{KSGNS}(\tau)_\otimes)}}"{description}, from=1-1, to=1-4]
	\arrow["{\pounds_1}"{description}, from=4-1, to=1-1]
	\arrow["{\Phi_{(f, \mathcal{Q}_2(\tau))}}"{description}, from=4-1, to=4-4]
	\arrow["{\pounds_2}"{description}, from=4-4, to=1-4]
\end{tikzcd}
\end{center}
\item the composition rule is given by composition rule of $\textnormal{EqCpl}_{*\textnormal{-iso}}$.

\end{itemize}

\end{definition}

\begin{remark}\

\noindent Using the functor $\mathcal{Q}_2$ as well as those coming from Theorem \ref{Theorem: functor for adjointable operators} and Theorem \ref{Theorem: functoriality of the RF construction}, it immediately follows that $\textnormal{HMRF}$ is a well-defined category. 

\end{remark}

\begin{proposition}\

\noindent Let $((G_1, \mathcal{F}_1), \pounds_1), ((G_2, \mathcal{F}_2), \pounds_2)\in \textnormal{HMRF}$. If $(f, \tau):((G_1, \mathcal{F}_1), \pounds_1)\to ((G_2, \mathcal{F}_2), \pounds_2)$ is a morphism in $\textnormal{HMRF}$, then every face of the following diagram commutes:
\begin{center}
    \begin{tikzcd}
	{\mathcal{L}(E_{A_1})} &&& {\mathcal{L}(E_{A_2})} & \\
	\\
	{\mathcal{L}(E_{A_1})^{G_1}} &&& {\mathcal{L}(E_{A_2})^{G_2}} \\
	& {\mathcal{L}(E_{A_1}\otimes_{\pi_{\phi_1}}F_{\phi_1})^{G_1}} &&& {\mathcal{L}(E_{A_2}\otimes_{\pi_{\phi_2}}F_{\phi_2})^{G_2}}
	\arrow["{\Phi_{(f, \mathcal{Q}_2(\tau))}}"{description}, from=1-1, to=1-4]
	\arrow["{\pounds_1}"{description}, from=1-1, to=4-2]
	\arrow["{\pounds_2}"{description}, from=1-4, to=4-5]
	\arrow[hook', from=3-1, to=1-1]
	\arrow["{\Phi_{(f, \mathcal{Q}_2(\tau))}}"{description}, from=3-1, to=3-4]
	\arrow[from=3-1, to=4-2]
	\arrow[hook', from=3-4, to=1-4]
	\arrow[from=3-4, to=4-5]
	\arrow["{\Phi_{(f, \textnormal{KSGNS}(\tau)_\otimes)}}"{description}, from=4-2, to=4-5]
\end{tikzcd}
\end{center}
    
\end{proposition}

\begin{proof}\

\noindent By assumption, the two triangular faces and the slanted front face commute. As $\Phi_{(f, \mathcal{Q}_2(\tau))}$ is equivariant (Proposition \ref{Prop: properties of the induce CP map}), it follows that the back square commutes. By a simple diagram chase and using that the other four faces commute, it follows that the bottom face commutes.

\end{proof}

\subsection{Transformations of Hilbert Module Reference Frames}

Given a Hilbert module dynamical system $(G_M, \mathcal{F}_M)$ representing a physical system, we restrict from the category $\textnormal{HMRF}$ to the subcategory whose objects represent reference frames for $(G_M, \mathcal{F}_M)$.

\begin{definition}\

\noindent Let $(G_M, \mathcal{F}_M)$ be a Hilbert module dynamical system. Define the full subcategory $\textnormal{HMRF}(G_M, \mathcal{F}_M)$ of $\textnormal{HMRF}$ as the subcategory where objects $((G, \mathcal{F}), \pounds)$ have the property
    \begin{itemize}
        \item $G=G_M$
        \item $\mathcal{Q}_2\circ\mathcal{F}=\mathcal{F}_M$.
    \end{itemize}
 
\end{definition}

To describe frame transformations for each object in $\textnormal{HMRF}(G_M, \mathcal{F}_M)$, we take, as a first pass, the groupoid/core of $\textnormal{HMRF}(G_M, \mathcal{F}_M)$ which we denote by $\mathcal{G}(G_M, \mathcal{F}_M)$. For each $(f, \tau)\in \mathcal{G}(G_M, \mathcal{F}_M)$, we obtain the commutative diagram
\begin{center}
    \begin{tikzcd}
	{\mathcal{L}(E_{A}\otimes_{\pi_{\phi}}F_{\phi})^{G_M}} &&& {\mathcal{L}(E_{A}\otimes_{\pi_{\phi'}}F_{\phi'})^{G_M}} \\
	\\
	\\
	{\mathcal{L}(E_{A})} &&& {\mathcal{L}(E_{A})}
	\arrow["{\Phi_{(f, \textnormal{KSGNS}(\tau)_\otimes)}}"{description}, from=1-1, to=1-4]
	\arrow["{\pounds_1}"{description}, from=4-1, to=1-1]
	\arrow["{\Phi_{(f, \mathcal{Q}_2(\tau))}}"{description}, from=4-1, to=4-4]
	\arrow["{\pounds_2}"{description}, from=4-4, to=1-4]
\end{tikzcd}
\end{center}
As $(f, \tau)$ is invertible, functoriality implies $\Phi_{(f, \textnormal{KSGNS}(\tau)_\otimes)}$ and $\Phi_{(f, \mathcal{Q}_2(\tau))}$ are invertible strict completely positive maps. 

While morphisms in $\mathcal{G}(G_M, \mathcal{F}_M)$ provide a notion of frame transformation at the level of observables, we want the ability to dualize the diagram above to transform states. Though $\pounds_1$ and $\pounds_2$ are unital completely positive maps and therefore can pullback states, the same cannot be said for the other two strict completely positive maps. As the next lemma shows, the maps $\Phi_{(f, \textnormal{KSGNS}(\tau)_\otimes)}$ and $\Phi_{(f, \mathcal{Q}_2(\tau))}$ must be unital to pullback states.

\begin{lemma}\

\noindent Let $A$ and $B$ be unital $C^*$-algebras, and let $\Phi:A\to B$ be a strict completely positive map. Then $\Phi^*(S(B))\subset S(A)$ if and only if $\Phi$ is unital.

\end{lemma}

\begin{proof}\

\noindent If $\Phi$ is unital, then for each $\phi\in S(B)$, $\Phi^*(\phi)$ is a positive linear functional with $$(\Phi^*\phi)(1)=\phi(\Phi(1))=\phi(1)=1.$$
Hence $\Phi^*\phi\in S(A)$ showing $\Phi^*(S(B))\subset S(A)$. Now if $\Phi^*(S(B))\subset S(A)$, then for all $\phi\in S(B)$, $\phi(\Phi(1))=1=\phi(1)$. As $S(B)$ separates points in $B$, it follows $\Phi(1)=1$ showing $\Phi$ is unital.
    
\end{proof}

A way of getting $\Phi_{(f, \textnormal{KSGNS}(\tau)_\otimes)}$ and $\Phi_{(f, \mathcal{Q}_2(\tau))}$ to be unital is to take $(f, \tau)\in \mathcal{G}(G_M, \mathcal{F}_M)$ for which the data $((\rho, \chi, \eta), (\rho, \xi))$ associated to $\tau$ has the property that $\xi$ and $\Lambda_{(f, \textnormal{KSGNS}(\tau))}$ are unitary maps. As the next theorem shows, this is both necessary and sufficient.

\begin{theorem}\label{Theorem: unitary result 1}\

\noindent Let $(G_M, \mathcal{F}_M)$ be a Hilbert module dynamical system. For each $(f, \tau)\in \mathcal{G}(G_M, \mathcal{F}_M)$, the maps $\Phi_{(f, \textnormal{KSGNS}(\tau)_\otimes)}$ and $\Phi_{(f, \mathcal{Q}_2(\tau))}$ are unital if and only if the data $((\rho, \chi, \eta), (\rho, \xi))$ associated to $\tau$ has the property that $\xi$ and $\Lambda_{(f, \textnormal{KSGNS}(\tau))}$ are unitary isomorphisms.

\end{theorem}

\begin{proof}\

\noindent Suppose $(f, \tau): (G_M, \mathcal{F})\to (G_M, \mathcal{F}')$ is in $\mathcal{G}(G_M, \mathcal{F}_M)$. Let
$$(((A, G_M, \alpha), (B, G_M, \beta), ((E_B, \phi), W)), ((A, G_M, \alpha), E_A, U))$$
be the $G_M$-equivariant coupling to $(G_M, \mathcal{F})$, and let
$$(((A, G_M, \alpha'
), (B', G_M, \beta'), ((E_{B'}, \phi'), W')), ((A, G_M, \alpha'), E_{A}, U'))$$
be the $G_M$-equivariant coupling associated to $(G_M, \mathcal{F}')$. 

If $\xi$ and $\Lambda_{(f, \textnormal{KSNGS}(\tau))}$ are unitary isomorphisms, then, from Proposition \ref{Prop: properties of the induce CP map}, it follows $\Phi_{(f, \textnormal{KSGNS}(\tau)_\otimes)}$ and $\Phi_{(f, \mathcal{Q}_2(\tau))}$ are $*$-algebra isomorphisms and thus unital. Now suppose $\Phi_{(f, \textnormal{KSGNS}(\tau)_\otimes)}$ and $\Phi_{(f, \mathcal{Q}_2(\tau))}$ are unital. First, we show $\xi$ is a unitary isomorphism. As $\Phi_{(f, \mathcal{Q}_2(\tau))}$ is unital, Proposition \ref{Prop: properties of the induce CP map} implies $\xi$ is a coisometry. We claim $\xi$ is bijective. Note, from this, it will follow $\xi$ is a unitary as $\xi$ would be a bijective coisometry.

To show $\xi$ is bijective, let $(\sigma, \zeta)$ be the inverse of $(\rho, \xi)$, then
$$\xi\circ (\zeta\otimes_\rho I)=\iota\circ U\quad\quad \zeta\circ (\xi\otimes_\sigma I)=\iota'\circ U'$$
where $\iota, \iota', U, U'$ are the unitaries in Proposition \ref{Prop: induced map from interior}. From the first equality, we see $\xi$ is surjective. 

To see $\xi$ is injective, we first show that for all non-zero $x\in E_{A}\otimes_\rho A$, there exists $a\in A$ such that $x\dot{\otimes}a\in (E_{A}\otimes_\rho A)\otimes_{\sigma}A$ is non-zero. Fix a non-zero $x\in E_A\otimes_\rho A$. For the sake of contradiction, suppose $x\dot{\otimes}a$ is zero for all non-zero $a\in A$. Therefore
$$0=\langle x\dot{\otimes}a, x\dot{\otimes}a\rangle=a^*\sigma(\langle x, x\rangle)a$$
for all $a\in A$. Taking $a=\sqrt{\sigma(\langle x, x\rangle)}$ implies $\sigma(\langle x, x\rangle)^2=0$. Hence $\sigma(\langle x, x\rangle)=0$. As $\sigma:A\to A$ is a $*$-isomorphism, it follows $\langle x, x\rangle=0$. Therefore $x=0$ which is a contradiction.

Now suppose $x\in \ker(\xi)$. For the sake of contradiction, suppose $x\neq 0$. By our prior claim, there exists a non-zero $a\in A$ for which $x\dot{\otimes} a\in (E_{A}\otimes_\rho A)\otimes_\sigma A$ is non-zero. By definition of $\xi\otimes_\sigma I$, we have 
$$(\xi\otimes_\sigma I)(x\dot{\otimes}a)=\xi(x)\dot{\otimes} a=0$$
 Therefore $x\dot{\otimes}a\in \ker(\iota'\circ U')=\{0\}$ so that $x\dot{\otimes}a=0$ which is a contradiction. Hence $x=0$ showing $\xi$ is injective.

Now we show $\Lambda_{(f, \text{KSGNS}(\tau))}$ is a unitary isomorphism. First, following a similar argument in the last three paragraphs, it follows $\Lambda_{(f, \text{KSGNS}(\tau))}$ is a bijection. As $\Phi_{(f, \textnormal{KSGNS}(\tau)_\otimes)}$ is unital, Proposition \ref{Prop: properties of the induce CP map} implies $\Lambda_{(f, \tau_\otimes)}$ is a coisometry. Therefore $\Lambda_{(f, \tau_\otimes)}$ is a unitary map.

\end{proof}

\begin{corollary}\label{Cor: properties of the induced maps from FT}\

\noindent Let $(G_M, \mathcal{F}_M)$ be a Hilbert module dynamical system. For each $(f, \tau)\in \mathcal{G}(G_M, \mathcal{F}_M)$, the following are equivalent:
\begin{enumerate}
    \item  $\Phi_{(f, \textnormal{KSGNS}(\tau)_\otimes)}$ and $\Phi_{(f, \mathcal{Q}_2(\tau))}$ are $*$-algebra isomorphisms.
    \item $\Phi_{(f, \textnormal{KSGNS}(\tau)_\otimes)}$ and $\Phi_{(f, \mathcal{Q}_2(\tau))}$ are unital.
    \item The data $((\rho, \chi, \eta), (\rho, \xi))$ associated to $\tau$ has the property that $\xi$ and $\Lambda_{(f, \textnormal{KSNGS}(\tau))}$ are unitary isomorphisms.
\end{enumerate}

\end{corollary}

\begin{definition}\

\noindent Let $(G_M, \mathcal{F}_M)$ be a Hilbert module dynamical system.
\begin{itemize}
    \item  A morphism $(f, \tau)\in \mathcal{G}(G_M, \mathcal{F}_M)$ is a unitary frame transformation if the data $((\rho, \chi, \eta), (\rho, \xi))$ associated to $\tau$ has the property that $\xi$ and $\Lambda_{(f, \textnormal{KSGNS}(\tau))}$ are unitary isomorphisms. 
    \item Define the unitary transformation groupoid of $(G_M, \mathcal{F}_M)$ as the subcategory $\mathcal{U}(G_M, \mathcal{F}_M)$ of $\mathcal{G}(G_M, \mathcal{F}_M)$ for which morphisms are unitary frame transformation.
\end{itemize}

\end{definition}

\begin{proposition}\

\noindent  Let $(G_M, \mathcal{F}_M)$ be a Hilbert module dynamical system. Then $\mathcal{U}(G_M, \mathcal{F}_M)$ is a sub-groupoid of $\mathcal{G}(G_M, \mathcal{F}_M)$.

\end{proposition}

As the next result shows, we can substantially improve the prior theorem from requiring $\Lambda_{(f, \tau)}$ being a unitary map to $\eta$ being a unitary map provided we know the Hilbert module associated to $\mathcal{F}_M(\textnormal{pt})$ is full.

\begin{theorem}\label{Theorem: unitary result 2}\

\noindent Let $(G_M, \mathcal{F}_M)$ be a Hilbert module dynamical system for which the Hilbert module associated to $\mathcal{F}_M(\textnormal{pt})$ is full. For each $(f, \tau)\in \mathcal{G}(G_M, \mathcal{F}_M)$, the maps $\Phi_{(f, \textnormal{KSGNS}(\tau)_\otimes)}$ and $\Phi_{(f, \mathcal{Q}_2(\tau))}$ are unital if and only if the data $((\rho, \chi, \eta), (\rho, \xi))$ associated to $\tau$ has the property that $\xi$ and $\eta$ are unitary isomorphisms.

\end{theorem}

\begin{proof}\

\noindent Suppose $(f, \tau): (G_M, \mathcal{F})\to (G_M, \mathcal{F}')$. Let
$$(((A, G_M, \alpha), (B, G_M, \beta), ((E_B, \phi), W)), ((A, G_M, \alpha), E_A, U))$$
be the $G_M$-equivariant coupling to $(G_M, \mathcal{F})$, and let
$$(((A, G_M, \alpha'
), (B', G_M, \beta'), ((E_{B'}, \phi'), W')), ((A, G_M, \alpha'), E_{A}, U'))$$
be the $G_M$-equivariant coupling associated to $(G_M, \mathcal{F}')$. Let $((\rho, \chi, \eta), (\rho, \xi))$ be the data associated to $\tau$.

If $\xi$ and $\eta$ are unitary, then $\xi$ and $\Lambda_{(f, \textnormal{KSGNS}(\tau))}$ are unitary isomorphisms. Thus, by the prior Theorem, $\Phi_{(f, \textnormal{KSGNS}(\tau)_\otimes)}$ and $\Phi_{(f, \mathcal{Q}_2(\tau))}$ are unital. Now suppose $\Phi_{(f, \textnormal{KSGNS}(\tau)_\otimes)}$ and $\Phi_{(f, \mathcal{Q}_2(\tau))}$ are unital, then we know $\xi$ and $\Lambda_{(f, \textnormal{KSGNS}(\tau))}$ are unitary isomorphisms. The data associated to $\textnormal{KSGNS}(\tau)$ is given by $((\rho, \chi, \tilde{\eta}\circ V^{-1}), (\rho, \xi))$ where 
$$\tilde{\eta}\circ V^{-1}:F_\phi\otimes_\chi B\to E_{B'}$$
is the induced adjointable map from Theorem $4.7$ in \cite{brady2026functorialityksgnsconstructionintertwiners}. Recall, 
$$\Lambda_{(f, \textnormal{KSGNS}(\tau))}=(\xi\otimes_{\pi_{\phi'}}I)\circ \sigma^*\circ (I\otimes(\tilde{\eta}\circ V^{-1}))\circ\mu.$$
Therefore, as $\xi\otimes_{\pi_{\pi'}}I, \sigma,$ and $\mu$ are unitary maps, it follows $I\otimes(\tilde{\eta}\circ V^{-1})$ is a unitary map.

We claim that $E_A$ being a full module enables us to conclude $\tilde{\eta}$ is a unitary map from $I\otimes(\tilde{\eta}\circ V^{-1})$ being a unitary map. To prove this, we first show that if $y\in F_\phi\otimes_\chi B'$ such that
$$0=\langle y, \pi_{\phi}(\langle x,x \rangle)y\rangle$$
for all $x\in E_A$, then $y=0$. If the equality holds, then, as $\langle E, E\rangle$ is dense in $A$, it follows
$$0=\langle y, \pi_\phi(a)y\rangle$$
for all $a\in A$. Let $(a_\lambda)_{\lambda\in\Lambda}$ be an approximate unit for $A$, then $(\pi_\phi(a_\lambda))_{\lambda\in\Lambda}$ converges strictly to the identity map. Therefore
$$0=\lim\limits_{\lambda\to\infty}\langle y, \pi_\phi(a_\lambda)y\rangle=\langle y, y\rangle$$
which implies $y=0$. From this claim, it follows that for each non-zero $y\in F_\phi\otimes_{\chi}B'$, there exists a non-zero $x\in E_A$ such that $x\dot{\otimes}y\in E_A\otimes_{\pi_\phi\otimes_{\chi}I}(F_\phi\otimes_{\chi}B')$ is non-zero. Additionally, by a similar argument, we have that for each non-zero $y\in F_{\phi'}$ there exists a non-zero $x\in E_A$ for which $x\dot{\otimes}y\in E_A\otimes_{\pi_{\phi'}\circ\rho}F_{\phi'}$ is non-zero.

Now, suppose for the sake of contradiction, $\tilde{\eta}\circ V^{-1}$ is not a unitary map, then either
$$(\tilde{\eta}\circ V^{-1})^*(\tilde{\eta}\circ V^{-1})\neq I\quad\text{ or }\quad(\tilde{\eta}\circ V^{-1})(\tilde{\eta}\circ V^{-1})^*\neq I.$$
If $(\tilde{\eta}\circ V^{-1})^*(\tilde{\eta}\circ V^{-1})\neq I$, then there exists a non-zero $y\in F_\phi\otimes_{\chi}B'$ for which
$$(\tilde{\eta}\circ V^{-1})^*(\tilde{\eta}\circ V^{-1})(y)-y\neq 0.$$
Let $x\in E_A$ for which $x\dot{\otimes}((\tilde{\eta}\circ V^{-1})^*(\tilde{\eta}\circ V^{-1})(y)-y)\neq 0$. However, as 
$$x\dot{\otimes}((\tilde{\eta}\circ V^{-1})^*(\tilde{\eta}\circ V^{-1})(y)-y)=x\dot{\otimes}(\tilde{\eta}\circ V^{-1})^*(\tilde{\eta}\circ V^{-1})(y)-x\dot{\otimes}y=x\dot{\otimes}y-x\dot{\otimes}y=0,$$
we have a contradiction. Thus $(\tilde{\eta}\circ V^{-1})^*(\tilde{\eta}\circ V^{-1})=I$. By a similar argument, we have $(\tilde{\eta}\circ V^{-1})(\tilde{\eta}\circ V^{-1})^*=I$. Hence $\tilde{\eta}\circ V^{-1}$ is a unitary map. As $V$ is unitary, it follows $\tilde{\eta}$ is a unitary map.

Now, we claim $\tilde{\eta}$ being a unitary map implies $\eta$ is a unitary map. Recall, the map $\tilde{\eta}$ on simple tensors of $F_{\phi}$ is given by $\tilde{\eta}(a\dot{\otimes}y)=\rho(a)\dot{\otimes}\eta(y)$. As $\phi\otimes_{\chi}I:A\to \mathcal{L}(E_B\otimes_\chi B')$ is a non-degenerate completely positive map, it follows that if  $y\in E_B\otimes_\chi B'$ such that
$$0=\langle y, (\phi\otimes_\chi I)(a^*a)y\rangle$$
for all $a\in A$, then $y=0$. Thus, for each non-zero $y\in E_B\otimes_{\chi}B'$, there exists a non-zero $a\in A$ for which $a\dot{\otimes}y\in A\otimes_{\phi\otimes_\chi I}(E_B\otimes_\chi B')$ is non-zero. Similarly, for each non-zero $y\in E_{B'}$, there exists $a\in A$ such that $a\dot{\otimes}y\in A\otimes_{\phi_2} E_{B'}$ is non-zero. Using both of these last two properties, a similar argument as that in the last paragraph implies $\eta$ is a unitary map.

\end{proof}

\begin{corollary}\label{Cor: construction unitary frame transformations}\

\noindent Let $(G_M, \mathcal{F}_M)$ be a Hilbert module dynamical system for which the Hilbert module associated to $\mathcal{F}_M(\textnormal{pt})$ is full. Let $((G_M, \mathcal{F}_1), \pounds_1)$ and $((G_M, \mathcal{F}_2), \pounds_2)$ be objects in  $\mathcal{U}(G_M, \mathcal{F}_M)$ with associated $G_M$-equivariant couplings
$$(((A, G_M, \alpha), (B_i, G_M, \beta_i), ((E_{B_i}, \phi_i), W_i)), ((A, G_M, \alpha), E_A, U))$$
Then every unitary frame transformation
$$((G_M, \mathcal{F}_1), \pounds_1)\to ((G_M, \mathcal{F}_2), \pounds_2)$$
determines and is determined by the data
\begin{enumerate}
    \item a group automorphism $f:G_M\to G_M$
    \item a $*$-algebra isomorphism $\rho:A\to A$
    \item a $*$-algebra isomorphism $\chi:B_1\to B_2$
    \item a unitary map $\xi:E_{A}\otimes_{\rho}A\to E_{A}$
    \item a unitary map $\eta:E_{B_1}\otimes_{\textnormal{inc}\circ\chi}B_2\to E_{B_2}$
\end{enumerate}
and the data satisfies the following conditions:
\begin{enumerate}
    \item for all $a\in A$, $\phi_2(\rho(a))\circ \eta=\eta\circ (\phi_1(a)\otimes_\rho I)$
    \item for all $g\in G_M$, $\alpha_{2, f(g)}\circ\rho=\rho\circ\alpha_{1, g}$
    \item for all $g\in G_M$, $\beta_{2, f(g)}\circ \chi=\chi\circ \beta_{1, g}$
    \item for all $g\in G_M$, $U_{f(g)}\circ \xi=\xi\circ (U_{g}\otimes_\rho \alpha_{1, g})$.
    \item  for all $g\in G_M$, $W_{2, f(g)}\circ \eta=\eta\circ (W_{1, g}\otimes_\chi \beta_{1, g})$.
    \item $\Phi_{(f, \textnormal{KSGNS}(\tau)_\otimes)}\circ\pounds_1=\pounds_2\circ \Phi_{(f, \mathcal{Q}_2(\tau))}$ where $(f, \tau):(G_M, \mathcal{F}_1)\to (G_M, \mathcal{F}_2)$ is the morphism associated to the data from Corollary \ref{Cor: building morphism in EqCpl}.
\end{enumerate}
    
\end{corollary}

\begin{proof}\

\noindent First, suppose we have a unitary frame transformation $(f, \tau)$. Let $((\rho, \chi',\eta), (\rho, \xi))$ be the data associated to $\tau$. From Theorem \ref{Theorem: unitary result 2}, we know $\eta$ and $\xi$ are unitary maps. Furthermore, as $(f, \tau)$ is invertible, $f:G_M\to G_M$ is a group automorphism of $G_M$, and both $\rho:A\to A$ and $\chi:B_1\to M(B_2)$ are invertible morphisms. Thus $\rho$ is a $*$-algebra isomorphism. As $\chi':B_1\to M(B_2)$ is invertible with the the property $\chi(B_1)\subset B_2$, we can uniquely write $\chi'=\textnormal{inc}\circ\chi$ for $\chi:B_1\to B_2$ a $*$-algebra isomorphism. Thus we obtain the data from $(f, \tau)$. Using Corollary \ref{Cor: building morphism in EqCpl} and the definition of a morphism in $\textnormal{HMRF}$, we know the data satisfies conditions $(1)-(6)$.

Now assume we have the data satisfying the five conditions. From conditions $(1)-(5)$, we can apply Corollary \ref{Cor: building morphism in EqCpl} to obtain a morphism
$$(f, \tau):(G_M, \mathcal{F}_1)\to (G_M, \mathcal{F}_2)$$ 
for which the associated data to $\tau$ is given by $((\rho, \textnormal{inc}\circ\chi, \eta), (\rho, \xi))$. From condition $(6)$, this morphism defines a morphism 
$$(f, \tau):((G_M, \mathcal{F}_1), \pounds_1)\to ((G_M, \mathcal{F}_2), \pounds_2)$$
in $\textnormal{HMRF}$. Since the data associated to $\tau$ is all invertible and $f$ is invertible, it follows $(f, \tau)$ is invertible. Thus $(f, \tau)$ is a morphism in $\mathcal{G}(G_M, \mathcal{F}_M)$. From Theorem \ref{Theorem: unitary result 2}, this is in fact a unitary frame transformation.

It is clear that the two constructions are inverses of each other.

\end{proof}

\subsection{Frame Transformations for a Classical Indistinguishable Two State System}

In our first example, we characterize the frame transformations when the main system and reference system are given by a classical indistinguishable two state system. Let $X=\{x_1, x_2\}$ be the state space of the main system, and let $Y=\{y_1, y_2\}$ be the state space of the reference system. We equip both spaces with the discrete topology and the unique transitive $\mathbb{Z}/2\mathbb{Z}$ action. Denote the actions on $X$ and $Y$, respectively, by $\tilde{\alpha}$ and $\tilde{\beta}$. Denote the corresponding actions on $C(X)$ and $C(Y)$, respectively, by $\alpha$ and $\beta$.

With respect to the actions on $X$ and $Y$, every $\mathbb{Z}/2\mathbb{Z}$ invariant weakly continuous Markov kernel $Y\times \textnormal{Bor}(X)\to [0, 1]$ is of the form
$$\kappa_p(y_j, B)=\begin{cases}
    (1-p)\delta_{x_1}(B)+p\delta_{x_2}(B) & j=1\\
    p\delta_{x_1}(B)+(1-p)\delta_{x_2}(B) & j=2\\
\end{cases}$$
where $p\in [0, 1]$. Letting $\phi_p:C(X)\to C(Y)$ be the induced $\mathbb{Z}/2\mathbb{Z}$ equivariant unital completely positive map from $\kappa_p$ (see Proposition 3.12 in \cite{brady2026hilbertmoduleapproachclassical}), we obtain a $\mathbb{Z}/2\mathbb{Z}$ equivariant coupling for each $p\in [0, 1]$:
$$(((C(X), \mathbb{Z}/2\mathbb{Z}, \alpha),(C(Y), \mathbb{Z}/2\mathbb{Z}, \beta), ((C(y), \phi_p), \beta)), ((C(X), \mathbb{Z}/2\mathbb{Z}, \alpha), C(X), \alpha)).$$
We denote the functorial representation of this equivariant coupling as $(\mathbb{Z}/2\mathbb{Z}, \mathcal{F}_p)$. 

Following Example 3.26 in \cite{brady2026hilbertmoduleapproachclassical}, the reference frame system is given by
$$\mathcal{L}(\Gamma_{\kappa_p})^{\mathbb{Z}/2\mathbb{Z}}\cong \begin{cases}
    M_2(\mathbb{C}) & 0<p<1\\
    \mathbb{C} & p=0, 1
\end{cases}.$$
By identifying $\mathbb{C}$ as the closed $*$-subalgebra given by the scalar matrices in $M_2(\mathbb{C})$, a relativization map for the equivariant coupling is given by a choice of completely positive map
$$\pounds:C(X)\to M_2(\mathbb{C})$$
which takes values on scalar matrices when $p=0$ or $p=1$. As $C(X)$ has a basis given by $\chi_1:=\chi_{\{x_1\}}, \chi_2:=\chi_{\{x_2\}}$ with $\chi_1+\chi_2=1$, it follows the map $\pounds$ is completely determined by the value of $\pounds(\chi_1)\in [0, I]\subset M_2(\mathbb{C})$.

\begin{proposition}\label{Prop: info1}\

\noindent If $(f, \tau):((\mathbb{Z}/2\mathbb{Z}, \mathcal{F}_p), \pounds_1)\to ((\mathbb{Z}/2\mathbb{Z}, \mathcal{F}_q), \pounds_2)$
is a unitary frame transformation with associated data $((\rho, \chi, \eta), (\rho, \xi))$ to $\tau$, then
\begin{enumerate}
    \item $f=\textnormal{id}$
    \item $\chi=\beta_{[0]}$ or $\chi=\beta_{[1]}$
    \item There exists a unique $\theta\in [0, 2\pi)$ such that on simple tensors of $C(X)\otimes_\rho C(X)$
    $$\xi(f\dot\otimes g)=e^{i\theta}f\rho(g).$$
    \item There exists a unique $\theta\in [0, 2\pi)$ such that on simple tensors of $C(Y)\otimes_\chi C(Y)$
    $$\eta(f\dot\otimes g)=e^{i\theta}f\chi(g).$$
    \item Either $p=q$ or $1=p+q$.
    \item If $p=q$ with $p\neq \frac{1}{2}$, then $\rho=\alpha_{[0]}$.
    \item If $1=p+q$ with $p\neq \frac{1}{2}$, then $\rho=\alpha_{[1]}$.
    \item If $p=\frac{1}{2}=q$, then $\rho=\alpha_{[0]}$ or $\rho=\alpha_{[1]}$.
\end{enumerate}

\end{proposition}

\begin{proof}\

\noindent First, as $(f, \tau)$ is invertible, it follows $$f:\mathbb{Z}/2\mathbb{Z}\to \mathbb{Z}/2\mathbb{Z}\quad\quad \rho:C(X)\to C(X)\quad\quad \chi:C(Y)\to C(Y)$$
are all automorphisms. Therefore $f=\textnormal{id}$. As $\textnormal{Aut}_{*\textnormal{-alg}}(C(X))\cong \mathbb{Z}/2\mathbb{Z}\cong \textnormal{Aut}_{*\textnormal{-alg}}(C(X))$, it follows that $\rho=\alpha_{[0]}$ or $\rho=\alpha_{[1]}$, and $\chi=\beta_{[0]}$ or $\chi=\beta_{[1]}$. This proves the claims in $(1)-(2)$.

We observe that we have the unitary isomorphism
$$\zeta:C(X)\otimes_\rho C(X)\to C(X)\quad\quad \zeta(f\dot{\otimes}g)=f\rho(g).$$ Additionally, we observe that all unitary maps $C(X)\to C(X)$ are of the form
$$L_h:C(X)\to C(X)\quad\quad L_h(f)=hf$$
where $h\in C(X)$ is a unitary element i.e. $h(X)\subset\mathbb{S}^1$. Thus, on simple tensors,  $\xi$ is uniquely given by $\xi(f\dot{\otimes}g)=hf\rho(g)$. Denote $\xi$ by $\xi_h$. 

Using Theorem \ref{Theorem: characterization of morphisms}, we have
\begin{enumerate}
    \item $\alpha_{[1]}\circ\rho=\rho\circ\alpha_{[1]}$
    \item $\beta_{[1]}\circ\chi=\chi\circ\beta_{[1]}$
    \item $\xi\circ (\alpha_{[1]}\otimes_\rho \alpha_{ [1]})=\alpha_{[1]}\circ \xi$
    \item $\eta\circ (\beta_{[1]}\otimes_\chi \beta_{[1]})=\beta_{[1]}\circ \eta$
\end{enumerate}
On simple tensors of $C(X)\otimes_{\rho}C(X)$, we have
\begin{align*}
    (\xi_h\circ (\alpha_{[1]}\otimes_\rho \alpha_{[1]}))(f\dot{\otimes}g)&=h\alpha_{[1]}(f)\rho(\alpha_{[1]}(g))=h\alpha_{[1]}(f)\alpha_{[1]}(\rho(g))\\
    &=\alpha_{[1]}(\alpha_{[1]}(h)f\rho(g))=(\alpha_{[1]}\circ \xi_{\alpha_{[1]}(k)})(f\dot{\otimes}g)
\end{align*}
Thus
$$\xi_h\circ (\alpha_{[1]}\otimes_\rho\alpha_{[1]})=\alpha_{[1]}\circ\xi_{\alpha_{[1]}(h)}.$$
For condition $(3)$ to hold, we require $\alpha_{[1]}(h)=h$; that is, $h$ is a constant function. As $h$ is a constant function, the function is uniquely determined by $\theta\in [0, 2\pi)$ such that $h(x_k)=e^{i\theta}$. Hence, there exists a unique $\theta\in [0, 2\pi)$ such that
$$\xi(f\dot{\otimes}g)=e^{i\theta}f\rho(g).$$
By a similar argument, there exists a unique $\theta\in [0, 2\pi)$ such that
$$\eta(f\dot{\otimes}g)=e^{i\theta}f\chi(g)$$
This prove the claims in $(3)-(4)$.

To prove $(5)-(8)$, we utilize $(\rho, \chi, \eta)$ is a morphism in $\textnormal{PosCor}$. In particular, we have for all $f\in C(X)$
$$\eta\circ (\phi_p(f)\otimes_\chi I)=\phi_q(\rho(f))\circ\eta.$$
On simple tensors, this equality implies
$$e^{i\theta}\phi_p(f)g\chi(h)=e^{i\theta}\phi_q(\rho(f))g\chi(h).$$
Thus, for all $f\in C(X)$, $\phi_p(f)=\phi_q(\chi(f))$.
Note,
$$\phi_p(f)(y_i)=\begin{cases}
    (1-p)f(x_1)+pf(x_2) & i=1\\
    pf(x_1)+(1-p)f(x_2) & i=2
\end{cases}$$
and
$$\phi_q(\chi(f))(y_i)=\begin{cases}
    (1-q)\chi(f)(x_1)+q\chi(f)(x_2) & i=1\\
    q\chi(f)(x_1)+(1-q)\chi(f)(x_2) & i=2
\end{cases}$$
Taking $f=\chi_1\in C(X)$, we have
$$\phi_p(\chi_1)=(1-p)\chi_1+p\chi_2\in C(Y),$$
while
$$\phi_q(\rho(\chi_1))=\begin{cases}
    (1-q)\chi_1+q\chi_2 & \rho=\alpha_{[0]}\\
    q\chi_1+(1-q)\chi_2 & \rho=\alpha_{[1]}
\end{cases}$$
In either case, we see that either $p=q$ or $1=p+q$. Additionally, we see that 
\begin{enumerate}
    \item when $p=q$ with $p\neq \frac{1}{2}$, $\rho=\alpha_{[0]}$,
    \item when $1=p+q$ with $p\neq \frac{1}{2}$, $\rho=\alpha_{[1]}$,
    \item when $p=\frac{1}{2}$, $\rho=\alpha_{[0]}$ or $\rho=\alpha_{[1]}$.
\end{enumerate}
This proves $(5)-(8)$.

\end{proof}

\begin{remark}\

\noindent We denote $\xi$ by $\xi_\theta$ where $\theta\in[0, 2\pi)$ such that $(3)$ holds. We use a similar notation for $\eta$.

\end{remark}

\begin{proposition}\label{Prop: info2}\

\noindent If $(\textnormal{id}, \tau):((\mathbb{Z}/2\mathbb{Z}, \mathcal{F}_p), \pounds_1)\to ((\mathbb{Z}/2\mathbb{Z}, \mathcal{F}_q), \pounds_2)$ is a unitary frame transformation with associated data $((\rho, \chi, \eta_{\theta_1}), (\rho, \xi_{\theta_2}))$ to $\tau$, then
\begin{enumerate}
    \item $\Phi_{(\textnormal{id}, \mathcal{Q}_2(\tau))}=1_{C(X)}$.
    \item Under the identification of the invariant relational observables with $M_2(\mathbb{C})$, the matrix formulation of the unitary map $\Lambda_{(\textnormal{id}, \textnormal{KSGNS}(\tau)_\otimes)}$ is given by the matrix
    $$\begin{cases}
        e^{i(\theta_1+\theta_2)}\begin{bmatrix}
            0 & 1\\
            1 & 0
        \end{bmatrix} & \textnormal{if }\chi=\alpha_{[1]}\\
        e^{i(\theta_1+\theta_2)}I & \textnormal{if }\chi=\alpha_{[0]} \textnormal{ or }p=0, 1
    \end{cases}$$
\end{enumerate}
  
\end{proposition}

\begin{proof}\

\noindent We begin by proving $(1)$. Recall, for each $f\in C(X)$, we view $f$ as an adjointable operator on $C(X)$ given by multiplication by $f$, denoted $M_f$. Therefore
$$\Phi_{(\textnormal{id}, \mathcal{Q}(\tau))}(M_f)=\xi_{\theta_2}(M_f\otimes_\rho I)\xi_{\theta_2}^*.$$
Given $g\in C(X)$, we have
\begin{align*}
    (\xi_{\theta_2}(M_f\otimes_\rho I)\xi_{\theta_2}^*)(g)&=\xi(e^{-i\theta_2}fg\dot{\otimes}1)=fg=M_f(g)
\end{align*}
Since this holds for all $g\in C(X)$, we have $\Phi_{(\textnormal{id}, \mathcal{Q}(\tau))}(M_f)=M_f$. Hence $\Phi_{(\textnormal{id}, \mathcal{Q}(\tau))}$ is the identity map proving $(1)$.

Now we prove $(2)$. To prove this claim, we begin by unpacking the construction $\Phi_{(\textnormal{id}, \textnormal{KSGNS}(\tau)_{\otimes})}$. Let $(F_{\phi_p}, \pi_{\phi_p}, V_{\phi_p})$ and $(F_{\phi_q}, \pi_{\phi_q}, V_{\phi_q})$ denote the triples obtained from applying the KSGNS construction to $(C(Y), \phi_p)$ and $(C(Y), \phi_q)$, respectively. Then the unitary map
$$\Lambda_{(\textnormal{id}, \textnormal{KSGNS}(\tau)_\otimes)}:(C(X)\otimes_{\pi_{\phi_p}}F_{\phi_p})\otimes_\chi C(Y)\to C(X)\otimes_{\pi_{\phi_q}} F_{\phi_q}$$
is given on simple tensors by
$$\Lambda_{(\textnormal{id}, \textnormal{KSGNS}(\tau)_\otimes)}((f\dot{\otimes}(g\dot{\otimes}h))\dot{\otimes}k)=e^{i(\theta_1+\theta_2)}f\dot{\otimes}(\rho(g)\dot{\otimes}h\chi(k)).$$
We have the unitary isomorphisms
$$C(X)\otimes_{\pi_{\phi_p}}F_{\phi_{p}}\to F_{\phi_p}\quad\quad f\otimes (g\dot{\otimes}h)\mapsto (fg)\dot{\otimes}h,$$
$$F_{\phi_p}\to \Gamma_{\kappa_p}\quad\quad \sum\limits_{i=1}^n\pi_{\phi_p}(f_i)V_{\phi_p}(g_i)\mapsto \sum\limits_{i=1}^n\pi_{\kappa_p}(f_i)V_{\kappa_p}(g_i),$$
and
$$F_{\phi_q}\to \Gamma_{\kappa_q}\quad\quad \sum\limits_{i=1}^n\pi_{\phi_q}(f_i)V_{\phi_q}(g_i)\mapsto \sum\limits_{i=1}^n\pi_{\kappa_q}(f_i)V_{\kappa_q}(g_i).$$
Under these isomorphisms, we obtain the unitary map
$$\Lambda:\Gamma_{\kappa_p}\otimes_{\chi}C(Y)\to \Gamma_{\kappa_q}\quad\quad \Lambda(\pi_{\kappa_p}(f)V_{\kappa_p}(g)\dot{\otimes}h)=e^{i(\theta_1+\theta_2)}\pi_{\kappa_q}(\rho(f))V_{\kappa_q}(g\chi(h))$$
Thus, on $\Gamma_{\kappa_p}$ and $\Gamma_{\kappa_q}$, the map $\Phi_{(\textnormal{id}, \textnormal{KSGNS}(\tau)_{\otimes})}$ is given by
$$\Phi_{(\textnormal{id}, \textnormal{KSGNS}(\tau)_{\otimes})}:\mathcal{L}(\Gamma_{\kappa_p})^{\mathbb{Z}/2\mathbb{Z}}\to \mathcal{L}(\Gamma_{\kappa_q})^{\mathbb{Z}/2\mathbb{Z}}\quad\quad \Phi_{(\textnormal{id}, \textnormal{KSGNS}(\tau)_{\otimes})}(T)=\Lambda(T\otimes_\rho I)\Lambda^*$$

In the case $0<p<1$ (therefore $0<q<1$), we have $\Gamma_{\kappa_p}=C(X\times Y)=\Gamma_{\kappa_q}$. With respect to the basis $\chi_{\{x_1\}}, \chi_{\{x_2\}}$, we have
$$C(X\times Y)\cong C(Y)\oplus C(Y).$$
Therefore 
$$\mathcal{L}(\Gamma_{\kappa_p})\cong M_2(C(Y))\cong \mathcal{L}(\Gamma_{\kappa_q}).$$
As $C(Y)\cong \mathbb{C}\oplus \mathbb{C}$, we have
$$\mathcal{L}(\Gamma_{\kappa_p})\cong M_2(\mathbb{C})\oplus M_2(\mathbb{C})\mathbb{C}\cong \mathcal{L}(\Gamma_{\kappa_q})$$
with
$$\mathcal{L}(\Gamma_{\kappa_p})^{\mathbb{Z}/2\mathbb{Z}}\cong M_2(\mathbb{C})\cong \mathcal{L}(\Gamma_{\kappa_p})^{\mathbb{Z}/2\mathbb{Z}}.$$
Under this identification, we have the matrix formulation of $\Lambda$ is given by
$$\Lambda=\begin{cases}
        e^{i(\theta_1+\theta_2)}\begin{bmatrix}
            0 & 1\\
            1 & 0
        \end{bmatrix} & \textnormal{if }\rho=\alpha_{[1]}\\
        e^{i(\theta_1+\theta_2)}I & \textnormal{if }\rho=\alpha_{[0]}
\end{cases}$$
as claimed.

In the case $p=0, 1$ (therefore $q=0, 1$), we have $\Gamma_{\kappa_p}\cong C(Y)\cong \Gamma_{\kappa_q}$ via the basis $\pi_{\kappa_p}(\chi_{\{x_1\}}+\chi_{\{x_2\}})V_{\kappa_p}1$ and $\pi_{\kappa_q}(\chi_{\{x_1\}}+\chi_{\{x_2\}})V_{\kappa_q}1$, respectively. Therefore 
$$\mathcal{L}(\Gamma_{\kappa_p})\cong M_q(C(Y))\cong \mathcal{L}(\Gamma_{\kappa_q}).$$
As $C(Y)\cong \mathbb{C}\oplus \mathbb{C}$, we have
$$\mathcal{L}(\Gamma_{\kappa_p})\cong M_1(\mathbb{C})\oplus M_1(\mathbb{C})\cong \mathcal{L}(\Gamma_{\kappa_q})$$
with
$$\mathcal{L}(\Gamma_{\kappa_p})^{\mathbb{Z}/2\mathbb{Z}}\cong \mathbb{C}\cong \mathcal{L}(\Gamma_{\kappa_p})^{\mathbb{Z}/2\mathbb{Z}}.$$
Under this identification, we have the matrix formulation of $\Lambda$ is given by $\Lambda=[e^{i(\theta_1+\theta_2)}]$. By identifying $\mathbb{C}$ as the diagonal $*$-subalgebra of $M_2(\mathbb{C})$, the matrix formulation of $\Lambda$ is given by
$$\Lambda=e^{i(\theta_1+\theta_2)}I$$
as claimed.

\end{proof}

\begin{theorem}\

\noindent Let $((\mathbb{Z}/2\mathbb{Z}, \mathcal{F}_p), \pounds_1)$ and $((\mathbb{Z}/2\mathbb{Z}, \mathcal{F}_q), \pounds_2)$ be Hilbert module reference frames. Assume the relational invariant observables are identified with $M_2(\mathbb{C})$.
\begin{enumerate}
    \item If $p=q$ with $p\neq \frac{1}{2}$, then there exists a unitary frame transformation 
    $$(\textnormal{id}, \tau):((\mathbb{Z}/2\mathbb{Z}, \mathcal{F}_p), \pounds_1)\to ((\mathbb{Z}/2\mathbb{Z}, \mathcal{F}_q), \pounds_2)$$
    if and only if $\pounds_1=\pounds_2$. When $\pounds_1=\pounds_2$, the data associated to $\tau$ is of the form
    $$((\alpha_{[0]}, \beta_{[0]}, \eta_{\theta_1}), (\alpha_{[0]}, \xi_{\theta_2}))\quad\textnormal{ or }\quad ((\alpha_{[0]}, \beta_{[1]}, \eta_{\theta_1}), (\alpha_{[0]}, \xi_{\theta_2}))$$
    where $\theta_1, \theta_2\in [0, 2\pi)$.
    \item If $1=p+q$ with $p=0, 1$, then there exists a unitary frame transformation 
    $$(\textnormal{id}, \tau):((\mathbb{Z}/2\mathbb{Z}, \mathcal{F}_p), \pounds_1)\to ((\mathbb{Z}/2\mathbb{Z}, \mathcal{F}_q), \pounds_2)$$
    if and only if $\pounds_1=\pounds_2$. When $\pounds_1=\pounds_2$, the data associated to $\tau$ is of the form
    $$((\alpha_{[0]}, \beta_{[0]}, \eta_{\theta_1}), (\alpha_{[0]}, \xi_{\theta_2}))\quad\textnormal{ or }\quad ((\alpha_{[0]}, \beta_{[1]}, \eta_{\theta_1}), (\alpha_{[0]}, \xi_{\theta_2}))$$
    where $\theta_1, \theta_2\in [0, 2\pi)$.
    \item If $1=p+q$ with $p\neq \frac{1}{2}, 0, 1$, then there exists a unitary frame transformation 
    $$(\textnormal{id}, \tau):((\mathbb{Z}/2\mathbb{Z}, \mathcal{F}_p), \pounds_1)\to ((\mathbb{Z}/2\mathbb{Z}, \mathcal{F}_q), \pounds_2)$$
    if and only if 
    $$\begin{bmatrix}
        0 & 1\\
        1 & 0
    \end{bmatrix}\pounds_1\begin{bmatrix}
        0 & 1\\
        1 & 0
    \end{bmatrix}=\pounds_2.$$
    In this case, the data associated to $\tau$ is of the form
    $$((\alpha_{[1]}, \beta_{[0]}, \eta_{\theta_1}), (\alpha_{[1]}, \xi_{\theta_2}))\quad\textnormal{ or }\quad ((\alpha_{[1]}, \beta_{[1]}, \eta_{\theta_1}), (\alpha_{[1]}, \xi_{\theta_2}))$$
    where $\theta_1, \theta_2\in [0, 2\pi)$.
    \item If $p=\frac{1}{2}=q$, then there exists a unitary frame transformation 
    $$(\textnormal{id}, \tau):((\mathbb{Z}/2\mathbb{Z}, \mathcal{F}_p), \pounds_1)\to ((\mathbb{Z}/2\mathbb{Z}, \mathcal{F}_q), \pounds_2)$$
    if and only if $\pounds_1=\pounds_2$ or  
    $$\begin{bmatrix}
        0 & 1\\
        1 & 0
    \end{bmatrix}\pounds_1\begin{bmatrix}
        0 & 1\\
        1 & 0
    \end{bmatrix}=\pounds_2.$$ 
    In the case $\pounds_1=\pounds_2$, the data associated to $\tau$ is of the form
    $$((\alpha_{[0]}, \beta_{[0]}, \eta_{\theta_1}), (\alpha_{[0]}, \xi_{\theta_2}))\quad\textnormal{ or }\quad ((\alpha_{[0]}, \beta_{[1]}, \eta_{\theta_1}), (\alpha_{[0]}, \xi_{\theta_2}))$$
    where $\theta_1, \theta_2\in [0, 2\pi)$. In the case
     $$\begin{bmatrix}
        0 & 1\\
        1 & 0
    \end{bmatrix}\pounds_1\begin{bmatrix}
        0 & 1\\
        1 & 0
    \end{bmatrix}=\pounds_2,$$
    the data associated to $\tau$ is of the form
    $$((\alpha_{[1]}, \beta_{[0]}, \eta_{\theta_1}), (\alpha_{[1]}, \xi_{\theta_2}))\quad\textnormal{ or }\quad ((\alpha_{[1]}, \beta_{[1]}, \eta_{\theta_1}), (\alpha_{[1]}, \xi_{\theta_2}))$$
    where $\theta_1, \theta_2\in [0, 2\pi)$.
\end{enumerate}

\end{theorem}

\begin{proof}\

\noindent The forward direction of each case as well as the identification of the data associated to $\tau$ directly follows from Proposition \ref{Prop: info1} and Proposition \ref{Prop: info2}. For the reverse directions, \ref{Prop: info2} and Corollary \ref{Cor: construction unitary frame transformations} implies the data defines unitary frame transformations.

\end{proof}

\begin{remark}\

\noindent For a large number of cases, the automorphism group of unitary frame transformations for $((\mathbb{Z}/2\mathbb{Z}, \mathcal{F}_p), \pounds)$ is given by $\mathbb{Z}/2\mathbb{Z}\times\mathbb{T}^2$. However, when $p=\frac{1}{2}$ and $$\pounds(\chi_1)=a\begin{bmatrix}
    1 & 1\\
    1 & 1
\end{bmatrix}$$
for $0\leq a\leq \frac{1}{2}$, both conditions in case (4) of the theorem are satisfied. As a result, the group of unitary frame transformation of this Hilbert module reference frame is $\mathbb{Z}/2\mathbb{Z}\times\mathbb{Z}/2\mathbb{Z}\times\mathbb{T}^2$.

\end{remark}

\subsection{Frame Transformations for Hilbert Spaces}

We now consider the example when both the main system and reference system have Hilbert modules given by Hilbert spaces. Let $H_S$ be the Hilbert space for the main system, and let $H_R$ be the Hilbert space for the reference system. Let $G$ be a group acting on $H_S$ and $H_R$ via the unitary representations $U_S$ and $U_R$, respectively. Letting $\phi_{R}$ denote the unique unital $*$-algebra map $\phi_R:\mathbb{C}\to \mathcal{B}(H_R)$, and letting $\alpha$ denote the unique group homomorphism $G\to \textnormal{Aut}_{*\textnormal{-alg}}(\mathbb{C})$, we obtain the G-equivariant coupling
$$(((\mathbb{C}, G, \alpha), (\mathbb{C}, G, \beta), ((H_R, \phi_R), U_R)), ((\mathbb{C}, G, \alpha), H_S, U_S))$$
Denote the functorial representation of this equivariant coupling as $(G, \mathcal{F}_{(H_R, U)})$. From Proposition 4.7 of \cite{brady2026hilbertmoduleapproachclassical}, the invariant relational observables can be identified with $\mathcal{B}(H_S\hat{\otimes}H_R)^G$. 

\begin{proposition}\

\noindent Let $((G, \mathcal{F}_{(H_{R_1}, U_{R_1})}), \pounds_1)$ and $((G, \mathcal{F}_{(H_{R_2}, U_{R_2})}), \pounds_2)$ be Hilbert module reference frames. Let
$$(f, \tau):((G, \mathcal{F}_{(H_{R_1}, U_{R_1})}), \pounds_1)\to ((G, \mathcal{F}_{(H_{R_2}, U_{R_2})}), \pounds_2)$$
be a unitary frame transformation with data $((\rho, \chi, \eta), (\rho, \xi))$ associated to $\tau$.
\begin{enumerate}
    \item $\rho=1_{\mathbb{C}}=\chi$.
    \item There exists a unique unitary map $U:H_S\to H_S$ such that $\xi(x\dot{\otimes}\lambda)=\lambda U(x)$.
    \item There exists a unique unitary map $V:H_{R_1}\to H_{R_2}$ such that $\eta(x\dot{\otimes}\lambda)=\lambda V(x).$
    \item For each $T\in \mathcal{B}(H_S)$, $\Phi_{(f, \mathcal{Q}_2(\tau))}(T)=UTU^*$
    \item Under the identification of the relational invariant observables with $\mathcal{B}(H_S\hat{\otimes}H_{R_1})^G$ and $\mathcal{B}(H_{S}\hat{\otimes}H_{R_2})^G$, for all $T\in \mathcal{B}(H_S\hat{\otimes}H_{R_1})^G$, $\Phi_{(f, \textnormal{KSGNS}(\tau)_\otimes)}(T)=(U\hat{\otimes}V)T(U\hat{\otimes}V)^*$.
\end{enumerate}

\end{proposition}

\begin{proof}\

\noindent Since $\textnormal{Aut}_{*\textnormal{-alg}}(\mathbb{C})=\{1_\mathbb{C}\}$, it follows $\rho=1_\mathbb{C}=\chi$. To prove $(2)$, we first observe the unitary isomorphism
$$H_S\otimes\mathbb{C}\to H_S\quad\quad x\otimes \lambda\mapsto \lambda x$$
Thus the unitary map $\xi:H_S\otimes\mathbb{C}\to H_S$ determines and is determined by a unitary map $U:H_S\to H_S$ where $\eta(x\otimes\lambda)=\lambda U(x)$. A similar argument proves $(3)$. To prove $(4)$, fix $T\in \mathcal{B}(H_S)$, then $\Phi_{(f, \mathcal{Q}_2(\tau))}(T)=\xi(T\otimes_{1_{\mathbb{C}}} I)\xi^*$. Given $x\in H_S$, we have
\begin{align*}
    \xi(T\otimes_{1_{\mathbb{C}}} I)\xi^*x&=\xi(T\otimes_{1_{\mathbb{C}}}I)(U^*(x)\otimes 1)=\xi((TU^*)(x)\otimes 1)=UTU^*x
\end{align*}
Therefore $\Phi_{(f, \mathcal{Q}_2(\tau))}(T)=UTU^*$.

Now we prove $(5)$. Let $(F_{\phi, R_1}, \pi_{\phi, R_1}, V_{\phi, R_1})$ and $(F_{\phi, R_2}, \pi_{\phi, R_2}, V_{\phi, R_2})$ be the triples obtain from applying the KSGNS construction to $(H_{R_1}, \phi_{R_1})$ and $(H_{R_2}, \phi_{R_2})$. Thus the unitary
$$\Lambda_{(f, \textnormal{KSGNS}(\tau)_\otimes)}:(H_S\otimes_{\pi_{\phi, R_1}} F_{\phi, R_1})\otimes\mathbb{C}\to H_S\otimes_{\pi_{\phi, R_2}}F_{\pi_{\phi, R_2}}$$
on simple tensors is given by
$$\Lambda_{(f, \textnormal{KSGNS}(\tau)_\otimes)}((x\dot{\otimes}(\lambda\dot{\otimes}y))\otimes\mu)=\mu(V(x)\dot{\otimes}(\lambda\dot{\otimes} U(y)))$$
We observe that we have the unitary isomorphisms
$$F_{\phi, R_1}\to H_{R_1}\quad\quad \lambda\dot{\otimes}x\mapsto \lambda x$$
$$F_{\phi, R_2}\to H_{R_1}\quad\quad \lambda\dot{\otimes}x\mapsto \lambda x$$
and
$$(H_S\hat{\otimes} H_{R_1})\otimes\mathbb{C}\to H_S\hat{\otimes}H_{R_1}\quad\quad (x\otimes y)\otimes\lambda\mapsto \lambda (x\otimes y)$$
Under these isomorphisms, we unitary map $\Lambda_{(f, \textnormal{KSGNS}(\tau)_\otimes)}$ uniquely determines the unitary map
$$H_S\hat{\otimes}H_{R_1}\to H_S\hat{\otimes}H_{R_2}\quad\quad x\otimes y\mapsto U(X)\otimes V(y)=(U\hat{\otimes}V)(x\otimes y)$$
Hence, the map $\Phi_{(f, \textnormal{KSGNS}(\tau)_\otimes)}$ corresponds the the $*$-algebra isomorphism
$$\mathcal{B}(H_S\hat{\otimes}H_{R_1})^G\to \mathcal{B}(H_S\hat{\otimes}H_{R_2})^G\quad\quad T\mapsto (U\hat{\otimes}V)T(U\hat{\otimes}V)^*$$

\end{proof}

\begin{remark}\

\noindent Since $\xi$ and $\eta$ are uniquely determined by the unitary maps $U$ and $V$, respectively, we denote the maps as $\xi_U$ and $\eta_V$, respectively.
    
\end{remark}

\begin{theorem}\

\noindent Let $((G, \mathcal{F}_{(H_{R_1}, U_{R_1})}), \pounds_1)$ and $((G, \mathcal{F}_{(H_{R_2}, U_{R_2})}), \pounds_2)$ be Hilbert module reference frames. A group automorphism $f:G\to G$ and unitary maps $U:H_S\to H_S$ and $V:H_{R_1}\to H_{R_2}$ determine a unitary frame transformation
$$(f, \tau):((G, \mathcal{F}_{(H_{R_1}, U_{R_1})}), \pounds_1)\to ((G, \mathcal{F}_{(H_{R_2}, U_{R_2})}), \pounds_2)$$
for which the associated data to $\tau$ is $((1_{\mathbb{C}}, 1_{\mathbb{C}}, \eta_V), (1_{\mathbb{C}}, \xi_U))$ if and only if
\begin{enumerate}
    \item for all $g\in G$, $U\circ U_{S, g}=U_{S, f(g)}\circ U$.
    \item for all $g\in G$, $V\circ U_{R_1 ,g}=U_{R_2, f(g)}\circ V$.
    \item under the identification of the relational invariant observables with $\mathcal{B}(H_S\hat{\otimes}H_{R_1})^G$ and $\mathcal{B}(H_{S}\hat{\otimes}H_{R_2})^G$, for all $T\in \mathcal{B}(H_S)$, $\pounds_2(UTU^*)=(U\hat{\otimes}V)\pounds_1(T)(U\hat{\otimes}V)^*$.
\end{enumerate}

\end{theorem}

\begin{proof}\

\noindent First suppose the unitary maps $U$ and $V$ determine the unitary frame transformation $(f, \tau)$ for which the associated data to $\tau$ is $((1_{\mathbb{C}}, 1_{\mathbb{C}}, \eta_V), (1_{\mathbb{C}}, \xi_U))$. Using Theorem \ref{Theorem: characterization of morphisms}, we know that for all $g\in G$
\begin{enumerate}
    \item $\xi_U\circ (U_{S, g}\otimes_{1_\mathbb{C}} I)=U_{S, f(g)}\circ \xi_U$
    \item $\eta_V\circ (U_{R_1, g}\otimes_{1_\mathbb{C}} I)=U_{R_2, f(g)}\circ \eta_V$
\end{enumerate}
Note, under the unitary isomorphisms
$$H_{S}\otimes_{1_\mathbb{C}}\mathbb{C}\xrightarrow{x\dot{\otimes}\lambda\mapsto \lambda x} H_{S}\quad\quad H_{R_1}\otimes_{1_\mathbb{C}}\mathbb{C}\xrightarrow{x\dot{\otimes}\lambda\mapsto \lambda x} H_{R_2}$$
the conditions are equivalent to
\begin{enumerate}
    \item $U\circ U_{S, g}=U_{S, f(g)}\circ U$
    \item $V\circ U_{R_1, g}=U_{R_2, f(g)}\circ V$
\end{enumerate}
Using the prior proposition, we also have for all $T\in \mathcal{B}(H_R)$, 
$$\pounds_2(UTU^*)=(U\hat{\otimes}V)\pounds_1(T)(U\hat{\otimes}V)^*.$$
 
Now we prove the reverse direction. From the first two properties, it follows that 
\begin{enumerate}
    \item $\xi_U\circ (U_{S, g}\otimes_{1_\mathbb{C}} I)=U_{S, f(g)}\circ \xi_U$
    \item $\eta_V\circ (U_{R_1, g}\otimes_{1_\mathbb{C}} I)=U_{R_2, f(g)}\circ \eta_V$
\end{enumerate}
Thus, using Corollary \ref{Cor: construction unitary frame transformations}, the data $((1_{\mathbb{C}}, 1_{\mathbb{C}}, \eta_V), (1_{\mathbb{C}}, \xi_U))$ determines a unitary frame transformation 
$$(f, \tau):(G, \mathcal{F}_{(H_{R_1}, U_{R_1})}))\to (G, \mathcal{F}_{(H_{R_2}, U_{R_2})})$$
 such that the data associated to $\tau$ is $((1_{\mathbb{C}}, 1_{\mathbb{C}}, \eta_V), (1_{\mathbb{C}}, \xi_U))$.
    
\end{proof}

\begin{example}\

\noindent Let $H_S=L^2(\mathbb{R})$ equipped with the strongly continuous unitary representation from $\mathbb{R}$ given by
$$U_{S, t}=e^{-itH_q}\quad\quad H_q=-\frac{\hbar^2}{2m}\frac{d^2}{dx^2}$$
Let $H_{R_1}=L^2(\mathbb{R})=H_{R_2}$. Note, we view $H_{R_1}$ as the position space of a reference particle, while $H_{R_2}$ represents the momentum space of a reference particle. As a result, we equip $H_{R_1}$ with the strongly continuous unitary representation of $\mathbb{R}$ given by $U_{R_1, t}=U_{S, t}$. We equip $H_{R_2}$ with the strongly continuous unitary representation of $\mathbb{R}$ given by
$$U_{R_2, t}=e^{-itH_p}\quad\quad H_p=\frac{p^2}{2m}.$$ 
Thus, we obtain the equivariant couplings
$$(\mathbb{R}, \mathcal{F}_{(H_{R_1}, U_{R_1})})\quad\quad (\mathbb{R}, \mathcal{F}_{(H_{R_2}, U_{R_2})}).$$

Fix $\sigma\in (0, \infty)$. Define the semi-spectral measure
$$E_\sigma:\textnormal{Bor}(\mathbb{R})\to \mathcal{B}(L^2(\mathbb{R}))\quad\quad (E_\sigma(B)\psi)(x)=\frac{1}{\sqrt{2\pi \sigma}}\int_{B}\exp\left(\frac{-(x-y)^2}{2\sigma}\right) \;dy \psi(x)$$
Observe $E_\sigma$ is covariant under translation on $\textnormal{Bor}(\mathbb{R})$ and the strongly continuous unitary representation $U_{R_1}$. Therefore, we obtain a covariant system $(\mathbb{R}, U_{R_1}, E)$. Let 
$$\yen_{1}:\mathcal{B}(H_{S})\to \mathcal{B}(H_{S}\hat{\otimes} H_{R_1})^{\mathbb{R}}\quad\quad\yen_1(T)=\int_{\mathbb{R}}U_{S, t}TU_{s, t}^*\hat{\otimes}dE_\sigma(dt)$$
be the induced relativization map from \cite{Loveridge_2018}. Thus, we obtain a Hilbert module reference frame
$((\mathbb{R}, \mathcal{F}_{(H_{R_1}, U_{R_1})}), \yen_1)$. 

Let $\mathcal{F}:L^2(\mathbb{R})\to L^2(\mathbb{R})$ denote the Fourier transform. As $U_{R_2, t}\circ \mathcal{F}=\mathcal{F}\circ U_{R_1, t}$ for all $t\in\mathbb{R}$, the semi-spectral measure
$$E_\sigma':\textnormal{Bor}(\mathbb{R})\to \mathcal{B}(L^2(\mathbb{R}))\quad\quad E'_\sigma(B)=\mathcal{F}E_\sigma(B)\mathcal{F}^*$$
defines a covariant system $(\mathbb{R}, U_{R_2}, E)$ with respect to the strongly continuous unitary representation $U_{R_2}$.  Let $$\yen_{2}:\mathcal{B}(H_{S})\to \mathcal{B}(H_{S}\hat{\otimes} H_{R_2})^{\mathbb{R}}\quad\quad \yen_2(T)=\int_{\mathbb{R}}U_{S, t}TU_{s, t}^*\hat{\otimes}dE'_\sigma(dt)$$
be the induced relativization map from \cite{Loveridge_2018}. Thus, we obtain a Hilbert module reference frame
$((\mathbb{R}, \mathcal{F}_{(H_{R_2}, U_{R_2})}), \yen_2)$. 

We claim for all $T\in \mathcal{B}(H_S)$, $(I\hat{\otimes}\mathcal{F})\yen_1(T)(I\hat{\otimes}\mathcal{F}^*)=\yen_2(T).$ Indeed, we have
\begin{align*}
    (I\hat{\otimes}\mathcal{F})\yen_1(T)(I\hat{\otimes}\mathcal{F}^*)&=\int_{\mathbb{R}}U_{S, t}TU_{S, t}^*\otimes \mathcal{F}dE_\sigma(dt) \mathcal{F}^*\\
    &=\int_{\mathbb{R}}U_{S, t}TU_{S, t}^*\otimes dE'_\sigma(dt)=\yen_2(T)
\end{align*}
Therefore, using the last theorem, the unitary maps $I:H_S\to H_S$ and $\mathcal{F}:H_{R_1}\to H_{R_2}$ along with the identity group homomorphism $\mathbb{R}\to \mathbb{R}$ define a unitary frame transformation
$$((\mathbb{R}, \mathcal{F}_{(H_{R_1}, U_{R_1})}), \yen_1)\to ((\mathbb{R}, \mathcal{F}_{(H_{R_2}, U_{R_2})}), \yen_2).$$ 

\end{example}

\bibliographystyle{amsalpha} 
\bibliography{bibliography} 

\end{document}